\documentclass[9pt,lineno]{elife}

\usepackage{lipsum} 
\usepackage[version=4]{mhchem}
\usepackage{siunitx}
\usepackage{float}
\DeclareSIUnit\Molar{M}

\graphicspath{ {./Figures/} }

\nolinenumbers
\microtypesetup{expansion=false}
\title{Bayesian inference-driven model parameterization and model selection for 2CLJQ fluid models}

\author[1]{Owen C. Madin}
\author[2]{Simon Boothroyd}
\author[3]{Richard A. Messerly}
\author[4]{John D. Chodera}
\author[5]{Josh Fass}
\author[1]{Michael R. Shirts}

\affil[1]{Department of Chemical \& Biological Engineering, University of Colorado Boulder, Boulder, CO 80309}

\affil[2]{Boothroyd Scientific Consulting Ltd., 71-75 Shelton Street, London, Greater London, United Kingdom, WC2H 9JQ}

\affil[3]{Theoretical Division, Los Alamos National Laboratory, Los Alamos, NM 87545}

\affil[4]{Computational \& Systems Biology Program, Sloan Kettering Institute, Memorial Sloan Kettering Cancer Center, New York, NY 10065}

\affil[5]{Tri-Institutional PhD Program in Computational Biology and Medicine, Weill Cornell Graduate School of Medical Sciences, New York, NY 10065, USA. Current address: Relay Therapeutics, Cambridge, MA 02139, USA}

\corr{michael.shirts@colorado.edu}{MRS}

\begin{document}
\maketitle

\begin{abstract}
A high level of physical detail in a molecular model improves its ability to perform high accuracy simulations, but can also significantly affect its complexity and computational cost.  In some situations, it is worthwhile to add additional complexity to a model to capture properties of interest; in others, additional complexity is unnecessary and can make simulations computationally infeasible.   
In this work we demonstrate the use of Bayes factors for molecular model selection, using Monte Carlo sampling techniques to evaluate the evidence for different levels of complexity in the two-centered Lennard-Jones + quadrupole (2CLJQ) fluid model. Examining three levels of nested model complexity, we demonstrate that the use of variable quadrupole and bond length parameters in this model framework is justified only sometimes.  We also explore the effect of the Bayesian prior distribution on the Bayes factors, as well as ways to propose meaningful prior distributions. This Bayesian Markov Chain Monte Carlo (MCMC) process is enabled by the use of analytical surrogate models that accurately approximate the physical properties of interest. This work paves the way for further atomistic model selection work via Bayesian inference and surrogate modeling.  
\end{abstract}

\section{Introduction}
\label{section:Introduction}

Parameterization of molecular force fields is a long-standing problem for the classical molecular simulation community~\cite{dauber-osguthorpeBiomolecularForceFields2019,haglerForceFieldDevelopment2019,rinikerFixedChargeAtomisticForce2018}, as the choice of models and their respective parameters are critical for achieving quantitative accuracy in modeling of molecular interactions. The treatment of electrostatic and dispersion-repulsion interactions, commonly referred to as \emph{non-bonded interactions}, is a difficult parameterization task because these interactions are described by relatively simple models that are crude approximations of the underlying electronic behavior, requiring mapping electronic probability distributions onto pairwise interactions between a small number of points. Dispersion-repulsion interactions, commonly modeled with a Lennard-Jones (LJ) potential are especially challenging to derive~\cite{monticelliBiomolecularSimulationsMethods2013} as values of the Lennard-Jones parameters are not directly extractable from  quantum mechanics (QM) calculations.
Most force fields train these parameters against macroscopic condensed phase physical properties, though there are some attempts to obtain them with reference to QM calculations~\cite{kantonenDataDrivenMappingGasPhase2020,tkatchenkoAccurateMolecularVan2009}.

Although most commonly employed classical force fields use electrostatic point charges and pairwise LJ potentials~\cite{wangDevelopmentTestingGeneral2004a,banksIntegratedModelingProgram2005}, other models for these interactions exist.  Atomic multipoles~\cite{ponderCurrentStatusAMOEBA2010,jiaoCalculationProteinLigand2008}, Drude oscillators~\cite{lemkulPolarizableForceField2017}, and fluctuating charges~\cite{patelCHARMMFluctuatingCharge2004} for electrostatic interactions and alternate functional forms~\cite{dauber-osguthorpeBiomolecularForceFields2019,messerlyMie16Force2019,chiuCoarseGrainedModelBased2010,halgrenMerckMolecularForce1996} for dispersion-repulsion interactions have been proposed over the years.  Even considering only LJ potentials, many different choices for atom typing~\cite{schauperlDatadrivenAnalysisNumber2020,boulangerOptimizedLennardJonesParameters2018a} and combination rules~\cite{halgrenRepresentationVanWaals1992a,waldmanNewCombiningRules1993} exist.
Choosing a potential energy function with a more complex functional form can improve agreement with experiment, however, increasing the complexity may also make the parameterization more difficult due to an increased likelihood of overfitting~\cite{harrisonReviewForceFields2018,frohlkingEmpiricalForceFields2020}.

This creates two distinct problems in designing force fields: (1) selecting among discrete choices of models and (2) optimizing continuous choice of parameters within models.  It is common, although often expensive, to compare the quantitative performance of different parameter sets within a single model, but it is difficult to quantitatively compare fitness between models.

A common statistical formalism across for comparing the overall fitness of discrete models is Bayesian inference~\cite{vontoussaintBayesianInferencePhysics2011a}, which allows users to evaluate models in a way that automatically penalizes unnecessary complexity~\cite{mackayInformationTheoryInferencea}.
By integrating over the marginal distribution of the parameters one can calculate \emph{Bayes factors} which are interpreted as odds ratios between the separate models~\cite{kassBayesFactors1995}.
The Bayesian framework combines specific information on a model's ability to reproduce target data, with more general prior knowledge based on data collected with previous parameter sets, physical constraints or invariances, or chemical intuition. In this way, the prior distribution generalizes and influences the comparison between models. In any parameterization process, the influence of the prior knowledge about the system is always present; using a Bayesian approach makes the influence of any prior information on the parameters explicit.  Bayesian inference has previously been used to incorporate uncertainty quantification~\cite{duttaBayesianCalibrationForcefields2018,angelikopoulosBayesianUncertaintyQuantification2012,farrellBayesianFrameworkAdaptive2015,wus.HierarchicalBayesianFramework2016} and model selection~\cite{bacalladoBayesianComparisonMarkov2009,farrellBayesianFrameworkAdaptive2015} into molecular models.

To compute Bayes factors, samples must be drawn from the model posterior distributions which involves comparing model outputs to macroscopic observables.  Typically this is done through a MCMC sampling scheme in the parameter space, since posteriors are generally non-analytical. 
Once sufficient posterior samples are obtained, a range of statistical techniques can be used to estimate the Bayes factors and make quantitative judgements on the relative fitness of models.
One established MCMC technique for cross-model sampling is reversible jump Monte Carlo (RJMC)~\cite{greenReversibleJumpMarkov1995a}, which simultaneously samples over several models and their respective parameter spaces.  This technique can be used to compare model posterior parameter distributions and directly compare the support for each model.
Bayes factors can also be computed with \emph{bridge sampling} methods which compute the probability ratio between a simple analytical reference distribution and the posterior distribution in order to calculate model evidences. This method can be enhanced by using intermediate probability distributions to bridge the gap between the reference distribution and the posterior, increasing the overlap between distributions being compared and thus improving the accuracy of the calculations~\cite{nealAnnealedImportanceSampling1998}.

These statistical approaches for calculating Bayes factors have close analogues in statistical physics techniques used to calculate free energy differences in molecular simulations. Specifically, Bayes factors are analogous to the ratio of partition coefficients of two models, but integrated over the \emph{parameter space} of each model, rather than over atomic coordinates. Bridge sampling techniques are analogous to free energy perturbation and reweighting methods~\cite{mengSIMULATINGRATIOSNORMALIZING1996,gelmanSimulatingNormalizingConstants1998}, and RJMC is analogous to expanded ensemble techniques or Hamilton replica exchange techniques~\cite{sugitaReplicaexchangeMulticanonicalAlgorithm2000}, where the multiple ensembles are different models rather than different potential functions or temperatures.

Estimating physical observables with molecular simulations is often computationally expensive, especially if some molecular degrees of freedom are slow. To perform MCMC sampling in parameter space, observables must be calculated many times.
To avoid this cost and efficiently sample the posterior distribution, surrogate models~\cite{sidkyMolecularLatentSpace2020,kadupitiyaMachineLearningSurrogates2020} can be used to cheaply approximate the response surface of the observables with respect to the force field parameters.

To demonstrate the Bayesian model comparison strategy for atomistic models, we have selected a problem for which a surrogate model is already well-defined. The two-center Lennard-Jones + quadrupole (2CLJQ) model ~\cite{vrabecSetMolecularModels2001} is a simple 4-parameter fluid model that serves as a useful test bed for this problem. Analytical surrogate models exist for predicting saturated densities ($\rho_l$), saturated vapor pressure ($P_{sat}$), and surface tension ($\gamma$) for a number of small nonpolar fluids.~\cite{stollComprehensiveStudyVapourliquid2009a,werthSurfaceTensionTwo2015a}. A previous parameterization study~\cite{stobenerParametrizationTwocenterLennardJones2016} of the model demonstrated the potential to optimize the 2CLJQ model using surrogate models. In this study we use quantitative evidence from Bayes factors to examine whether including an adjustable bond length and/or quadrupole parameter are justified in this model. 

\section{Methods}
\label{section:Methods}
\subsection{Molecular Model}
\label{subsection:Molecular_Model}
\begin{figure}[H]
    \centering
        \includegraphics[width=0.4\textwidth]{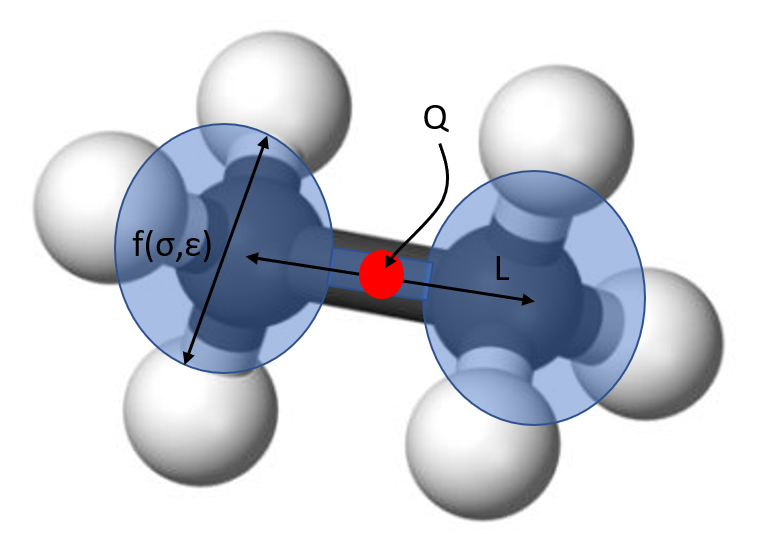}
    \caption{\textbf{The 2CLJQ model consists of two LJ sites parameterized by $\sigma$ and $\epsilon$, a bond length $L$, and a quadrupole interaction strength parameter $Q$.} 
    In this example, an ethane ($\mathrm{C_2H_6}$) molecule is parameterized as two identical sites, each consistent of a carbon atom and three hydrogens.}
    \label{fig:2cljq_model}
\end{figure}

We test our Bayesian model selection strategy on the 2CLJQ fluid model~\cite{vrabecSetMolecularModels2001}, a simple 2-site fluid model that fits diatomic and other similar molecules~\cite{stollSetMolecularModels2003,mollerDeterminationEffectiveIntermolecular1994,cheungPropertiesLiquidNitrogen1976,murthyInteractionSiteModels1981}.  In particular we consider this model for a range of compounds, 4 diatomic ($\mathrm{O_2, N_2, Br_2, F_2}$) and 4 ``diatomic-like'' hydro/fluoro carbon compounds \\ ($\mathrm{C_2H_2}$, $\mathrm{C_2H_4}$, $\mathrm{C_2H_6}$, $\mathrm{C_2F_4}$), which are both well parameterized by the 2CLJQ model~\cite{vrabecSetMolecularModels2001}.  Interactions within the 2CLJQ model are controlled by four parameters, as illustrated in Figure \ref{fig:2cljq_model}: a Lennard-Jones $\sigma$ (nm) and $\epsilon$ (K), a constant bond length $L$ (nm) between the two LJ sites, and a quadrupole interaction strength parameter $Q(\mathrm{D} \cdot \mathrm{nm})$.  The LJ $\sigma$ represents the distance at which the potential energy between two particles is equal to zero; the LJ $\epsilon$ represents the depth of the potential well and the strength of the attraction between two particles. The functional form of the molecular model is similar to the Lennard-Jones potential, but adapted for a 2-center model and with a quadrupole interaction at the model's geometric center~\cite{vrabecSetMolecularModels2001}:

\begin{equation}
    u_{2CLJQ}(\mathbf{r}_{ij}, \mathbf{\omega}_{i}, \mathbf{\omega}_{j}, L, Q) =  u_{2CLJ}(\mathbf{r}_{ij}, \mathbf{\omega}_{i}, \mathbf{\omega}_{j}, L) +  u_{Q}(\mathbf{r}_{ij}, \mathbf{\omega}_{i}, \mathbf{\omega}_{j}, Q)
\end{equation}
\begin{equation}
    u_{2CLJ}(\mathbf{r}_{ij}, \mathbf{\omega}_{i}, \mathbf{\omega}_{j}, L) = \sum_{a=1}^2 \sum_{b=1}^2 4\epsilon \left[\left(\frac{\sigma}{r_{ab}}\right)^{12} - \left(\frac{\sigma}{r_{ab}}\right)^{6} \right]
\end{equation}
\begin{equation}
    u_{Q}(\mathbf{r}_{ij}, \mathbf{\omega}_{i}, \mathbf{\omega}_{j}, Q) = \frac{3}{4}\frac{Q^2}{|\mathbf{r}_{ij}|^5}f(\mathbf{\omega}_{i}, \mathbf{\omega}_{j})
\end{equation}

\subsubsection{Defining models of varying complexity}
To investigate which parameters are justified to reproduce the physical properties of interest, we split the model into three levels of complexity, as shown in Table \ref{tbl:Model Definitions}: united atom (UA), anisotropic united atom (AUA), and anisotropic united atom + quadrupole (AUA+Q).
\begin{table}[h]
    \centering

\begin{tabular}[t]{|c|c|c|c|c|}
\hline
 Model & $\sigma$ (nm) & $\epsilon $ (K) & $L$ (nm) & $Q$ ($\mathrm{D}\cdot \mathrm{nm}$) \\
\hline
  UA & Variable & Variable & Fixed & Fixed (at 0)\\ 
  AUA & Variable & Variable & Variable & Fixed (at 0)\\ 
  AUA+Q & Variable & Variable & Variable & Variable \\
\hline
\end{tabular}

\caption{{\bf Different 2CLJQ models considered in this study.}  
While all models share the same function form, some models have parameters/interactions fixed, while others let them vary.
{\bf UA}: united atom; {\bf AUA}: anisotropic united atom; {\bf AUA+Q}: anisotropic united atom + quadrupole.
}
\label{tbl:Model Definitions}
\end{table}

The models here are nested---for example, the AUA model is the same as the AUA+Q model, but with the quadrupole moment permanently set to zero.  The UA model is a subset of the AUA model, but with a fixed bond length chosen from literature (typically taking an ``experimental'' value)~\cite{johnsonrusselld.NISTComputationalChemistry2018}.  This construction allows us to test whether the variable quadrupole and bond length parameters are useful in reproducing the chosen physical properties.
\subsubsection{Surrogate Models for 2CLJQ output}
Analytical surrogate models, which predict molecular properties as a function of molecular parameters, were developed for saturated liquid density ($\rho_l$) and saturated vapor pressure ($P_{sat}$) by Stoll~\cite{stollComprehensiveStudyVapourliquid2009a}; similar surrogate models for surface tension ($\gamma$) were developed by Werth~\cite{werthSurfaceTensionTwo2015a} by fitting to 2CLJQ simulation results at a variety of parameter and temperature conditions. 
St\"{o}bener~\cite{stobenerParametrizationTwocenterLennardJones2016} previously optimized parameters for the 2CLJQ model by using surrogate models to drive fast optimization and a Pareto optimization approach to choose parameter sets.  Parameter set fitness in this study was based on comparison to data from the Design Institute for Physical Properties (DIPPR) correlations; here we train against NIST experimental data for similar properties instead.  In this work, we expand the idea of extensive parameter sampling through surrogate models to include comparisons between disparate models.

To apply this technique to an arbitrary force field, one will usually need to construct such surrogate models for different properties.  Common techniques to build these surrogate models might include reweighting~\cite{messerlyConfigurationSamplingBasedSurrogateModels2018a}, Gaussian processes~\cite{befortMachineLearningDirected2021}, and machine learning methods~\cite{kadupitiyaMachineLearningSurrogates2020}.
Since the methods needed to construct such a model depend substantially on the property and the parameters of interest, that question is beyond the scope of this study; we focus only on applying Bayesian inference given a surrogate model.
\subsection{Bayesian inference}
\label{subsection:Bayesian_Inference}
Bayesian inference characterizes the fitness of models and parameters by combining specific information about a set of experiments with more general prior information.  The core of Bayesian inference is \emph{Bayes' rule}:
\begin{equation}
    \underbrace{P(\theta | D, M)}_{\mathrm{Posterior}} \propto \underbrace{P(D| \theta, M)}_{\mathrm{Likelihood}} \underbrace{P(\theta | M)}_{\mathrm{Prior}}
\end{equation}
where $\theta$ represents a set of parameters belonging to a model $M$, and $D$ represents a dataset (which can be compared to experimental data) generated by the model $M$ and parameter set $\theta$.
 Bayes factors, computed as the ratio of the model marginal likelihood ($P(D|M)$) between separate models, facilitate comparisons between those models.
\begin{equation}
B_{1/2} = \frac{P(D|M_1)}{P(D|M_2)} = \frac{\int_{\theta_1}P(D|\Theta_1,M_1)P(\theta_1|M_1)\mathrm{d}\theta_1}{\int_{\theta_2}P(D|\theta_2,M_2)P(\theta_2|M_2)\mathrm{d}\theta_2} = \frac{P(M_1|D)}{P(M_2|D)}\frac{P(M_2)}{P(M_1)}
\end{equation} 
In order to compute the model evidence $P(D|M)$, one must integrate the parameter posterior over all parameter space in the model (as shown in eq. 2). For most molecular systems this Bayes factor integral cannot be calculated analytically and must be estimated.
Common interpretations for levels of significance of Bayes Factor evidence were suggested by Kass and Raftery~\cite{kassBayesFactors1995} and are listed, with labels renamed for clarity, in Table \ref{tab:bayes-evidence}. 
\begin{table}[h]
\caption{\label{tab:bayes-evidence}
{\bf The interpretation of Bayes factors for model 0 over model 1, $B_{01}$, due to Kass and Raftery~\cite{kassBayesFactors1995}.} 
Bayes factors provide a quantitative measure of the \emph{model evidence} for choosing one model over another. }
    \centering
\begin{tabular}[t]{c|c|c}
 $\ln{\left(B_{01} \right)}$ & $B_{01}$ & Evidence in favor of model 0\\
\hline
  0 -1 & 1 - 3 & Inconclusive \\ 
  1-3 & 3 - 20 & Significant  \\
  3-5 & 20 - 150 & Strong \\
  $>$ 5 &$>$ 150 & Very Strong
\end{tabular}

\label{tbl:Bayes factors}
\end{table}

The advantage of using Bayesian inference to choose between models is that it captures both model performance and model parsimony, striking a balance between accuracy and generality.
More precisely, more complex models are penalized because their additional complexity allows them to make a wider range of predictions.  Unless this additional complexity makes the predictions uniformly more accurate, many of these predictions will be extraneous to the properties of interest.  The lower the proportion of useful prediction to total predictions, the more a model will be penalized.
\subsection{Construction of Posterior}
Since model evidence is based on the model posterior, one must consider the choice of the prior distributions and likelihood function that form that posterior carefully.
\subsubsection{Priors enable model parsimony}
The use of Bayesian priors allows us to consider model parsimony in the evaluation of models. The prior distribution defines a multi-dimensional probability landscape in parameter space containing all values of the prior to be considered, weighted by our prior estimates of which values are most likely. As the model's complexity grows, so does the volume of this parameter space.  Since priors are normalized probability distributions and must integrate to one, growing the parameter space volume \emph{must} reduce the probability of any specific parameter set.  

It is also important to emphasize that Bayesian model selection chooses the best model \emph{given the parameter space}.  If parameter values that produce good models are excluded from this space, or assigned very low probability, those values will not factor into the calculated Bayes factors.

\begin{figure}[H]
    \centering
        \includegraphics[width=0.6\textwidth]{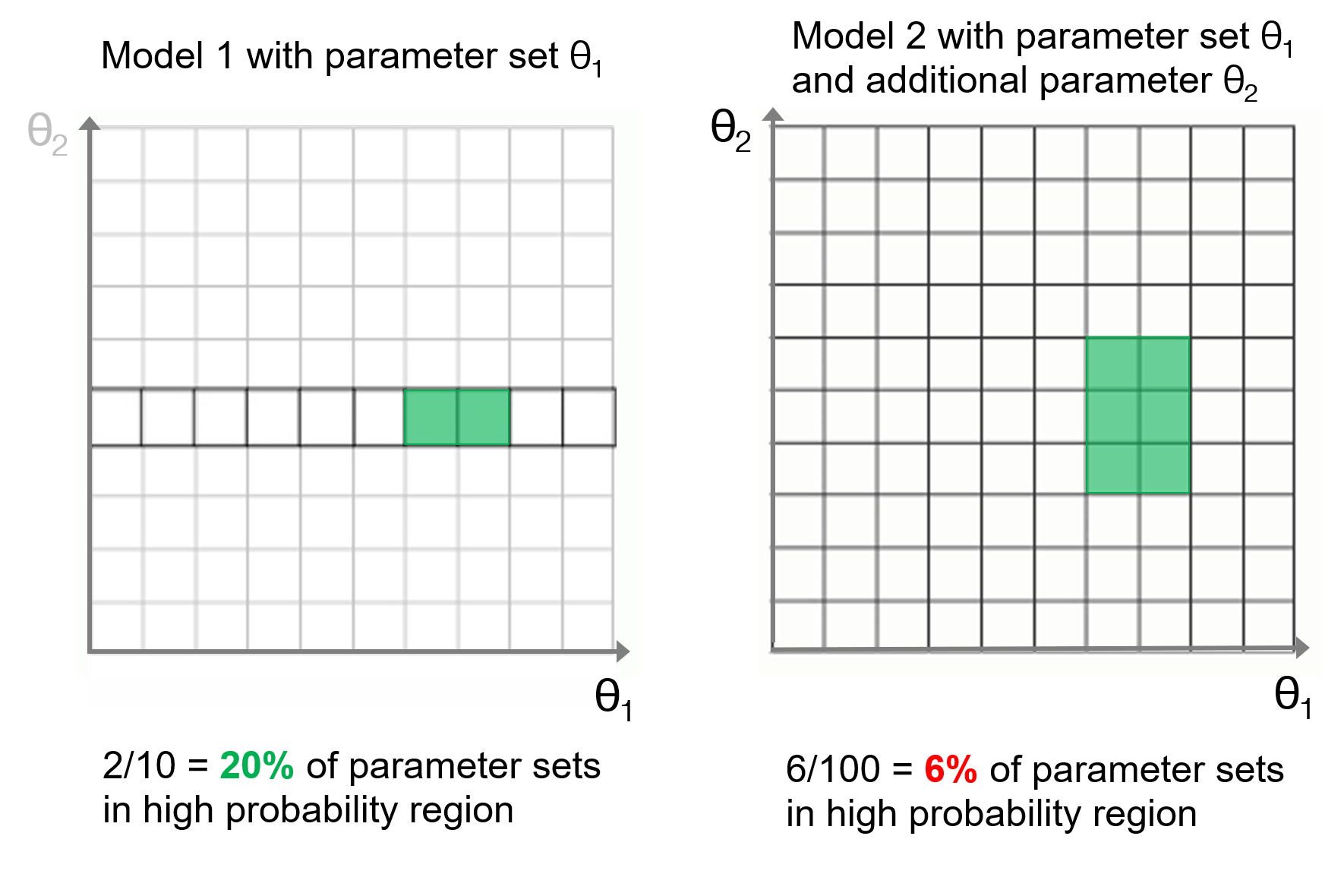}
    \caption{\textbf{ Demonstration of how increased model complexity incurs a penalty.} In the left panel, with the model only having the set $\theta_1$ as variable parameters, a higher proportion of parameter sets are high probability.  When the additional parameter $\theta_2$ is added, even though more high probability parameter sets are available, they represent a lower proportion of the total parameter space available to the model, which incurs a complexity penalty in the Bayesian context.}
    \label{fig:qual_prior}
\end{figure}

In this way, demonstrated in figure \ref{fig:qual_prior} the prior distribution creates a complexity penalty that the model must overcome with increased predictive power in order to justify increased complexity. Setting a prior is crucial because it encodes this penalty, which depends on the amount of initial information about the parameter.

\subsubsection{Choosing priors}
\label{subsubsection:Choosing_priors}
In order to avoid excluding useful parameter space, or including too much extraneous parameter space, we form the prior distribution using a training sample method.  In this method, data is split into a ``training sample'', used to inform the prior, and a ``test sample'' used to calculate the Bayes factor. This method of including partial information in training samples is a relatively simple method for calculating priors that are effective for each model and is established in the statistical literature~\cite{bergerIntrinsicBayesFactor1996,bergerRobustBayesianAnalysis1990}.

In this case, we start from a wide, non-negative uniform distribution, and simulate a posterior distribution (using a likelihood calculated as in section 2.3.4) with training samples of varying amounts of experimental data. We then fit a simple analytical distribution (Gaussian for $\epsilon,\sigma,L$; exponential or gamma for $Q$) to this \emph{training sample posterior}.  This analytical distribution becomes the prior for calculation of Bayes factors under a separate set of experimental data points.    We set the priors using 3 levels of information in the training sample (low, medium, high; 3, 5, 8 data points per property respectively). These data points are selected for each property in the target criteria and are distributed across the available temperature range, so they are not excluding information based on temperature or property.

\subsubsection{Evaluating the likelihood}
\label{subsubsection:Likelihood_Evaluations}
Likelihood evaluation requires a probabilistic model of observations in terms of model parameters (referred to in statistical literature as a "forward model"), as well as error model for comparing those observations with experimental ones. In this case the forward model is composed of the surrogate models for physical properties ($\rho_l$, $P_{sat}$, $\gamma$) as a function of the model parameters ($\sigma, \epsilon, L, Q$) using the functional forms as defined in Stoll et. al~\cite{stollComprehensiveStudyVapourliquid2009a} (density, saturation pressure) and Werth et al.~\cite{werthSurfaceTensionTwo2015a} (surface tension). More details of the surrogate models are available in Supporting Information, section 1.1.

Experimental measurements for the diatomic ($\mathrm{O_2,N_2,Br_2,F_2}$) and hydro/fluorocarbon compounds ($\mathrm{C_2H_2}$, $\mathrm{C_2H_4}$, $\mathrm{C_2H_6}$, $\mathrm{C_2F_4}$) for saturated liquid densities, saturation pressure, and surface tension between 55\% and 95\% of critical temperature (some training sets have slightly different temperature ranges due to data availability, full information available in Supporting Information section 1.3) are taken from the NIST ThermoData Engine database~\cite{frenkelThermoDataEngineTDE2005a}, with uncertainties assigned as a typical class uncertainty for each compound and property.  The use of class uncertainties does not heavily impact the uncertainty as experimental uncertainties are much smaller than correlation uncertainties. 
These surrogate models are used to calculate the likelihood $P(D|\Theta)$ by estimating each physical property at the temperatures corresponding to experimental physical property data points and comparing those values to experimental with an error model.
\subsubsection{Error Model}
\label{subsubsection:Error_Model}
We us a Gaussian error model to compare properties calculated from the surrogate models to corresponding experimental properties.
This error model is chosen based on the assumption that the experimental value $x$ is the ``true'' value and that the surrogate model produces a measurement of that value, $\hat{x}$, with some error.  We assume this error is Gaussian distributed, so the probability of observing a measurement $\hat{x}$ is given by:
\begin{equation}
 P(\hat{x} | x) = \frac{1}{u_{tot}\sqrt{2\pi}}\exp\left( -\frac{1}{2}\left(\frac{x-\hat{x}}{u_{tot}} \right)^2  \right)    
\end{equation}
where the total uncertainty $u_{tot}$ is given by a sum of squares of the average experimental uncertainty $u_{exp}$ and the surrogate model error $u_{surr}$.
\begin{equation}
u_{tot}^2 = u_{exp}^2 + u_{surr}^2
\end{equation}
This surrogate model is an analytical form fitted to several chemical species at a range of thermodynamic conditions, so there is systematic uncertainty when the model is applied to any specific compound.  While the creators of the surrogate models used in this study did not provide analytical correlations for the uncertainty, they provided uncertainty estimates at difference temperature and parameter values.  We used these estimates to assign model uncertainties $u_{surr}$ in our process.  This model is piecewise, depending on the temperature regime (defined as fraction of critical temperature) of the system. Details of this error model are listed in the Supporting Information section 1.2.

The full forms of the likelihood function (eq. 10) and prior distribution (eq. 11) are as follows:
\begin{equation}
    P(D | \Theta, M) =\left( \prod_{k=1}^3 \left( \prod_{i=1}^{n} \frac{1}{u_{tot}\sqrt{2\pi}} \exp \left( -\frac{1}{2}\left( \frac{\Vec{D_i}_k-\mathbf{f(\theta, T)}}{u_{tot}}\right)^2\right) \right)\right).
\end{equation}
The likelihood function evaluates agreement with experiment over a vector of $i$ temperature data points from $k$ different properties. $\mathbf{f(\theta, T)}$ is the output of the analytical surrogate model for the $k^{th}$ physical property at the temperatures corresponding to the data vector $\Vec{D_i}_k$.  It is important to note that for the saturation pressure, the data vector $\Vec{D_i}_{P_{sat}}$ is $\ln{P_{sat}}$ instead of $P_{sat}$ due to the wide range of values at different temperatures. The prior function is defined as:
\begin{equation}
    P(\Theta) = \left( \prod_{j=1}^3\frac{1}{\sigma_j \sqrt{2\pi}} \exp \left( -\frac{1}{2}\left( \frac{\theta_j - \mu_j}{\sigma_j}\right)^2\right) \right) \left( \frac{\beta^{\alpha}}{\Gamma(\alpha)} \theta_4^{\alpha-1}e^{-\beta \theta_4} \right).
\end{equation}
 The first term of the total prior is the prior for ($\sigma=\theta_1, \epsilon=\theta_2, L=\theta_3)$, and the second term is the prior for $Q=\theta_4$, either an exponential ($\alpha=1, \beta$ determined by prior fitting) or gamma prior ($\alpha, \beta$ determined by prior fitting).  Distributions are fit to samples using the SciPy \texttt{distributions} module and the \texttt{fit} functions.  For compounds with quadrupole distributions centered at values larger than zero, gamma priors are chosen; for compounds with quadrupole support near zero, exponential distributions are chosen. For the UA and AUA models, the prior terms for $L,Q$ (UA) or $Q$ (AUA) are set to 1, as the values are fixed. 
Combined, they form the posterior distribution as described in eq. 1.
\subsection{Sampling of Posteriors}
\label{subsection:Sampling}
For any method of calculating Bayes factors obtaining samples from the parameter posterior distribution is essential. To draw samples from complex distributions without simple closed-form distributions Markov Chain Monte Carlo (MCMC) methods are used~\cite{chipmanPracticalImplementationBayesian2001}.  
\subsubsection{MCMC parameter proposals}
\label{subsubsection:MCMC_param_proposals}
Within a particular model, we propose moves using component-wise Metropolis Hastings Monte Carlo~\cite{johnsonComponentWiseMarkovChain2013}, where at each step an individual variable parameter is chosen with uniform probability and perturbed. This perturbation is done by proposing a new value of the parameter based on a normal distribution centered at the current value, with any proposed negative values rejected.  The standard deviation of this distribution is initially set to be 1/100 of the initial value of the variable (or a minimum of 0.001 if the initial value is 0) and then tuned to obtain a between model acceptance ratio between 20--50\% during a ``burn-in'' / tuning simulation ran for 1/5th the length of the production simulation.
After the burn-in period, the simulation proceeds with fixed proposal distributions, as is required to satisfy detailed balance. Tuning steps are discarded when computing model evidences. 
\subsubsection{Reversible jump Monte Carlo}
\label{subsubsection:RJMC}
We next turn to the more complicated question of choosing moves between models. The reversible jump Monte Carlo (RJMC) technique~\cite{greenReversibleJumpMarkov1995a} can be employed to sample between multiple models with different numbers of parameters.
RJMC is an extension of traditional MCMC sampling from parameter space to combined parameter \emph{and} model space. It allows for the simultaneous sampling of parameter probability distributions from several models, as well as the discrete model probability distribution of the collection of models. RJMC is a powerful tool to simultaneously sample multiple models, but requires careful construction in cases with dissimilar models. Good chain mixing requires (1) the construction of high-quality mappings between the parameters in each model and (2) efficient ways of proposing new values when the number of parameters between models is uneven~\cite{karagiannisAnnealedImportanceSampling2013b,brooksClassicalModelSelection2003,brooksEfficientConstructionReversible2003,hastieModelChoiceUsing2012}. 

For the implementation presented in this paper, proposals for variables not present in current model, such as a quadrupole moment parameter when proposing a move from a model without a quadrupole (UA, AUA) to the model with a quadrupole (AUA+Q), are proposed independently from a Gaussian ``proposal distribution'' that approximates the 1-D posterior distribution of that parameter in the model they are moving to. These distributions are obtained by fitting Gaussians to short simulations of the model posterior in order to facilitate acceptances between inter-model moves.
\subsubsection{Parameter mappings}
\label{subsubsection:Parameter Mappings}
\begin{figure}[H]
    \centering
        \includegraphics[width=\textwidth]{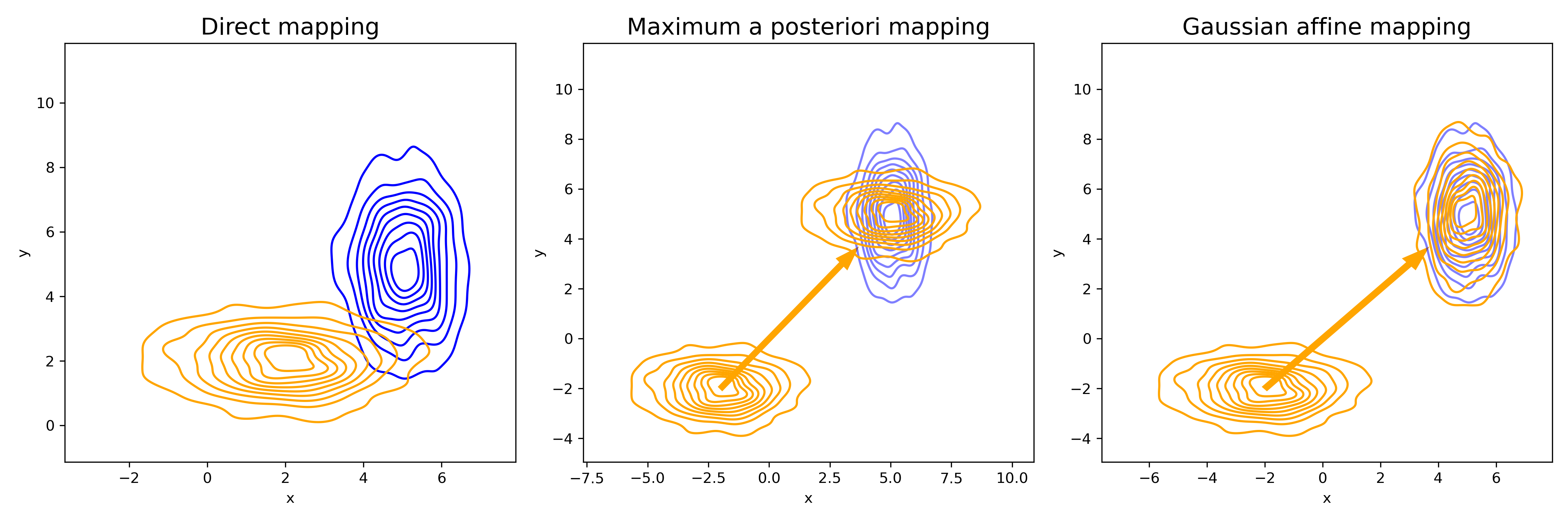}
    \caption{\textbf{Mappings between distributions with poor overlap, visualized. } In the first panel, there is enough overlap between the distributions to map without any transformation. Second panel shows maximum a posteriori mapping, improving overlap, but not matching shape. Third panel shows Gaussian affine mapping, transforming location and matching shape.}
    \label{fig:distribution_mapping}
\end{figure}

 To efficiently propose inter-model moves, one must define a \emph{parameter mapping function} that transforms a point in one parameter space to another point in a different parameter space (e.g. $\epsilon$ in the AUA model $\longrightarrow$ $\epsilon$ in the AUA+Q model). These mapping functions may range from a simple direct mapping to a highly complex one, depending on the situation. The form of the mapping function is often an important factor in model chain mixing and convergence; as long as there is \emph{some} probability overlap sampling between models is theoretically possible, but may be impossibly inefficient.  We explored several simple strategies for mapping common parameters between models, described below and illustrated in figure \ref{fig:distribution_mapping}; although others exist, such as non-equilibrium candidate Monte Carlo~\cite{nilmeierNonequilibriumCandidateMonte2011b}. 
\begin{itemize}
\item \textbf{Direct mapping:} In the simplest cases, parameters can be taken directly from one model and plugged into in the other model. This is only useful in the case where the distribution of the parameter in the two models overlap significantly.
\item \textbf{Maximum a posteriori (MAP) mapping:} Another possibility is to run a short MCMC simulation of each model, and then map the highest probability (MAP) values of each common parameter to each other, either by translation (additive mapping) or rescaling (multiplicative mapping).  This technique can improve overlap but will ultimately fail when the shapes of the distributions differ significantly, where only parameter sets near the MAP value will have good overlap.
\item \textbf{Gaussian affine mapping:} One can also run short MCMC simulations of each model, fit Gaussian distributions to each model posterior distribution, and then map these distributions on top of each other using an affine map that transforms the shape of one distribution to the shape of the other. This has an advantage of capturing differently shaped posterior distributions, allowing for more consistent quality mapping, especially away from the posterior maximum, but can still fail if the distributions are multi-modal (or more precisely, differently modal).
\end{itemize}

\subsubsection{Biasing Factors}
\label{subsubsection:BiasingFactors}
Even with perfect mapping between the shape of the distributions between models, inter-model moves will be rejected at a high rate if one model has much stronger unnormalized evidence than the other. In this case a biasing factor may be used to equalize the magnitude of the evidences.  After achieving improved sampling with the biased probabilities, one can easily unbias the results if the bias is constant across the model.  The biasing factors that result in even sampling of models are the Bayes factors between models, similar to how the weights leading to equal probability sampling in expanded ensemble or simulated tempering simulations are the free energies of each state.  However, even approximate biasing factors can enable sampling such model swaps are feasible and  Bayes factors may be estimated.

\subsection{Calculation of Bayes Factors}
\label{subsection:CalculationBayesFactors}
There are several approaches that can be used to calculate the model evidences and Bayes factors including analysis of the reversible jump Monte Carlo model posteriors ~\cite{bartolucciEfficientBayesFactor2006} and bridge sampling~\cite{mengSIMULATINGRATIOSNORMALIZING1996} from the prior to the posterior.
\subsubsection{Bridge sampling with intermediate states}
\label{subsubsection:BridgeSampling}

One method for computing Bayes factors involves calculating the marginal likelihoods (model normalizing constants) for each model $M$ individually, and then calculating the Bayes factor as in eq. 5.  The calculation of normalizing constants using Monte Carlo methods is a well-known problem in statistical inference and many methods are described in the literature~\cite{gewekeBayesianInferenceEconometric1989,nealProbabilisticInferenceUsing1993,diciccioComputingBayesFactors1997,gelmanSimulatingNormalizingConstants1998}.

A technique common to statistical inference and statistical physics is bridge sampling~\cite{mengSIMULATINGRATIOSNORMALIZING1996}, which calculates the ratio of normalizing constants between two distributions by utilizing information from Monte Carlo samples from both distributions.
The Bennett acceptance ratio (BAR)~\cite{bennettEfficientEstimationFree1976} is a specific case of bridge sampling where $\alpha$ is chosen to be unbiased and minimize variance of the estimator.

The BAR estimator is limited in cases when overlap between the two distributions is poor.  In this case, one can introduce intermediate distributions designed to create a path between the target distributions with sufficient overlap.  In this case, which we call \emph{bridge sampling with intermediates}, the ratio of normalizing constants can be estimated with the unbiased, lowest-variance MBAR estimator~\cite{shirtsStatisticallyOptimalAnalysis2008a}, which extends BAR to data collected over multiple states.  The process for calculating log Bayes factors from MBAR is analogous to the process of calculating free energies as described by Shirts and Chodera~\cite{shirtsStatisticallyOptimalAnalysis2008a}, with the Boltzmann distributions replaced with the prior $P(\theta|M)$, posterior $P(\theta|D,M)$, and intermediate distributions.  For a detailed discussion of how Bayes factors are calculated with MBAR, see Supporting Information section 2.

\begin{figure}[H]
    \centering
        \includegraphics[width=0.6\textwidth]{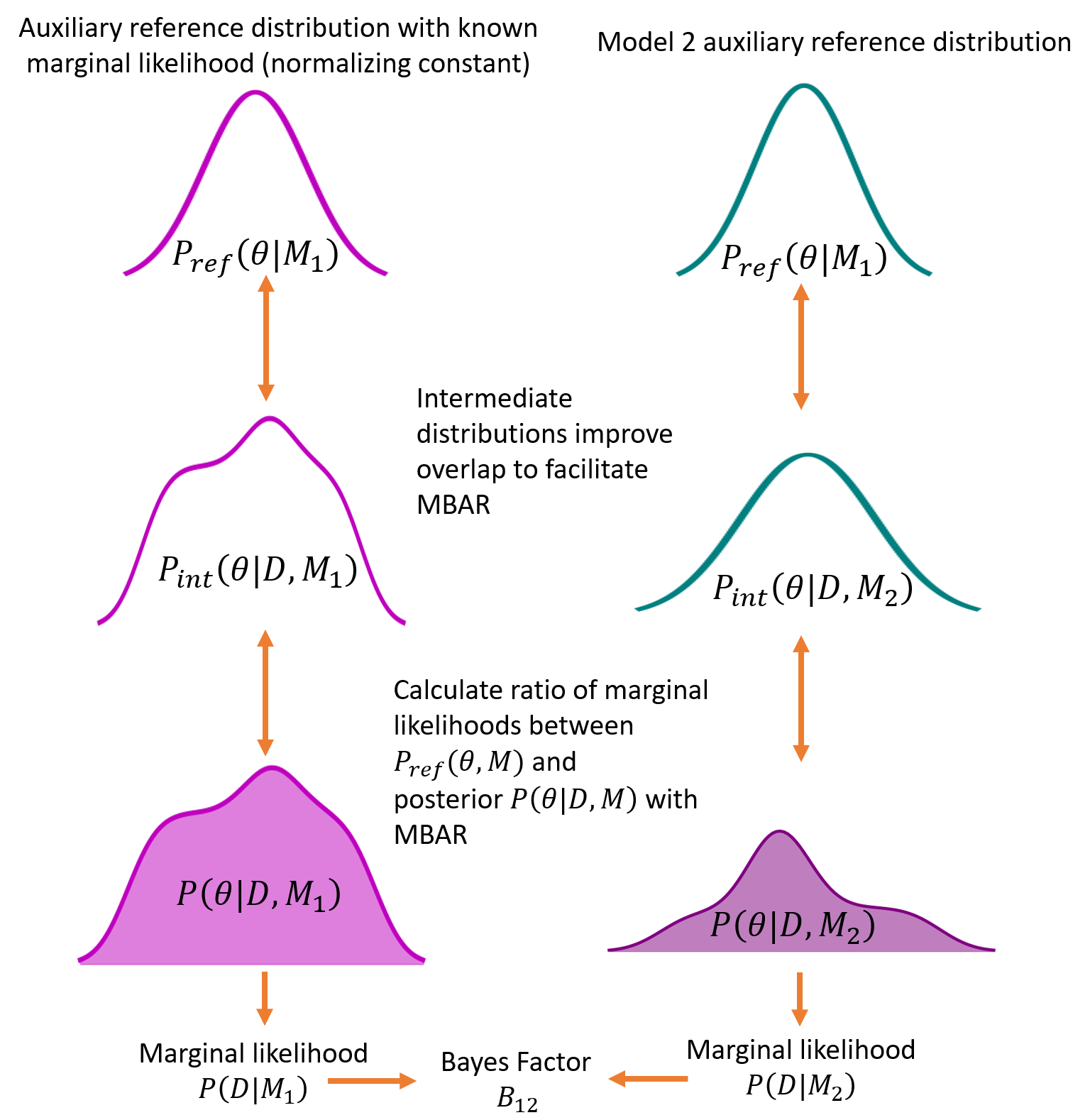}
    \caption{\textbf{ Bayes factor calculation scheme for 2CLJQ models.} Bridge sampling is used to calculate ratio of marginal likelihoods (normalizing constants) between analytical reference distribution (with known normalizing constant) and posterior distribution.  Posterior distribution marginal likelihoods are then calculated and used to compute Bayes factors.}
    \label{fig:mbar_calculations}
\end{figure}

Our strategy, described visually in figure \ref{fig:mbar_calculations}, is to calculate the model evidence (normalizing constant) of each model posterior $M$ by using MBAR to calculate the ratio of normalizing constants between that model's prior and posterior. Since all priors used here are normalized analytical distributions, the prior normalizing constant is known and we can calculate the posterior normalizing constant in combination with the ration from MBAR.  With this information we can calculate the Bayes factors by comparing the posterior normalizing constants between models.
Preliminary testing showed that the prior and posterior did not generally have enough overlap to produce reliable estimates of normalizing constant ratios with BAR.
In order to improve overlap, we replace the prior in this sampling process with a more informed auxiliary distribution.
By running a short MCMC simulation of the posterior distribution, we fit a simple multivariate normal distribution to the posterior sample; with more complex posteriors, Gaussian mixture models could be used, although we do not investigate them here.  This approach is not common in molecular simulation, with very complicated distributions of samples as a functions of the coordinates, with the possible exception of simple systems like the 2CLJQ model or crystals with near harmonic behavior~\cite{abrahamThermalGradientApproach2018}.  
Testing also showed that intermediate distributions were required to get sufficient overlap for bridge sampling.  Distributions $P_{int}$ are defined as:
\begin{equation}
P_{int} = P_{ref}(\theta)^{f(\lambda)}\times P(\theta|D,M)^{f(1-\lambda)}
\end{equation}
where $P_{ref}$ is the analytical auxiliary distribution.
Through testing we determined that using one intermediate state between the reference distribution and the posterior is enough to consistently achieve sufficient overlap for Bayes factor calculations.
Using intermediate distributions has an analog in free energy simulations with $\lambda$-dynamics simulations~\cite{knightLDynamicsFreeEnergy2009}, and in staged methods of calculating host/guest binding affinities through multiple windows.
\subsubsection{Model distributions from RJMC}
\label{subsubsection:ModelDistributionsRJMC}
RJMC allows for the direct estimation of Bayes factors through two methods: (1) estimating the ratio of normalizing constants by taking the sampling ratio (SR) of the visits to each model and (2) estimating this ratio with what is called in statistics the warped bridge sampling, using the BAR estimator.
The sampling ratio estimate is simply the ratio of the number of Monte Carlo samples from each model; assuming each model’s parameter space is sampled well, this will converge to the model Bayes factor. The warped bridge sampling approach involves using bridge sampling on a set of samples from two models, then using the proposal distributions from section 2.4.2 and parameter mappings from section 2.4.3 to "warp" the distributions to improve overlap, similar to mappings in configurational space. ~\cite{paliwalMultistateReweightingConfiguration2013,schieberConfigurationalMappingSignificantly2019}.
\subsubsection{Biased RJMC}
\label{subsubsection:BiasedRJMC}
As discussed in section 2.4.4, biasing factors can be used to equalize the evidence between models and facilitate cross model jumps in cases where there is a large difference in model evidences. In the case of our data, UA models often have very low model evidence for this data set, compared to AUA or AUA+Q. We can validate Bayes factor calculations as described in section 2.5.1 by performing RJMC with model evidences estimated from the importance sampling method as biasing factors.  If the biasing factors from the importance sampling method in 2.5.1 are accurate, then biased RJMC should yield  roughly equal sampling between models. In practice, the sampling will not be perfectly equal due to the stochastic nature of the simulation, but large deviations from equal sampling would indicate that these biasing factors are not accurate.

\subsection{Measuring Posterior Predictive Accuracy}
\label{subsection:MeasuringELPPD}

To perform an benchmark of the models independent of the posterior performance, we can calculate the \emph{expected log pointwise predictive density} for a new dataset (ELPPD), a metric introduced by Gelman \emph{et al.}~\cite{gelmanUnderstandingPredictiveInformation2014}.  This quantity is a sum of the \emph{expected log predictive density} (ELPD), which evaluates the expectation value of a new data point's ($\Tilde{D}_i$) log objective function over the previous posterior distribution $P(\theta | D)$.  For the objective function, we choose our standard likelihood function with the new data point as the target, $\log P(\Tilde{D}_i | \theta)$ . 

\begin{equation}
\mathrm{ELPPD (\tilde{D} | D)} = \sum_{i=1}^{n} \int_{\theta} \log P(\Tilde{D}_i | \theta) P(\theta | D) \mathrm{d}\theta 
\end{equation}

In practice, we estimate this quantity by simulating the posterior of each model, then drawing $k$ points at random from the posterior samples:

\begin{equation}
\mathrm{ELPPD}_k (\tilde{D}|D) \approx \sum_{i=1}^{n} \sum_{j=1}^{k} \log P(\Tilde{D}_i | \theta_j) \, ; \, \, \, \theta_i \sim P(\theta | D)
\end{equation}

These quantities are useful in that they assess the model's performance on independent datasets and do not include the prior information, so they are a more ``traditional'' metric to assess model performance.
\subsection{Chosen methods for this investigation}
\label{subsection:ChosenMethods}
Model priors are set using the training sample method described in section 2.3.3.    To calculate Bayes factors, we used the approach of MBAR with three intermediate states (section \ref{subsubsection:BridgeSampling}) is used, with three total $\lambda$ windows (analytical distribution, intermediate distribution, posterior distribution) simulated for $4\times 10^6$ steps each.  This method was chosen over the RJMC method due to its more accurate Bayes factor calculations.
To validate these results, RJMC with biasing factors taken as the model evidences from the MBAR calculation for $2\times10^6$ steps. In all cases, the biased RJMC calculations agree with the MBAR calculations. ELPPD and average deviation calculations are taken from benchmarking posterior simulations performed for $10^6$ steps for each model.
\section{Results \& Discussion}
\label{section:ResultsDiscussion}
\subsection{Effect of priors on Bayes factors}
\label{subsection:EffectPriors}
In order to compute meaningful Bayes factors between 2CLJQ models, we must address the effect of priors on the model posteriors.
\subsubsection{Illustration with uniform priors}
\label{subsubsection:IllustrationUniformPrior}
\begin{figure}[H]
    \centering
        \includegraphics[width=0.8\textwidth]{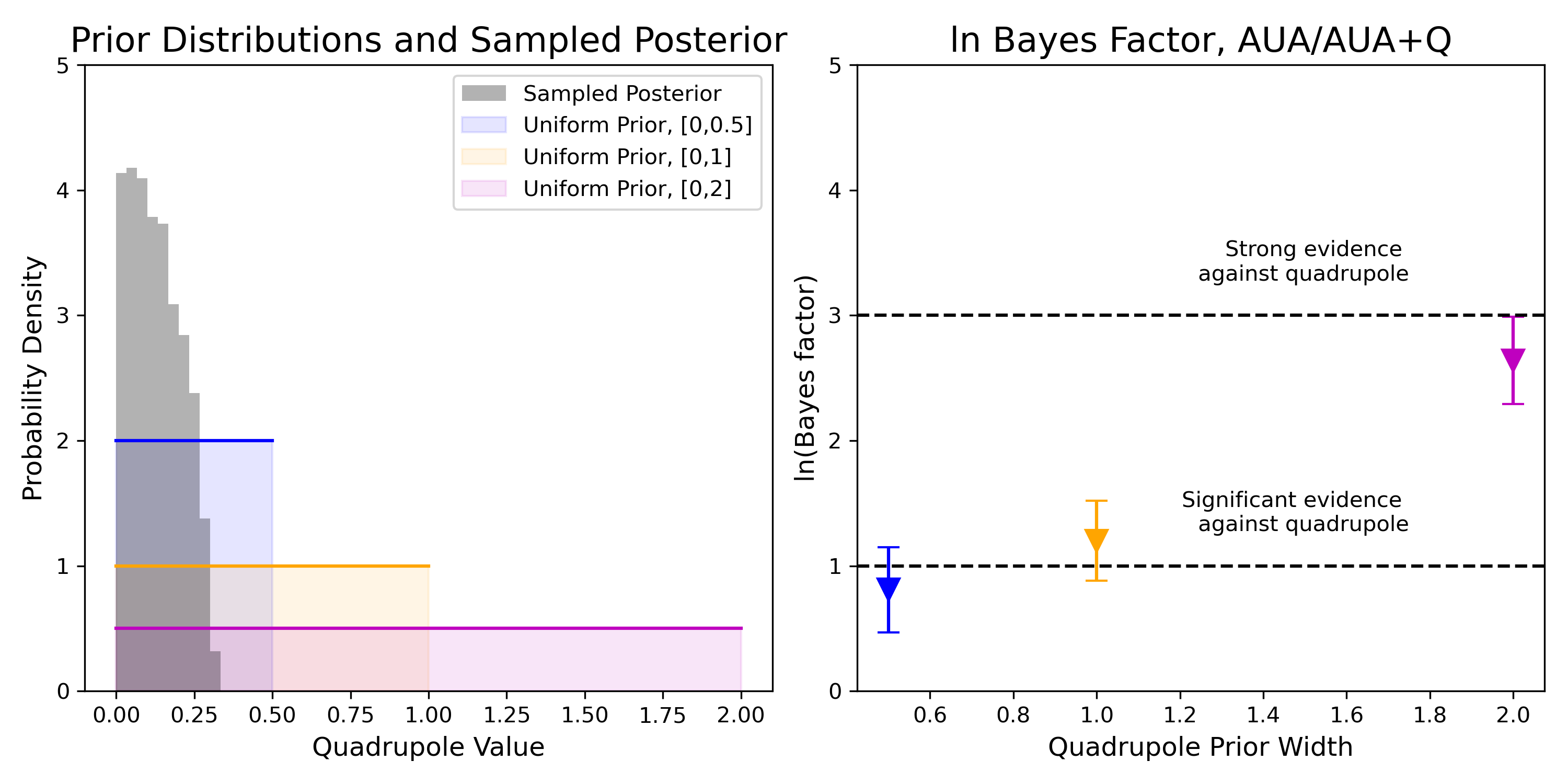}
    \caption{\textbf{Bayes factor comparison of AUA and AUA+Q models with varied quadrupole priors.} 
    Left: Bayes factors and uncertainties between AUA and AUA+Q, tested against C$_2$H$_6$ ($\rho_L$+$P_{sat}$) for an uniform prior distribution on 2CLJQ quadrupole parameter $Q$ with several different parameter values. 
    Wider prior distributions over the quadrupole decreases evidence in favor of AUA+Q model.  Right: corresponding in color uniform prior distributions, with sampled quadrupole posterior distribution superimposed over priors.}
    \label{fig:disjoint_prior}
\end{figure}

One of the most important effects of the choice of prior distribution is that the amount of \emph{parameter uncertainty} that the prior distribution encodes. In the case of a Bayes factor between the model without a quadrupole (AUA) and the model including a quadrupole (AUA+Q), different uniform priors over the quadrupole have a strong effect on the Bayes Factor.  This Bayes factor, shown in figure \ref{fig:disjoint_prior} is calculated with a target of ethane (C$_2$H$_6$) $\rho_l$ and $P_{sat}$ data.

As the uniform prior is widened, covering more parameter space, its probability density decreases proportionally, which in turn, lowers the model evidence for the AUA+Q model.  This illustrates that a less certain prior over a parameter (encoding more parameter uncertainty) will penalize a model including that parameter. This simple example illustrates how wider uniform priors penalize models with higher parameter uncertainty. This prior uncertainty penalty is an intended part of Bayesian model selection. In Bayes factors calculated in this work, we fit priors for each model in any given comparison to the same set of training sample, so any differences in prior uncertainty should be due to model differences rather than prior information.

\subsubsection{Using more data in prior fitting increases prior certainty}
\label{subsubsection:PriorFittingTrainingSample}
To evaluate the effect of different amounts of prior information on the Bayes factors between 2CLJQ models, we calculated Bayes factors with priors fit to training samples with different amounts of experimental data.  Our priors become more certain as more training data is included. While there is no objectively correct Bayes Factor due to its dependence on the prior, the prior should effectively constrain the parameter space, so that extraneous parameter values do not affect the model comparison.  To ensure that the priors for the 2CLJQ model were effectively constraining the parameter space, we constructed them with three levels of data in the training sample: n=3, 5, or 8 data points per property respectively (referred to as low, medium, and high information priors).
\begin{figure}[H]
    \centering
        \includegraphics[width=\textwidth,trim={0 0 0 1.5cm},clip]{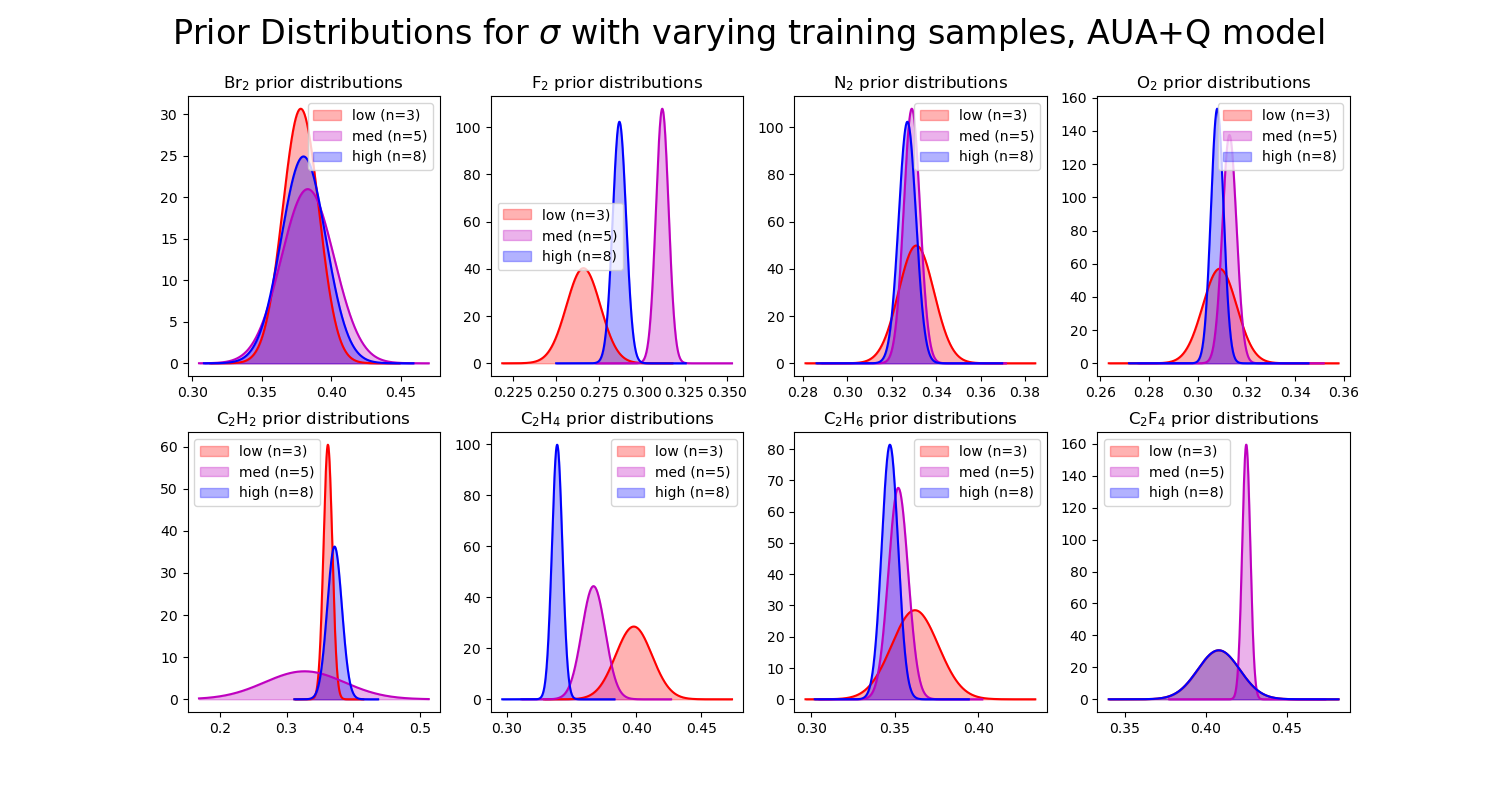}
    \caption{\textbf{ Changes in prior distribution shape of AUA+Q model with different amounts of data in training set.}  Lack of overlap between low data (n=3) training sample and higher data training samples indicate that prior is not well constrained with n=3 data points.}
    \label{fig:prior_changes}
\end{figure}
In this case, shown in figure \ref{fig:prior_changes}, the overlap between the low information prior (n=3 data points) and the medium and high information priors (n=5, n=8 respectively) is poor, indicating that the low information prior does not constrain the prior in the same way as the other priors.  In our Bayes factor calculations, the high information prior is used, since it has excellent overlap with the medium prior, and the increased amount of data in the training sample increases our confidence that the prior represents parameters that can model the whole range of experimental data points.  We varied the amount of information in this experiment to determine the effect of the amount of data included on the prior and therefore the Bayes factor; in general, we would use as much information as reasonably possible to inform the prior.

\subsection{Results for $\rho_l$, $P_{sat}$ targets}
\label{subsection:2crit}

For the $\rho_l, P_{sat}$ target, the preferred model varies significantly based on the molecule that it is applied to.
\subsubsection{Diatomic molecules}
\label{subsubsection:2crit_diatomic}
\begin{figure}[H]
    \centering
        \includegraphics[width=1.0\textwidth]{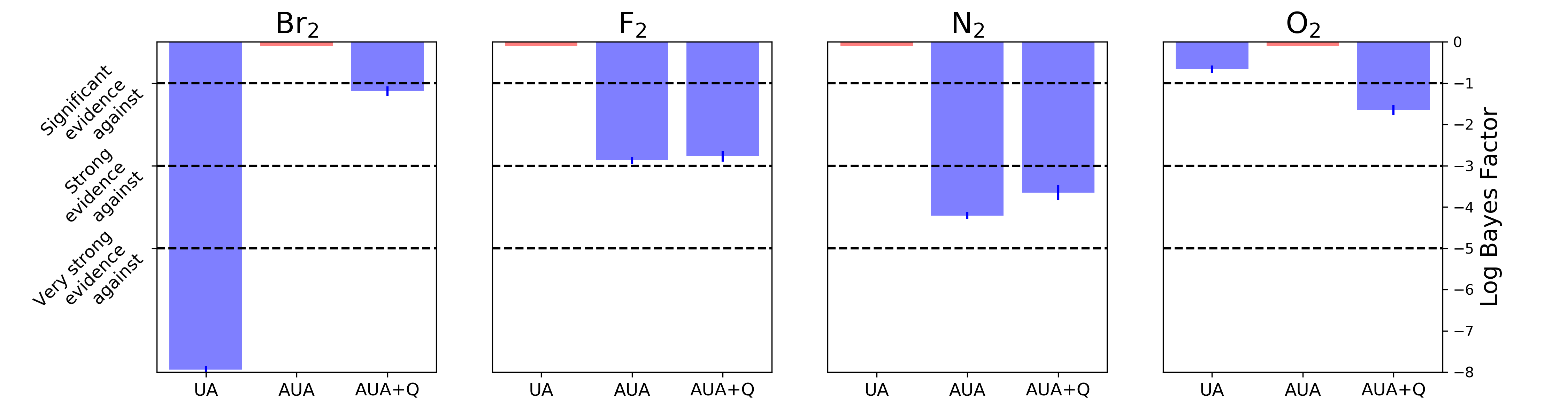}
    \caption{\textbf{Bayes factors between UA, AUA, AUA+Q models for diatomic molecules, tested against $\rho_l, P_{sat}$ data.}  Red bar indicates favored model, blue bars indicate log Bayes factor (strength of evidence) against corresponding models.  Simpler models (UA/AUA) are favored in these cases.}
    \label{fig:diatomic_2crit}
\end{figure}

For the set of diatomic molecules in figure \ref{fig:diatomic_2crit}, Bayes factor evidence favors either the simplest UA model or the AUA model.  For bromine, the AUA model produces slightly different LJ parameters than the UA model, which yield improved physical property accuracy for this target, but are incompatible with the fixed bond length value of the UA model (sampling bond lengths of $\sim$ 0.18 nm, rather than 0.23 nm, as in the UA model).  While the value in the UA model reflects a physical bond length, the anisotropic model's assumption that the interaction sites are not necessarily separated by the equilibrium bond length produces better physical property estimates.  This is similar to the behavior of Oxygen, which also selects a different bond length then the experimental value, but with weaker evidence for the AUA model.

\subsubsection{Hydro/fluorocarbons}
\label{subsubsection:3crit_hfc}
For the hydrocarbons/fluorocarbons in figure \ref{fig:hfc_3crit}, the evidence generally points towards the more complex models (AUA/AUA+Q), with the strength of the experimental quadrupole moment generally dictating whether the $Q$ parameter is required.  C$_2$H$_2$ and C$_2$F$_4$ have the largest sampled quadrupole moments (Q $\approx$ 0.5, 0.83), and overwhelming evidences in favor of the AUA+Q model.  C$_2$H$_4$ and C$_2$H$_6$ have lower sampled quadrupole moments (Q $\approx$ 0.41, <0.25), and less evidence in favor of including this parameter; C$_2$H$_4$ and C$_2$H$_6$ have significant evidence in favor of the AUA model.
\begin{figure}[H]
    \centering
        \includegraphics[width=1.0\textwidth]{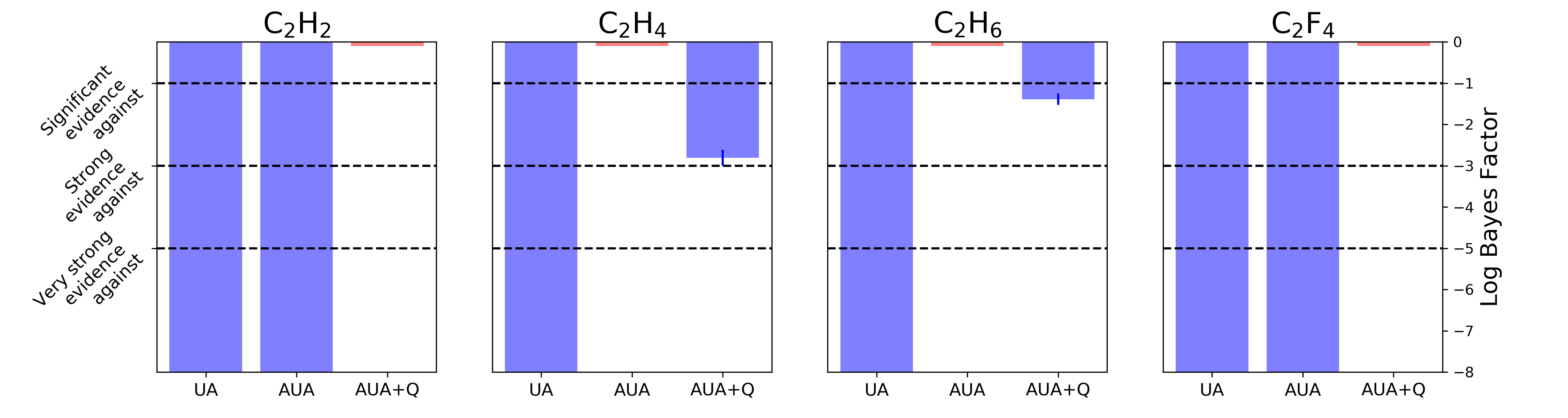}
    \caption{\textbf{ Top panel: Bayes factors between UA, AUA, AUA+Q models for hydro/fluorocarbon molecules, tested against $\rho_l, P_{sat}$ data.}  Red bar indicates favored model, blue bars indicate log Bayes factor (strength of evidence) between the favored model and the named model. Size of blue bar indicates the strength of the evidence for the favored model (a larger bar indicates stronger evidence for the favored model).}
    \label{fig:hfc_2crit}
\end{figure}
\subsection{Results for $\rho_l, P_{sat}, \gamma $ targets}
\label{subsection:3crit}
The addition of the surface tension target, tends to push the evidence in favor of the more complex models, which could be expected with the need to reproduce a very different type of molecular data.
\subsubsection{Diatomics}
\label{subsubsection:diatomic_3crit}
\begin{figure}[H]
    \centering
        \includegraphics[width=1.0\textwidth]{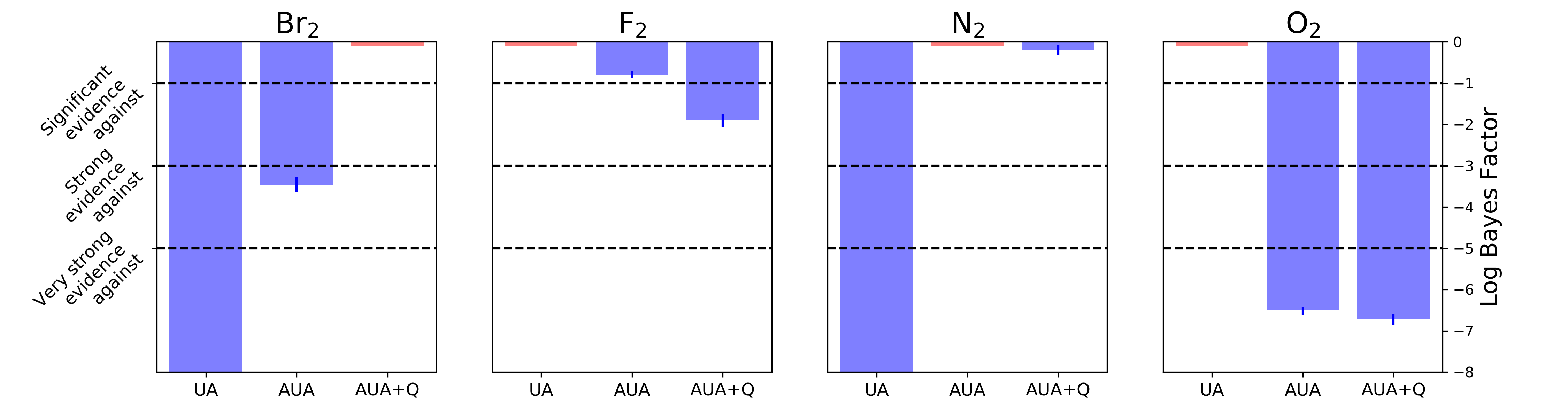}
    \caption{\textbf{ Bayes factors between UA, AUA, AUA+Q models for diatomic molecules, tested against $rho_l, P_{sat}, \gamma$ data.}  Red bar indicates favored model, blue bars indicate log Bayes factor (strength of evidence) against corresponding models.  Addition of $\gamma$ data pushes model decision for Br$_2$ and N$_2$ towards more complex models (AUA/AUA+Q)}
    \label{fig:diatomic_3crit}
\end{figure}

Among diatomics in figure \ref{fig:diatomic_3crit}, both Br$_2$ and N$_2$ require more complex models, with very strong evidence against the UA model, with weak evidence in favor of AUA over AUA+Q for N$_2$, and strong evidence in favor of AUA+Q in the case of Br$_2$.  In the cases of F$_2$ and O$_2$, there is very little difference in model performance (before accounting for parsimony) between UA and the more complex models, leading Bayesian inference to favor UA due to parsimony.  In the cases of Br$_2$ and N$_2$, the more complex models (AUA and AUA+Q) correct a systematic overprediction of both $P_{sat}$ and $\gamma$ compared to the UA model, overcoming its parsimony advantage.

\subsubsection{Hydro/fluorocarbons}
\label{subsubsection:3crit_hfc}
\begin{figure}[H]
    \centering
        \includegraphics[width=0.85\textwidth]{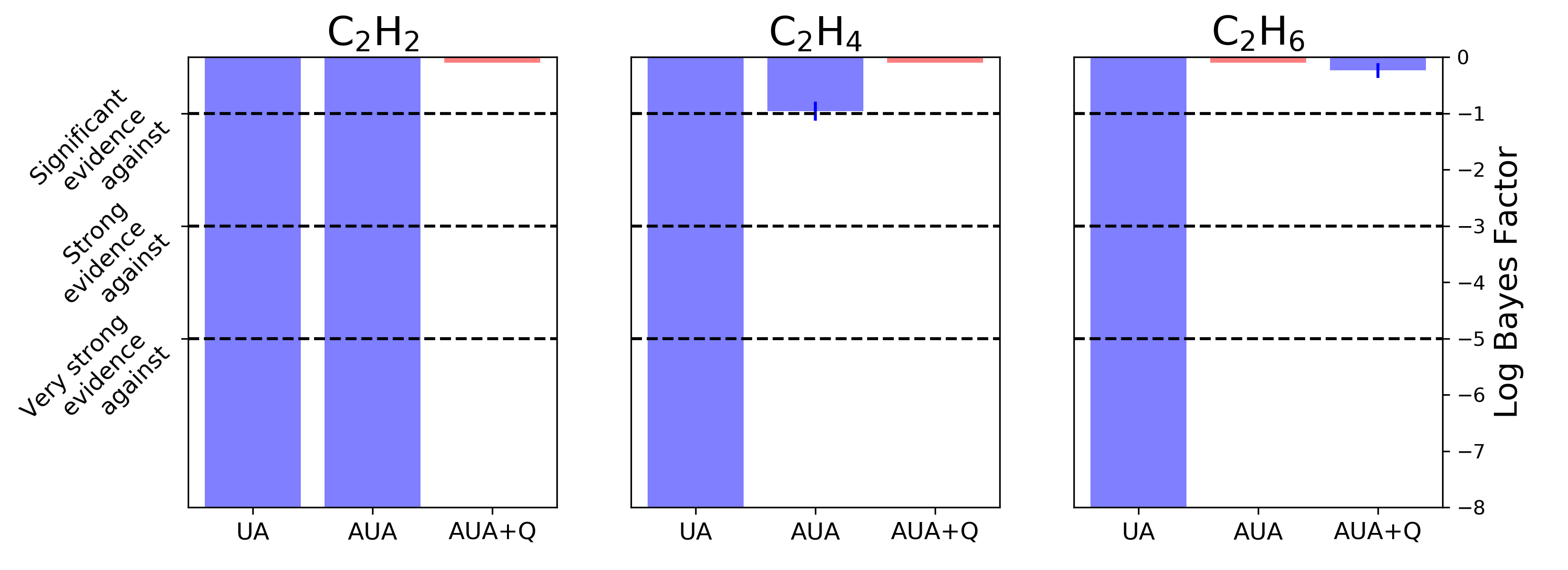}
    \caption{\textbf{ Bayes factors between UA, AUA, AUA+Q models for diatomic molecules, tested against $rho_l, P_{sat}, \gamma$ data.  Red bar indicates favored model, blue bars indicate log Bayes factor (strength of evidence) against corresponding models.  C$_2$F$_4$ omitted due to lack of $\gamma$ data.}}
    \label{fig:hfc_3crit}
\end{figure}

For the hydro/fluorocarbons in figure \ref{fig:hfc_3crit} (C$_2$F$_4$ omitted due to lack of experimental $\gamma$ data), C$_2$H$_2$ still has extremely strong evidence in favor of AUA+Q, and the inclusion of $\gamma$ in the target yields weak evidence in support of the AUA+Q model for C$_2$H$_4$.  C$_2$H$_6$ has very weak evidence against the inclusion of a quadrupole as the AUA and AUA+Q model sample very similar values.  All of these molecules have very strong evidence against the simplest UA model.

\subsection{Parameter Correlations}
\label{subsection:ParameterCorrelations}

An advantage of this Bayesian inference process is the information about parameter probability distributions obtained from the MCMC process.  This information is valuable for examining parameter correlations and understanding parameter sensitivity.
\subsubsection{Correlation of Lennard-Jones parameters}
\label{subsubsection:LJCorrelation}

\begin{figure}[H]
    \centering
        \includegraphics[width=\textwidth]{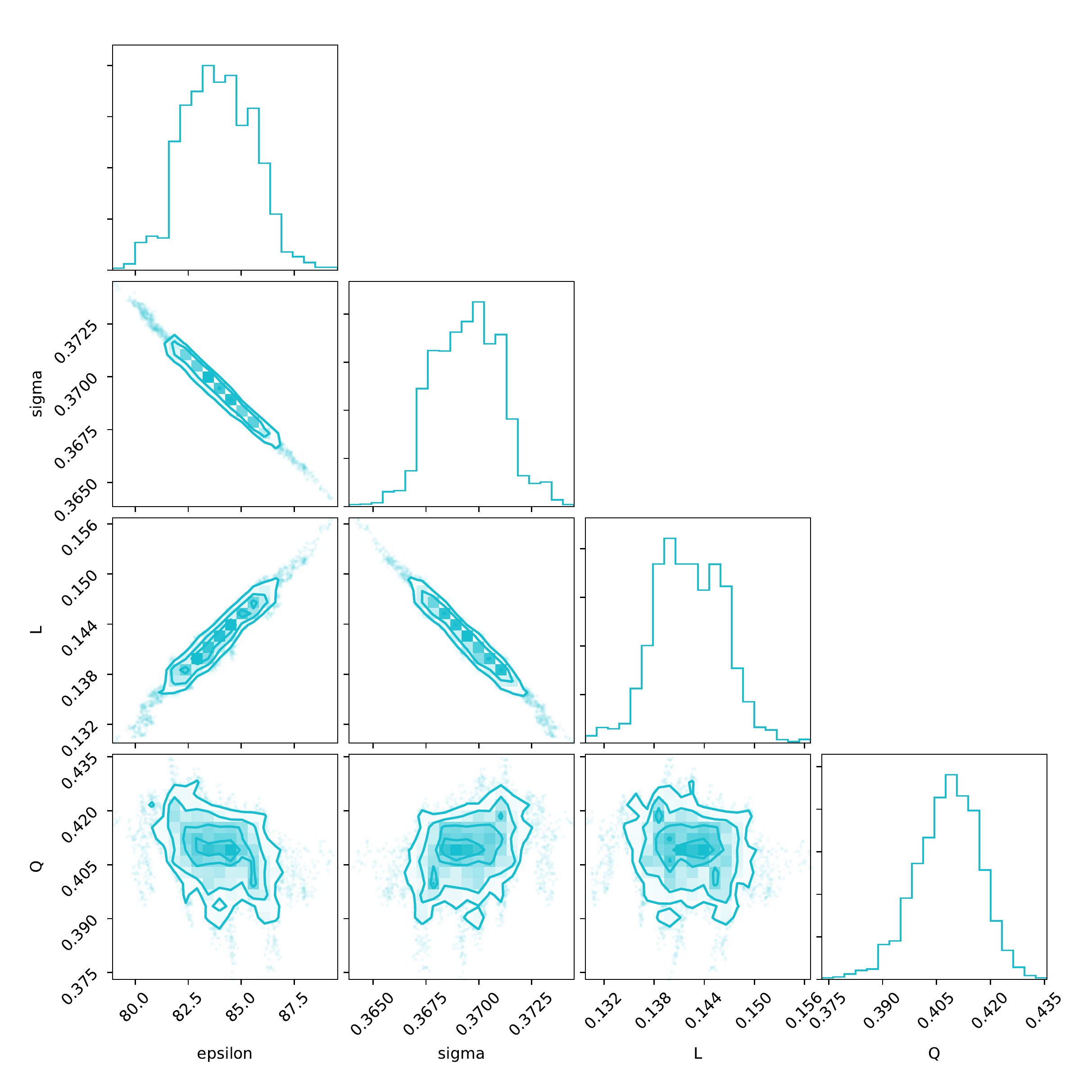}
    \caption{\textbf{ Parameter correlations shown in triangle plot of AUA+Q model's 2-d marginal parameter distributions for C$_2$H$_4$, $\rho_l+P_{sat}$ target.}
    Plot shows high degree of correlation between $\epsilon,\sigma,L$, and weaker correlation with $Q$.}
    \label{fig:LJ_Q_param_correlation}
\end{figure}

One of notable parameter trends in this model, shown in figure \ref{fig:LJ_Q_param_correlation} is the high degree of correlation between the $\epsilon,\sigma, L$ that represent the Lennard-Jones interactions.  A degree of correlation between LJ $\epsilon$ and $\sigma$ has been observed previously~\cite{messerlyConfigurationSamplingBasedSurrogateModels2018a,shirtsSolvationFreeEnergies2005}, but the linear/planar nature of the $(\epsilon,\sigma, L)$ probability surface suggests that the 2CLJQ model has a number of degenerate/nearly degenerate parameter sets that do a relatively good job of reproducing experimental data.  Since the sampling is based on the surrogate models, the correlation is probably overestimated compared to the full model.
The quadrupole parameter is also correlated with the other parameters, but more weakly.  Corner plots for all models and targets are available in the Supporting Information, section 7.

Notably, we do not find strong multi-modalities in any of these situations; most have a single maximum of parameter probability.  This is probably due to the simplicity of the model and its surrogate models; we do not necessarily expect this to be true for more complex models.

\subsection{Benchmarking}
\label{subsection:Benchmarking}

Benchmarking based on ELPPD results was performed for all compounds with enough experimental data available after prior fitting and Bayes factor calculation.  In general, measures of model performance based on Bayes factor and likelihood-only ELPPD benchmarking agree, especially when the evidence in favor of one model is strong. There are some situations where these methods disagree; this is where the parsimonious nature of the Bayesian approach is important.

\begin{figure}[H]
    \centering
        \includegraphics[width=\textwidth]{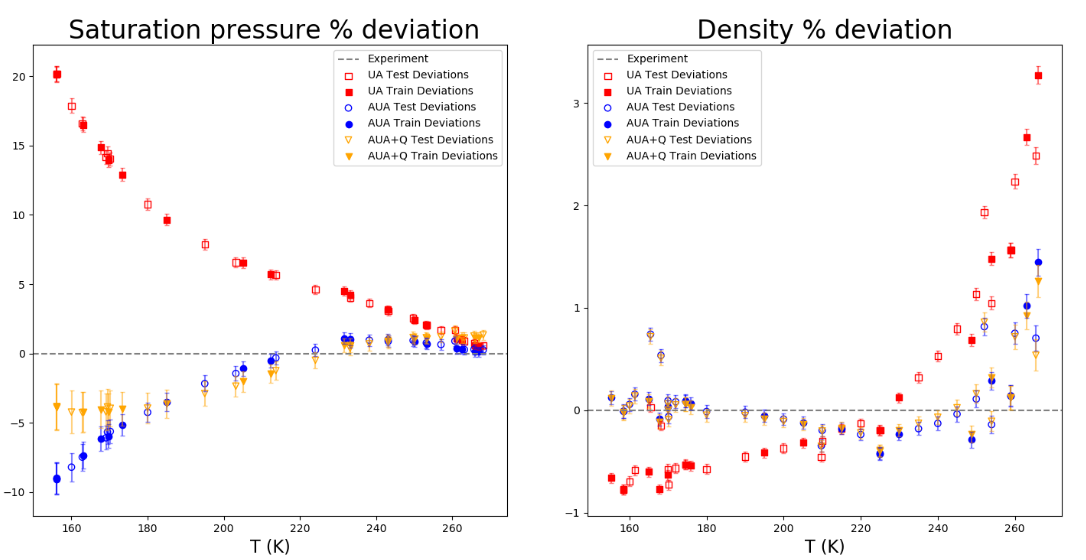}
    \caption{\textbf{ Average $\rho_l$ (left panel), $P_{sat}$ (right panel) \% deviation plots for C$_2$H$_4$.} Parameter sets drawn from posterior probability distribution, evaluated against separate benchmark data points (open points) as well as points used in calculated Bayes factor (filled points). Data shows AUA+Q model (yellow triangles) improves performance in reproducing $P_{sat}$ data at low temperatures, but Bayes factor evidence supports AUA model (blue circles) due to parsimony.}
    \label{fig:C2H4_benchmark}
\end{figure}

This pattern is illustrated in figure \ref{fig:C2H4_benchmark}, in the $\rho_l, P_{sat}$ criteria for C$_2$H$_4$.  Although the performance of the AUA/AUA+Q models are very similar at reproducing $rho_l$ (ELPPD average over test points is 1.78 for AUA vs 1.75 for AUA+Q), and $P_{sat}$ (0.81 for AUA vs 1.02 for AUA+Q), the Bayes factor favors the AUA model due to the complexity penalty that the Bayesian method assesses.

\subsection{Challenges of implementation}
\label{subsection:ImplementationChallenges}
A common challenge of using MCMC to sample points from a probability distribution is ensuring good sampling and mixing within the MCMC chains. In the posteriors sampled here, exhaustive sampling can be achieved with simple MCMC techniques and long chains. In higher-dimension spaces with rougher probability landscapes, directed sampling techniques like Hamiltonian Monte Carlo~\cite{nealMCMCUsingHamiltonian2012} (HMC), Langevin Dynamics~\cite{leimkuhlerRobustEfficientConfigurational2013a} or a No U-turn Sampler (NUTS)~\cite{hoffmanNoUTurnSamplerAdaptively2011} may be required to efficiently sample the space.  Running multiple chains from different starting points may also be important in situations with significant multi-modality; advanced sampling techniques to ``bridge the gap'' between disparate basins may also be helpful.  

A more practical challenge to MCMC sampling is the implementation of surrogate models used to sample parameter sets.  For most parameterization problems, implementing surrogate models will be much more difficult that for the 2CLJQ model.  For more complex models like biomolecular force fields, surrogate models will not be general enough to capture model responses for large classes of molecules or physical properties, so surrogate models will need to be purpose-built for a process like this. This is a challenging task, but should be possible by simulating properties of interest for multiple parameter sets, and then using modeling techniques well suited to sparse data, such as Gaussian processes (GP)~\cite{liuWhenGaussianProcess2019,oliverKrigingMethodInterpolation1990} regression. Befort et al.~\cite{befortMachineLearningDirected2021} recently had success using GPs to build physical property surrogate models based on small molecule force fields. These methods could be enhanced with reweighting techniques like MBAR on simulated points to add gradient information to the surrogate models.~\cite{messerlyConfigurationSamplingBasedSurrogateModels2018a}.

Another implementation challenge is selecting a technique to compute Bayes factors from these MCMC chains. As discussed in section \ref{subsubsection:RJMC}, RJMC, while attractive due to its simplicity and simultaneous sampling, will be difficult to implement in large model spaces, requiring much fine-tuning to facilitate model jumps.  The MBAR-based technique we use is more attractive in a large model space, but requires some testing to ensure a good reference model, and probability overlap between the reference model and the posterior.

\section{Conclusions}
\label{section:Conclusions}

In this work, we introduce Bayesian inference with surrogate modeling as a paradigm for making model decisions for non-bonded molecular interactions.  By testing this strategy on the 2CLJQ model, we demonstrate that the technique can choose between models in a way that balances model performance with parsimony.  We anticipate that this will be useful in force field parameterization, where tradeoffs between model complexity and performance are extremely common. We also assess the role the Bayesian prior has on the choice of model, and offer insight into how these choices may be made in a force field parameterization context.

The model decisions made in the Bayesian inference process vary widely due to the compound and data targets considered, which highlights the variability of the targets.  Although this does not yield a blanket recommendation for a level of model complexity, it identifies which compounds and properties require a more complex model.  This problem-specific parsimony is of particular interest to the molecular modeling community, because the goal for molecular model selection is choosing the simplest and computationally cheapest model that describes the data of interest with sufficient experimental accuracy.

The most significant challenge of applying this technique to more complex systems will be creating the surrogate models required for this process.  For biomolecular force fields, it is possible to build problem-specific Gaussian process surrogate models built from simulation data and MBAR reweighting  to explore the parameter space of combinations of Lennard-Jones parameters, though the sampling of parameter space required to construct them is a significant challenge. This technique is also applicable to other common dispersion-repulsion model selection problems such as LJ typing schemes, as well as LJ combination rules.  It is also potentially useful for choices between fixed-charge electrostatic models, and could help identify where virtual sites or polarizable sites could be most useful.  Overall, this technique appears to provide a systematic method of evaluating models based on performance and parsimony applied to molecular simulations.

\section{Data and Software Availability}
Datasets and code used to generate the results shown in this study are available from \url{https://github.com/SimonBoothroyd/bayesiantesting/tree/combined_run}.

\section{Author Contributions}
Contributions based on CRediT taxonomy:

\noindent O.C.M.:  Conceptualization, data curation, formal analysis, investigation, methodology, software, visualization, original draft, review and editing \\
S.B.: Methodology, software, investigation, visualization, review and editing \\
R.A.M.: Conceptualization, data curation, resources, software, methodology, review and editing \\
J.F.: Methodology, review and editing \\
J.D.C.: Resources, funding acquisition, review and editing \\
M.R.S.: Conceptualization, methodology, funding acquisition, resources, project administration, supervision, review and editing. \\

\section{Acknowledgements and Funding}
We thank the Open Force Field Consortium for funding, including our industry partners as listed at the Open Force Field website, and Molecular Sciences Software Institute (MolSSI) for its support of the Open Force Field Initiative.  We gratefully acknowledge along with all current and former members of the Open Force Field Initiative and the Open Force Field Scientific Advisory Board. Research reported in this publication was in part supported by National Institute of General Medical Sciences of the National Institutes of Health under award number R01GM132386, specifically partial support of OCM, MRS, and JDC.  OCM, MRS, JDC, and JF acknowledge support from NSF CHE-1738975 for parts of the project.  These findings are solely of the authors and do not necessarily represent the offical views of the NIH or NSF.

\section{Disclosures}
The Chodera laboratory receives or has received funding from multiple sources, including the National Institutes of Health, the National Science Foundation, the Parker Institute for Cancer Immunotherapy, Relay Therapeutics, Entasis Therapeutics, Silicon Therapeutics, EMD Serono (Merck KGaA), AstraZeneca, Vir Biotechnology, XtalPi, Foresite Labs, the Molecular Sciences Software Institute, the Starr Cancer Consortium, the Open Force Field Consortium, Cycle for Survival, a Louis V. Gerstner Young Investigator Award, and the Sloan Kettering Institute. A complete funding history for the Chodera lab can be found at \url{http://choderalab.org/funding}.
JDC is a current member of the Scientific Advisory Boards of OpenEye Scientific Software, Interline, and Redesign Science, and holds equity interests in Interline and Redesign Science.
MRS is an Open Science Fellow for Silicon Therapeutics.
SB is a director of Boothroyd Scientific Consulting Ltd.


% ...





\bibliography{paper}

\clearpage

\clearpage

\end{document}


\title{Supporting Information for "Bayesian inference-driven model parameterization and selection for 2CLJQ fluid models"}

\author{Owen C. Madin, Simon Boothroyd, Richard A. Messerly,\\ Josh Fass, John D. Chodera, Michael R. Shirts}
\maketitle

\section{Surrogate models}
\subsection{Functional forms}

Surrogate models used in this paper are analytical correlations of the following functional forms:
\begin{equation}
    \rho = \left(\rho^{*}_c + C_1(T^*_c-T^*)^{1/3} + C_2(T^*_c-T^*) + C_3 (T^*_c-T^*)^{3/2}\right)\sigma^{-3}
\end{equation}
\begin{equation}
    \ln P_{sat}(\sigma, \epsilon, L, Q) = c_1(\sigma, \epsilon, L, Q) + \frac{c_2(\sigma, \epsilon, L, Q)}{T^*} + \frac{c_3(\sigma, \epsilon, L, Q)}{T^{*4}} \\
\end{equation}
\begin{equation}
    \gamma = A(\sigma, \epsilon, L, Q)\left( 1- \frac{T}{T_c}\right)^B
\end{equation}
\begin{equation}
    T^* = Tk_B/\epsilon
\end{equation}
\begin{equation}
    T^*_c = f(\sigma, \epsilon, L, Q)
\end{equation}
For full details and values of constants, see Stoll~\cite{stollComprehensiveStudyVapourliquid2009a} and Werth~\cite{werthSurfaceTensionTwo2015a}
\subsection{Model uncertainty/error estimates}
\begin{figure}[h]
    \centering

\begin{tabular}[t]{c|c|c}

 Property & Temperature Range ($\%$ of $T_c$) & $\%$ error\\
\hline
   & $< 0.9 $ & $0.3$ \\
   $\rho_l$ & $0.9 - 0.95$ & $ 0.3 + \frac{1-0.3}{0.95-0.9}\times (T - 0.9) $\\
   & $>0.95$ & $1.0$\\
   \hline
   & $< 0.55 $ & $20$ \\
   $P_{sat}$ & $0.55 - 0.7$ & $ 20 + \frac{2-20}{0.7-0.55}\times (T - 0.55) $\\
   & $>0.7$ & $2.0$\\
   \hline
   & $< 0.75 $ & $4$ \\
   $\gamma$ & $0.75 - 0.95$ & $ 4 + \frac{12-4}{0.95-0.75}\times (T - 0.75) $\\
   & $>0.95$ & $12.0$\\
\end{tabular}

\caption{Piecewise uncertainty $u_{surr}$ developed for 2CLJQ surrogate models by Stoll and Werth \cite{stollComprehensiveStudyVapourliquid2009,werthSurfaceTensionTwo2015a} from those authors' simulation results. Piecewise behavior attempts to capture the temperature dependency of uncertainty without adding unjustified complex functions.}

\label{tbl:Uncertainty}
\end{figure}
\newpage
\section{Data temperature ranges for all Bayes factor calculations}
If data from 55-95 \% of $T_c$  was available for all properties in a given target, then data points were selected in that range.  However, some properties had limited data ranges, and in those cases temperature ranges were selected so that all property data was within the same range.  Temperature ranges are listed in table \ref{tbl:TempRanges}

\begin{table}[h]
    \centering
    \begin{tabular}[t]{c|c|c}
Compound & $rho_l, P_{sat}$ temperature range & $\rho_l,P_{sat},\gamma$ temperature range \\
\hline
 $\mathrm{Br_2}$ & $(0.55, 0.95)\times T_c$ & $(0.47, 0.55)\times T_c$ \\
 
 $\mathrm{F_2}$ & $(0.5, 0.6)\times T_c$ & $(0.5, 0.55)\times T_c$ \\
 
 $\mathrm{N_2}$ & $(0.55, 0.95)\times T_c$ & $(0.55, 0.95)\times T_c$ \\
 
 $\mathrm{O_2}$ & $(0.55, 0.95)\times T_c$ & $(0.55, 0.95)\times T_c$ \\
 
 $\mathrm{C_2H_2}$ & $(0.55, 0.95)\times T_c$ & $(0.62, 0.7)\times T_c$ \\
 
 $\mathrm{C_2H_4}$ & $(0.55, 0.95)\times T_c$ & $(0.41, 0.65)\times T_c$ \\
 
 $\mathrm{C_2H_6}$ & $(0.55, 0.95)\times T_c$ & $(0.55, 0.95)\times T_c$ \\
 
 $\mathrm{C_2F_4}$ & $(0.55, 0.95)\times T_c$ & --- \\
\end{tabular}
    \caption{Temperature ranges used to select property data points for Bayes factor calculation.  Temperature ranges chosen so that all data points from all properties fall within temperature range.}
    \label{tbl:TempRanges}
\end{table}
\newpage
\section{Bayes factor calculation with MBAR}
Bayes factors in the ``Bridge sampling with intermediates'' method are calculated using MBAR.  For a given model posterior, first the normalizing constant $c_{ref}$ of the auxiliary reference distribution is calculated. This is trivial because these distributions are analytical.

Then, MBAR is used to calculate the ratio of normalizing constants $c_{post}/c_{ref}$ between the posterior distribution $P(D|\theta, M)$ and the reference distribution $P_{ref}(\theta | M)$ by finding the ratio of MBAR normalizing constants $\hat{c}_{post}/\hat{c}_{ref}$. We note that only the ratio is defined, since the normalizing constants are only known up to a multiplicative constant.  $\hat{c}_{post}$ and $\hat{c}_{ref}$ are calculated in equations \ref{equation:MBARCpost} and \ref{equation:MBARCref}.  In these equations, the variables $j$ and $k$ iterate over the probability distributions which samples are taken from, and the variable $n$ iterates over the $N_j$ samples from the unnormalized probability distributions labeled by $j$. The $K$ total distributions include all unnornalized distributions from which samples are collected from, specifically the posterior $P(D|\theta,M)$, the reference distribution $P_{ref}(\theta,M)$, as well as any auxiliary intermediate distributions.  \begin{equation}
\label{equation:MBARCpost}
\hat{c}_{post} = -\sum_{j=1}^{K} \sum_{n=1}^{N_j} \frac{P(D|\theta_{jn},M)}{\sum_{k=1}^K N_k \hat{c}_k^{-1} P_k(\theta_{jn})}
\end{equation}

\begin{equation}
\label{equation:MBARCref}
\hat{c}_{ref} = -\sum_{j=1}^{K} \sum_{n=1}^{N_j} \frac{P_{ref}(\theta_{jn}|M)}{\sum_{k=1}^K N_k \hat{c}_k^{-1} P_k(\theta_{jn})}
\end{equation}

These equations must be solved self-consistently, since $\hat{c}_{post}$ and $\hat{c}_{ref}$ are included in $c_k$.
We note that $P_{ref}$ may be normalized or unnormalized, as long as the actual normalization constant of the version used is what is used as $c_{ref}$ in eq.~\ref{equation:PosteriorNormConstant}. These equations are only unique up to a multiplicative constant, so one of the constants must be set (usually to 1) rather than estimated, and all ratios can then be calculated uniquely. 

Self-consistent solution is performed by converting into log probability space to produce effective "energies" and then using the python package \texttt{pymbar} (\url{https://github.com/choderalab/pymbar}). For more details on the MBAR equations, see~\cite{shirtsStatisticallyOptimalAnalysis2008a}.
The posterior normalizing constant $c_{post}$ is then calculated by multiplication as in equation \ref{equation:PosteriorNormConstant}.
\begin{equation}
\label{equation:PosteriorNormConstant}
c_{post} = c_{ref} \times \hat{c}_{post}/\hat{c}_{ref}
\end{equation}

At this point we note that the posterior normalizing constant $c_{post}$ is the model marginal likelihood $P(D|M)$.  So, for models 1 and 2, we can estimate the Bayes factor $B_{1/2}$ as in equation \ref{equation:BayesFactorCalculate}.

\begin{equation}
\label{equation:BayesFactorCalculate}
    B_{1/2} \approx \frac{P(D|M_1)}{P(D|M_2)} = \frac{c_{post,1}}{c_{post,2}}
\end{equation}
\newpage
\section{ln Bayes Factor values for prior training samples (n=3, n=5, n=8)}
\subsection{Low information prior (n=3 data points per property)}

\subsubsection{$\rho_l, P_{sat}$ target}
\begin{table}[h]
\centering
\begin{tabular}[t]{c|c|c|c}
Compound & UA & AUA & AUA+Q \\
\hline
 $\mathrm{Br_2}$ & -10.91 $\pm$ 0.08 & 0 & -0.45 $\pm$ 0.13 \\
 $\mathrm{F_2}$ & 0 & -6.64 $\pm$ 0.12 & -10.49 $\pm$ 0.18 \\
 $\mathrm{N_2}$ & 0  & -4.53 $\pm$ 0.12 & -0.85 $\pm$ 0.2 \\
 $\mathrm{O_2}$ & -4.19 $\pm$ 0.08 & 0 &  -1.32 $\pm$ 0.14\\
 $\mathrm{C_2H_2}$ & -488.13 $\pm$ 0.20 & -43.13 $\pm$ 0.20 & 0\\
 $\mathrm{C_2H_4}$ & -93.21 $\pm$ 0.11 & 0 & -3.01 $\pm$ 0.21\\
 $\mathrm{C_2H_6}$ & -43.40 $\pm$ 0.10 & 0 & -1.01 $\pm$ 0.24 \\
 $\mathrm{C_2F_4}$ & -630.54 $\pm$ 0.20 & -18.91 $\pm$ 0.20 & 0\\
\end{tabular}
\caption{$\ln$~ (Bayes factors) relative to the most favored model, tested against $\rho_l, P_{sat}$ data, with low information (n=3 data points per property) training sample.}
\end{table}

\begin{figure}[h]
    \includegraphics[width=0.85\textwidth]{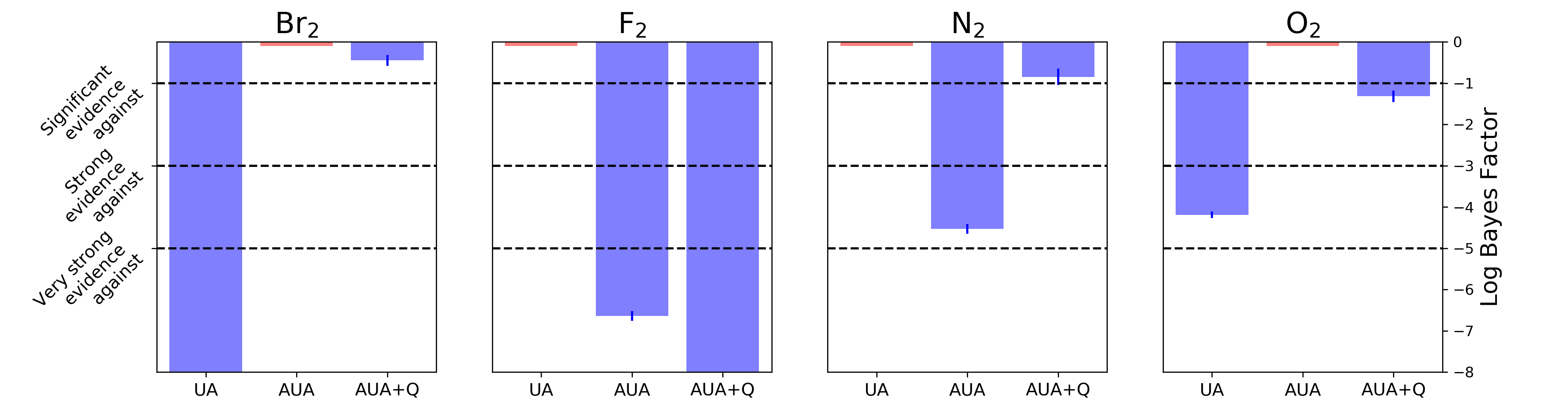}
    \includegraphics[width=0.85\textwidth]{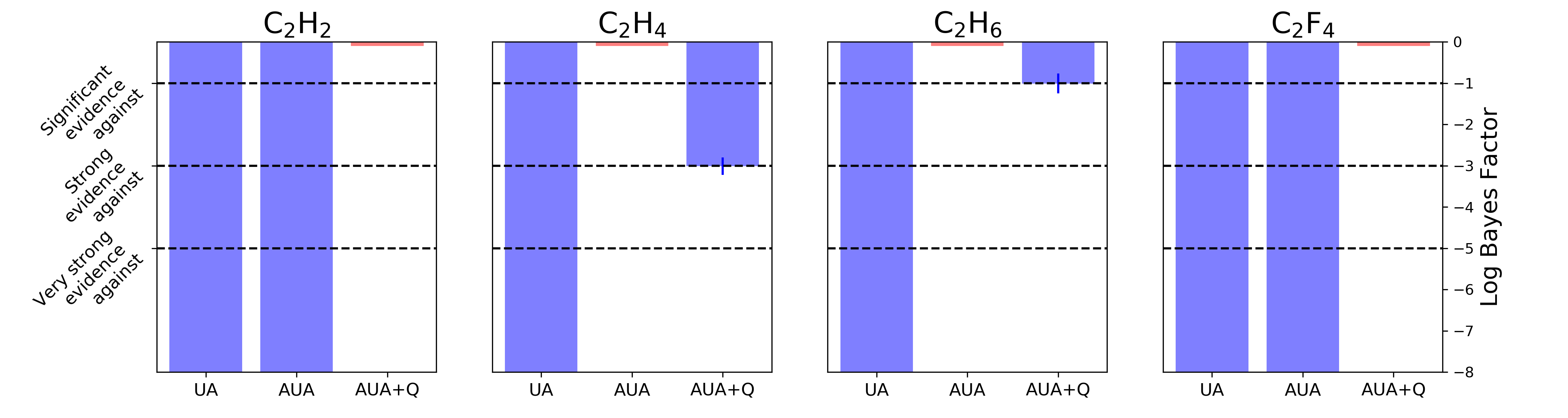}
    \caption{Bayes factors between UA, AUA, AUA+Q models for all molecules, tested against $\rho_l, P_{sat}$ data, with low information (n=3 data points per property) training sample.}
    \label{fig:2crit_low}
\end{figure}
\newpage
\subsubsection{$\rho_l, P_{sat}, \gamma$ target}
\begin{table}[h]
\centering
\begin{tabular}[t]{c|c|c|c}
Compound & UA & AUA & AUA+Q \\
\hline
 $\mathrm{Br_2}$ & -63.37 $\pm$ 0.18 & -10.98 $\pm$ 0.18 & 0 \\
 $\mathrm{F_2}$ & -3.25 $\pm$ 0.19 & -1.54 $\pm$ 0.19 & 0 \\
 $\mathrm{N_2}$ & -22.51 $\pm$ 0.12  & 0 & -1.00 $\pm$ 0.14 \\
 $\mathrm{O_2}$ & 0 & -4.54 $\pm$ 0.10 & -5.62 $\pm$ 0.12\\
 $\mathrm{C_2H_2}$ & -346.15 $\pm$ 0.08 & 0 & -142.68 $\pm$ 0.11\\
 $\mathrm{C_2H_4}$ & -130.41 $\pm$ 0.12 & 0 & -0.72 $\pm$ 0.18\\
 $\mathrm{C_2H_6}$ & -28.27 $\pm$ 0.09 & 0 & -1.32 $\pm$ 0.12\\
\end{tabular}
\caption{$\ln$ Bayes factors relative to the most favored model, tested against $\rho_l, P_{sat}, \gamma$ data, with low information (n=3 data points per property) training sample.}
\end{table}
\begin{figure}[h]
    \includegraphics[width=0.85\textwidth]{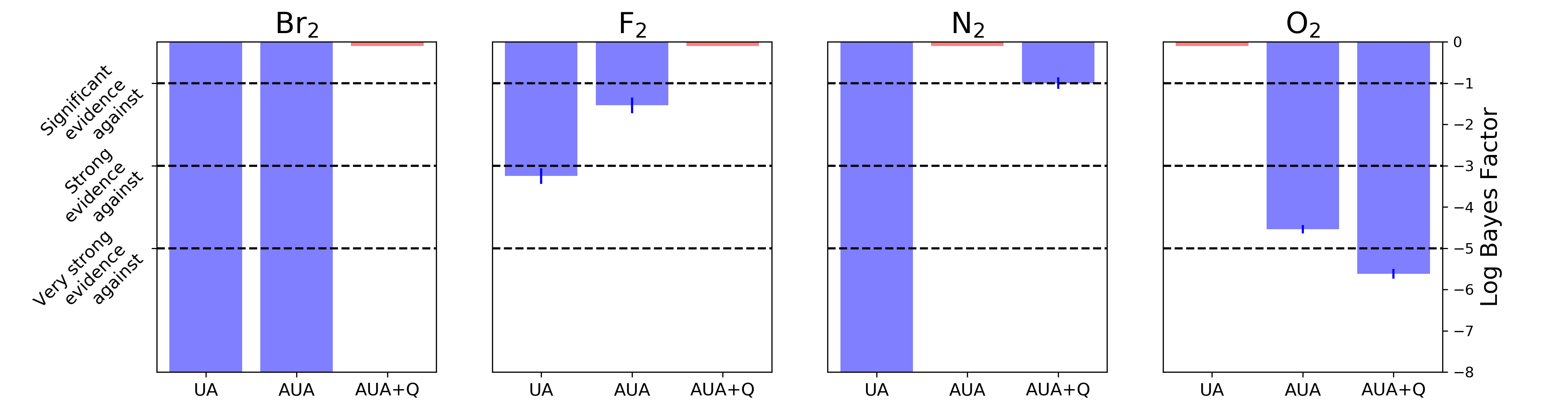}
    \includegraphics[width=0.64\textwidth]{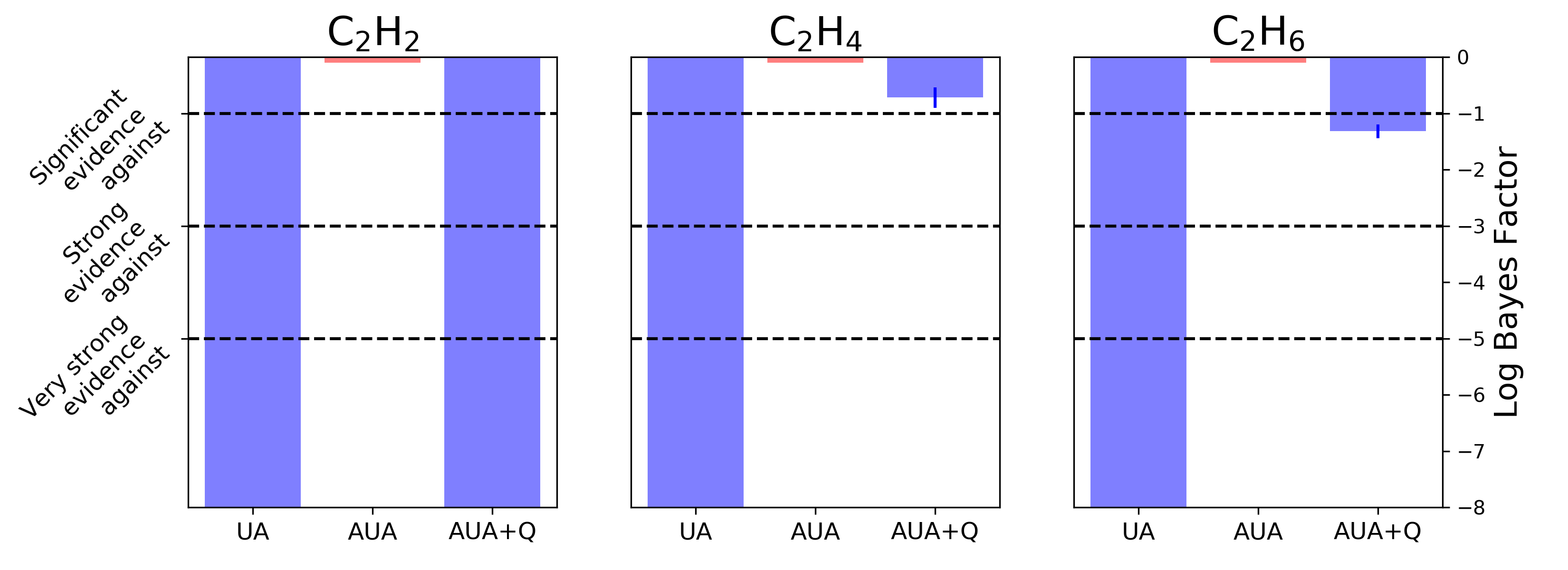}
    \caption{Bayes factors between UA, AUA, AUA+Q models for all molecules, tested against $\rho_l, P_{sat}, \gamma$ data, with low information (n=3 data points per property) training sample.}
    \label{fig:3crit_low}
\end{figure}
\newpage
\subsection{Medium information prior (n=5 data points per property)}
\subsubsection{$\rho_l, P_{sat}$ target}
\begin{table}[h]
\centering
\begin{tabular}[t]{c|c|c|c}
Compound & UA & AUA & AUA+Q \\
\hline
 $\mathrm{Br_2}$ & -9.19 $\pm$ 0.08  & 0 & -1.00 $\pm$ 0.15 \\
 $\mathrm{F_2}$ & 0 & -6.34 $\pm$ 0.09 & -6.61 $\pm$ 0.24 \\
 $\mathrm{N_2}$ & 0 & -3.33 $\pm$ 0.09 &  -2.33 $\pm$ 0.18 \\
 $\mathrm{O_2}$ & 0 & -3.99 $\pm$ 0.09 & -2.85 $\pm$ 0.15\\
 $\mathrm{C_2H_2}$ & -591.29 $\pm$ 0.19 & -74.29 $\pm$ 0.19 & 0\\
 $\mathrm{C_2H_4}$ & -117.64 $\pm$ 0.09 & 0 & -4.77 $\pm$ 0.19\\
 $\mathrm{C_2H_6}$ & -41.55 $\pm$ 0.11 & 0 & -1.82 $\pm$ 0.19\\
 $\mathrm{C_2F_4}$ & -489.93 $\pm$ 0.17 & -96.63 $\pm$ 0.17 & 0\\
\end{tabular}
\caption{$\ln$ Bayes factors relative to the most favored model, tested against $\rho_l, P_{sat}$ data, with medium information (n=5 data points per property) training sample.}
\end{table}
\begin{figure}[h]
    \includegraphics[width=0.85\textwidth]{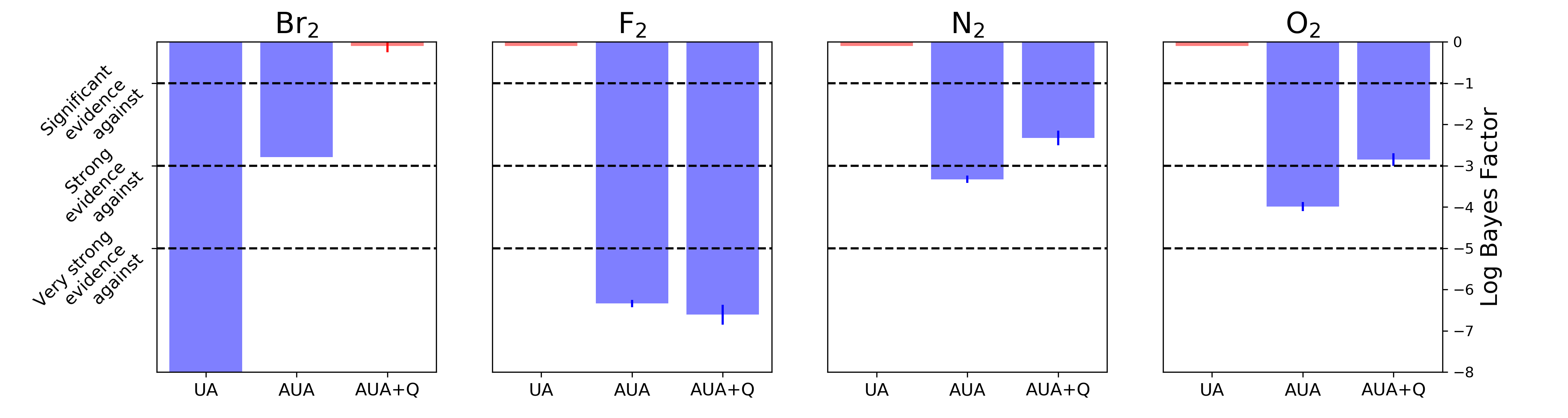}
    \includegraphics[width=0.85\textwidth]{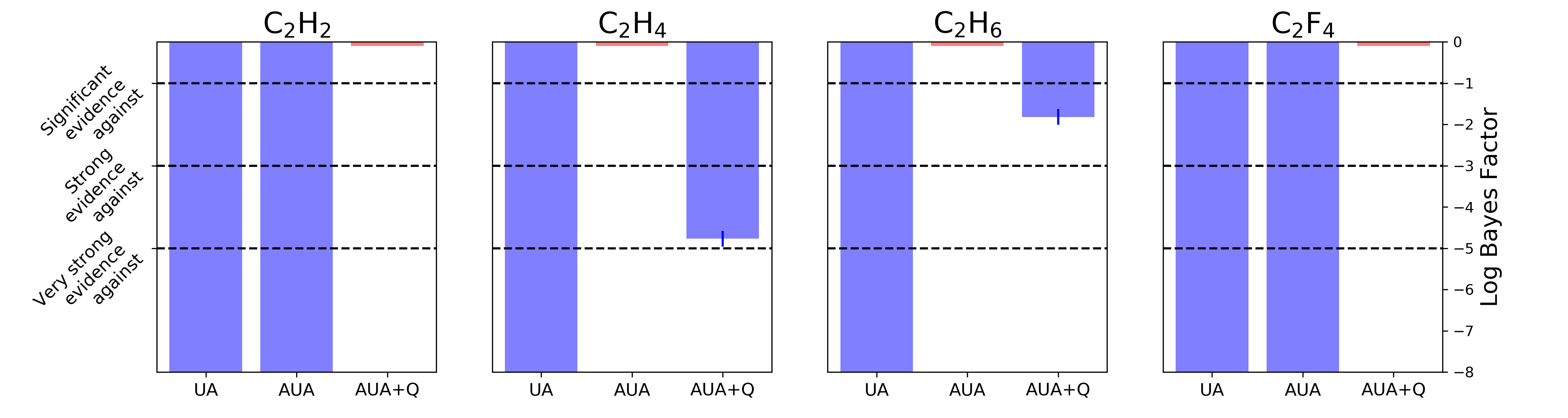}
    \caption{Bayes factors between UA, AUA, AUA+Q models for all molecules, tested against $\rho_l, P_{sat}$ data, with medium information (n=5 data points per property) training sample.}
    \label{fig:2crit_med}
\end{figure}
\newpage
\subsubsection{$\rho_l, P_{sat}, \gamma$ target}
\begin{table}[h]
\centering
\begin{tabular}[t]{c|c|c|c}
Compound & UA & AUA & AUA+Q \\
\hline
 $\mathrm{Br_2}$ & -48.98 $\pm$ 0.17  & -8.87 $\pm$ 0.17 & 0 \\
 $\mathrm{F_2}$ & -2.00 $\pm$ 0.16 & -0.02 $\pm$ 0.16 & 0 \\
 $\mathrm{N_2}$ & -17.51 $\pm$ 0.10 & 0 & -0.39 $\pm$ 0.12 \\
 $\mathrm{O_2}$ & 0 & -6.81 $\pm$ 0.09 & -6.40 $\pm$ 0.14\\
 $\mathrm{C_2H_2}$ & -287.72 $\pm$ 0.08 & 0 & -132.83 $\pm$ 0.10\\
 $\mathrm{C_2H_4}$ & -118.62 $\pm$ 0.12 & 0 & -0.47 $\pm$ 0.15\\
 $\mathrm{C_2H_6}$ & -29.98 $\pm$ 0.10 & 0 & -0.51 $\pm$ 0.12\\
\end{tabular}
\caption{$\ln$ (Bayes factors) relative to the most favored model, tested against $\rho_l, P_{sat}, \gamma$ data, with medium information (n=5 data points per property) training sample.}
\end{table}
\begin{figure}[h]
    \includegraphics[width=0.85\textwidth]{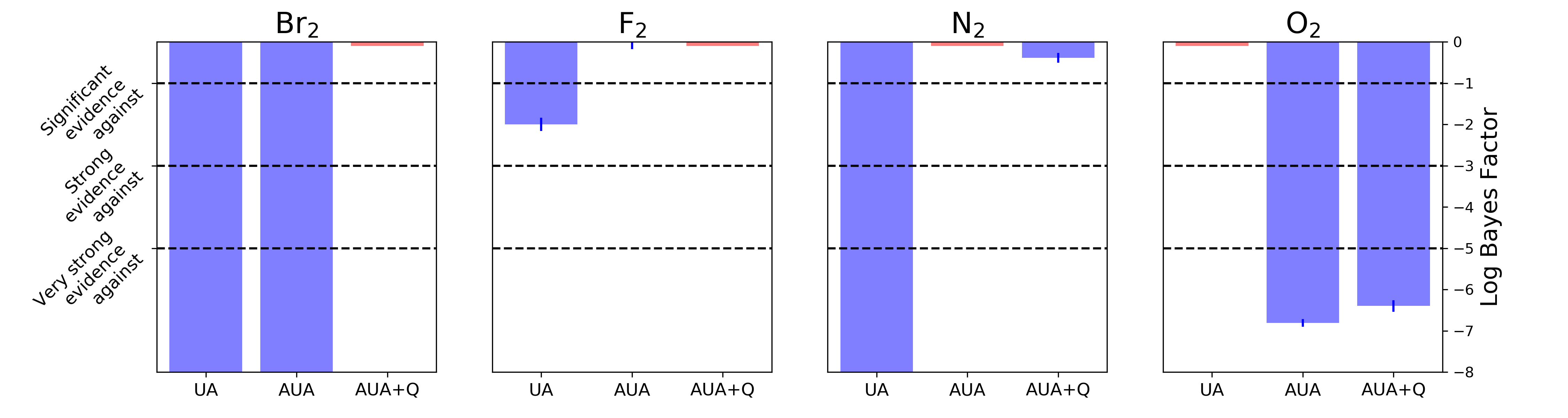}
    \includegraphics[width=0.64\textwidth]{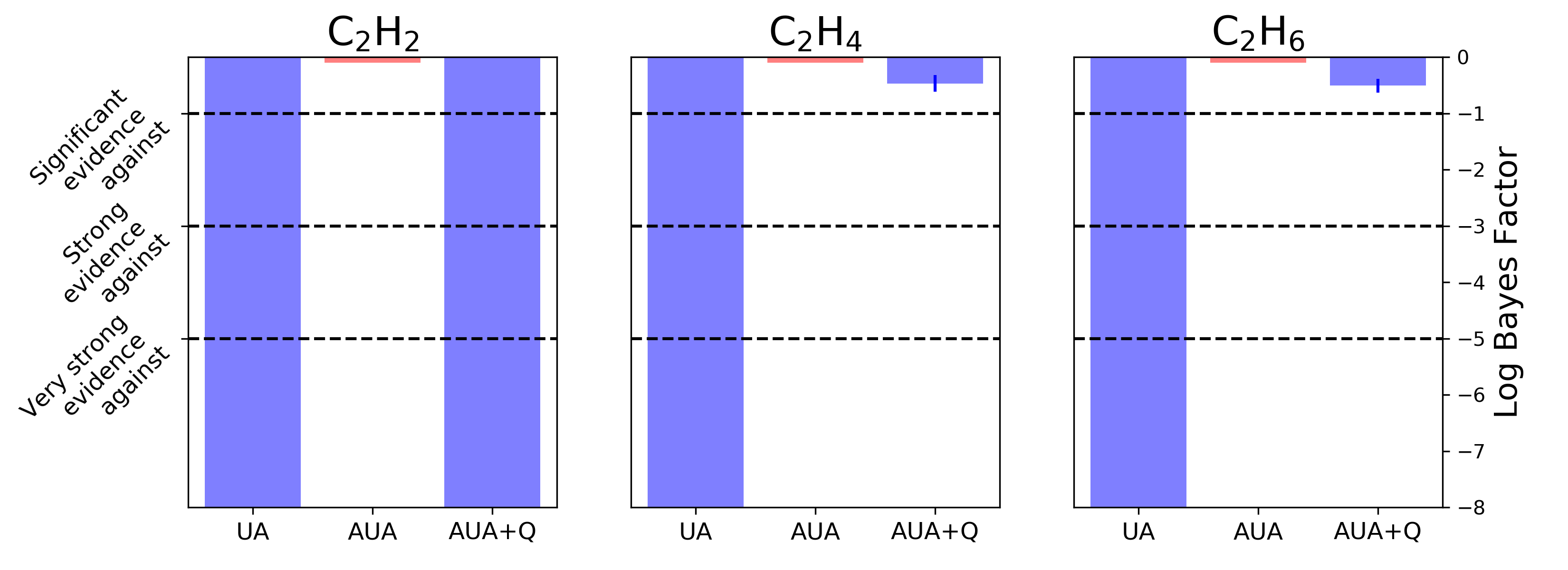}
    \caption{Bayes factors between UA, AUA, AUA+Q models for all molecules, tested against $\rho_l, P_{sat}, \gamma$ data, with medium information (n=5 data points per property) training sample.}
    \label{fig:3crit_med}
\end{figure}
\newpage
\subsection{High Information Prior (n=8 data points per property), used in final Bayes factor calculations.}
\subsubsection{$\rho_l, P_{sat}$ target}
\begin{table}[h]
\centering
\begin{tabular}[t]{c|c|c|c}
Compound & UA & AUA & AUA+Q \\
\hline
 $\mathrm{Br_2}$ & -7.94 $\pm$ 0.08 & 0 & -1.20 $\pm$ 0.12 \\
 $\mathrm{F_2}$ & 0 & -2.87 $\pm$ 0.08 & -2.77 $\pm$ 0.13 \\
 $\mathrm{N_2}$ & 0 & -4.21 $\pm$ 0.08 & -3.65 $\pm$ 0.18 \\
 $\mathrm{O_2}$ & -0.66 $\pm$ 0.09 & 0 & -1.65 $\pm$ 0.12\\
 $\mathrm{C_2H_2}$ & -382.28 $\pm$ 0.21 & -38.07 $\pm$ 0.21 & 0\\
 $\mathrm{C_2H_4}$ & -115.57 $\pm$ 0.08 & 0 & -2.81 $\pm$ 0.19\\
 $\mathrm{C_2H_6}$ & -38.46 $\pm$ 0.08 & 0 & -1.39 $\pm$ 0.14\\
  $\mathrm{C_2F_4}$ & -424.41 $\pm$ 0.23 & -84.85 $\pm$ 0.23 & 0 \\
\end{tabular}
\caption{$\ln$ (Bayes factors relative to the most favored model, tested against $\rho_l, P_{sat}$ data, with high information (n=8 data points per property) training sample.}
\end{table}
\subsubsection{$\rho_l, P_{sat}, \gamma$ target}
\begin{table}[h]
\centering
\begin{tabular}[t]{c|c|c|c}
Compound & UA & AUA & AUA+Q \\
\hline
 $\mathrm{Br_2}$ & 19.76 $\pm$ 0.18 & -3.46 $\pm$ 0.18 & 0 \\
 $\mathrm{F_2}$ & 0 & -0.79 $\pm$ 0.08 & -1.90 $\pm$ 0.16 \\
 $\mathrm{N_2}$ & -16.33 $\pm$ 0.10 & 0 & -0.19 $\pm$ 0.12 \\
 $\mathrm{O_2}$ & 0 & -6.51 $\pm$ 0.10 & -6.72 $\pm$ 0.13\\
 $\mathrm{C_2H_2}$ & -206.30 $\pm$ 0.10 & -50.14 $\pm$ 0.10 & 0\\
 $\mathrm{C_2H_4}$ & -78.90 $\pm$ 0.13 & -0.96 $\pm$ 0.17 & 0\\
 $\mathrm{C_2H_6}$ & -23.50 $\pm$ 0.10 & 0 & -0.24 $\pm$ 0.13\\
\end{tabular}
\caption{$\ln$ Bayes factors relative to the most favored model, tested against $\rho_l, P_{sat}, \gamma$ data, with high information (n=8 data points per property) training sample.}
\end{table}
\newpage
\section{ELPPD Benchmarking Results}
\subsection{$\rho_l, P_{sat}$ target}
\begin{table}[h]
\centering
\begin{tabular}[t]{|c|c|c|c|c|c|c|}
\hline 
\multicolumn{1}{|c|}{} & \multicolumn{3}{|c|}{ELPPD Avg. over test points} & \multicolumn{3}{|c|}{Avg. Stdev. from Exp.} \\
\hline
Compound & UA & AUA & AUA+Q & UA & AUA & AUA+Q \\
\hline
\multicolumn{7}{|c|}{$\rho_l$}\\
\hline
 $\mathrm{Br_2}$ &--- & --- & --- & --- & --- & --- \\
 $\mathrm{F_2}$ & --- & --- & --- & --- & --- & --- \\
 $\mathrm{N_2}$ & 1.51 & 1.13 & 0.94 & 1.74 & 1.50 & 1.37\\
 $\mathrm{O_2}$ & 1.36 & 1.26 & 1.22 & 1.65 & 1.59 & 1.57\\
 $\mathrm{C_2H_2}$ & --- & --- & --- &--- & --- & --- \\
 $\mathrm{C_2H_4}$ & 9.59 & 1.59 & 1.53 & 4.38 & 1.78 & 1.75\\
 $\mathrm{C_2H_6}$ & 3.07 & 0.81 & 0.82 & 2.48 & 1.27 & 1.28\\
 $\mathrm{C_2F_4}$ & ---  & --- & --- &--- & --- & --- \\
 \hline
\multicolumn{7}{|c|}{$P_{sat}$}\\
\hline
 $\mathrm{Br_2}$ & 0.31 & 0.10 & 0.09 & 0.79 & 0.44 & 0.43\\
 $\mathrm{F_2}$ & 0.03 & 0.06 & 0.06 & 0.26 & 0.33 & 0.35 \\
 $\mathrm{N_2}$ & 0.15 & 0.20 & 0.19 & 0.54 & 0.63 & 0.62 \\
 $\mathrm{O_2}$ & 0.87 & 0.25 & 0.38 & 1.32 & 0.70 & 0.88\\
 $\mathrm{C_2H_2}$ & 12.98 & 0.95 & 0.14 & 5.09 & 1.38 & 0.54\\
 $\mathrm{C_2H_4}$ & 3.78 & 0.33 & 0.52 & 2.75 & 0.81 & 1.02\\
 $\mathrm{C_2H_6}$ & 1.95 & 0.19 & 0.21 & 1.98 & 0.62 & 0.64\\
 $\mathrm{C_2F_4}$ & 17.84 & 2.83 & 0.21 & 5.97 & 2.38 & 0.64\\
 \hline
\end{tabular}
\caption{ELPPD Benchmarking for the $\rho_l, P_{sat}$ target with high information priors. ELPPD averaged over test points is (total ELPPD value/number of test points).  While this is not a true average due to the nature of the ELPPD, it allows for comparison when numbers of test data points are different. Average standard deviations from experimental value over test points also shown.  Larger values indicate worse overall model performance.  ELPPD measurements omitted for properties with insufficient (n$<$10) measurements not already used in prior fitting or Bayes factor calculations.}
\end{table}
\newpage
\subsection{$\rho_l, P_{sat}, \gamma$ target}
\begin{table}[h]
\centering
\begin{tabular}[t]{|c|c|c|c|c|c|c|}
\hline 
\multicolumn{1}{|c|}{} & \multicolumn{3}{|c|}{ELPPD} & \multicolumn{3}{|c|}{Avg. Stdev. from Experiment} \\
\hline
Compound & UA & AUA & AUA+Q & UA & AUA & AUA+Q \\
\hline
\multicolumn{7}{|c|}{$\rho_l$}\\
\hline
 $\mathrm{Br_2}$ & --- & --- & --- &--- & --- & ---\\
 $\mathrm{F_2}$ & --- & --- & --- &--- & --- & ---\\
 $\mathrm{N_2}$ & 0.81 & 0.67 & 0.66 & 1.27 & 1.16 & 1.15 \\
 $\mathrm{O_2}$ & 1.94 & 2.21 & 2.18 & 1.97 & 2.10 & 2.09\\
 $\mathrm{C_2H_2}$ & --- & --- & --- &--- & --- & ---\\
 $\mathrm{C_2H_4}$ & 0.59 & 1.20 & 0.89 & 1.09 & 1.55 & 1.33\\
 $\mathrm{C_2H_6}$ & 2.01 & 0.88 & 0.88 & 2.01 & 1.33 & 1.32\\
 \hline
\multicolumn{7}{|c|}{$P_{sat}$}\\
\hline
 $\mathrm{Br_2}$ & 1.63 & 0.75 & 0.50 & 1.81 & 1.22 & 1.00 \\
 $\mathrm{F_2}$ & --- & --- & --- &--- & --- & ---\\
 $\mathrm{N_2}$ & 1.52 & 1.18 & 1.17 & 1.74 & 1.54 & 1.53 \\
 $\mathrm{O_2}$ & 1.50 & 1.75 & 1.72 & 1.73 & 1.87 & 1.85\\
 $\mathrm{C_2H_2}$ & 9.01 & 0.73 & 0.08 & 4.25 & 1.21 & 0.41\\
 $\mathrm{C_2H_4}$ & 6.85 & 2.70 & 3.43 & 3.70 & 2.33 & 2.62\\
 $\mathrm{C_2H_6}$ & 3.97 & 1.09 & 1.18 & 2.82 & 1.47 & 1.53\\
 \hline
 \multicolumn{7}{|c|}{$\gamma$}\\
\hline
 $\mathrm{Br_2}$ & --- & --- & --- &--- & --- & ---\\
 $\mathrm{F_2}$ & --- & --- & --- &--- & --- & ---\\
 $\mathrm{N_2}$ & 9.03 & 6.99 & 7.04 & 4.25 & 3.74 & 3.75\\
 $\mathrm{O_2}$ & 8.12 & 7.71 & 7.76 & 4.03 & 3.93 & 3.94\\
 $\mathrm{C_2H_2}$ & --- & --- & --- &--- & --- & ---\\
 $\mathrm{C_2H_4}$ & --- & --- & --- &--- & --- & ---\\
 $\mathrm{C_2H_6}$ & 2.97 & 4.18 & 4.11 & 2.44 & 2.89 & 2.87\\
 \hline
\end{tabular}
\caption{ELPPD Benchmarking for the $\rho_l, P_{sat}, \gamma$ target with high information priors. ELPPD averaged over test points is (total ELPPD value/number of test points).  Average standard deviations from experimental value over test points also shown.  Larger values indicate worse overall model performance.  ELPPD measurements omitted for properties with insufficient (n$<$10) measurements not already used in prior fitting or Bayes factor calculations.}
\end{table}
\newpage
\section{Benchmarking Figures}
\subsection{$\rho_l, P_{sat}$ target}
\subsubsection{F$_2$}
\begin{figure}[h]
\centering
    \includegraphics[width=0.48\textwidth]{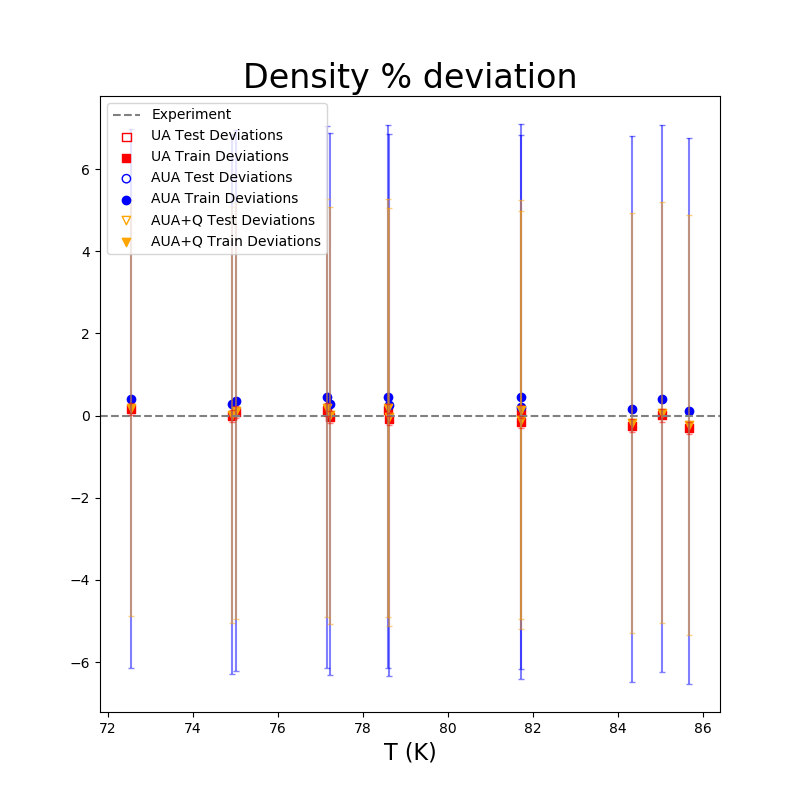}
    \includegraphics[width=0.48\textwidth]{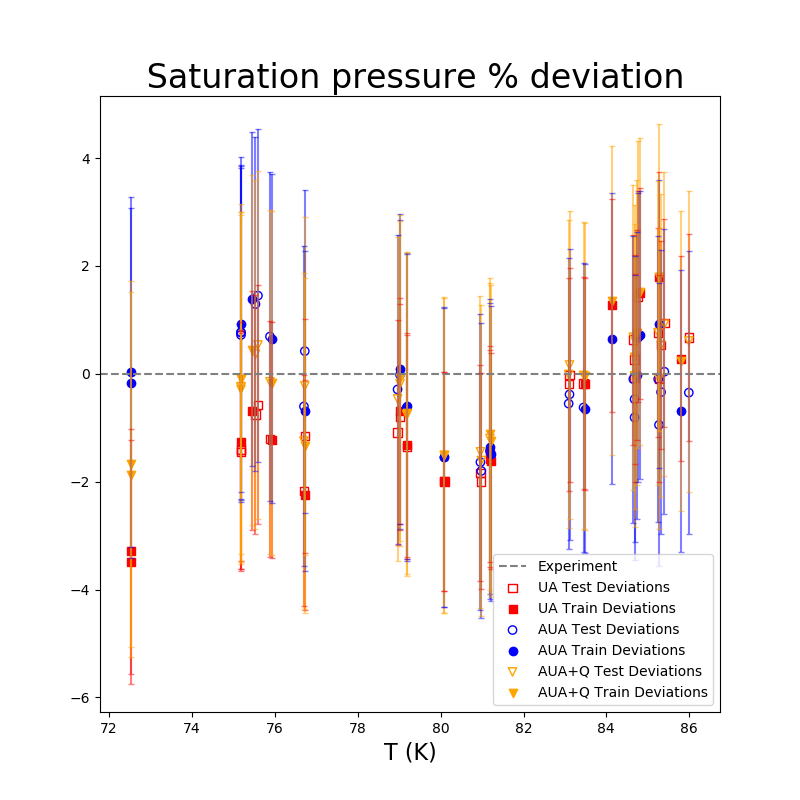}
    \caption{ Average $\rho_l$ (left panel), $P_{sat}$ (right panel) \% deviation plots for F$_2$. Parameter sets drawn from posterior probability distribution, evaluated against separate benchmark data points (open points) as well as points used in calculated Bayes factor (filled points).}
\end{figure}
\newpage
\subsubsection{Br$_2$}
\begin{figure}[h]
\centering
    \includegraphics[width=0.48\textwidth]{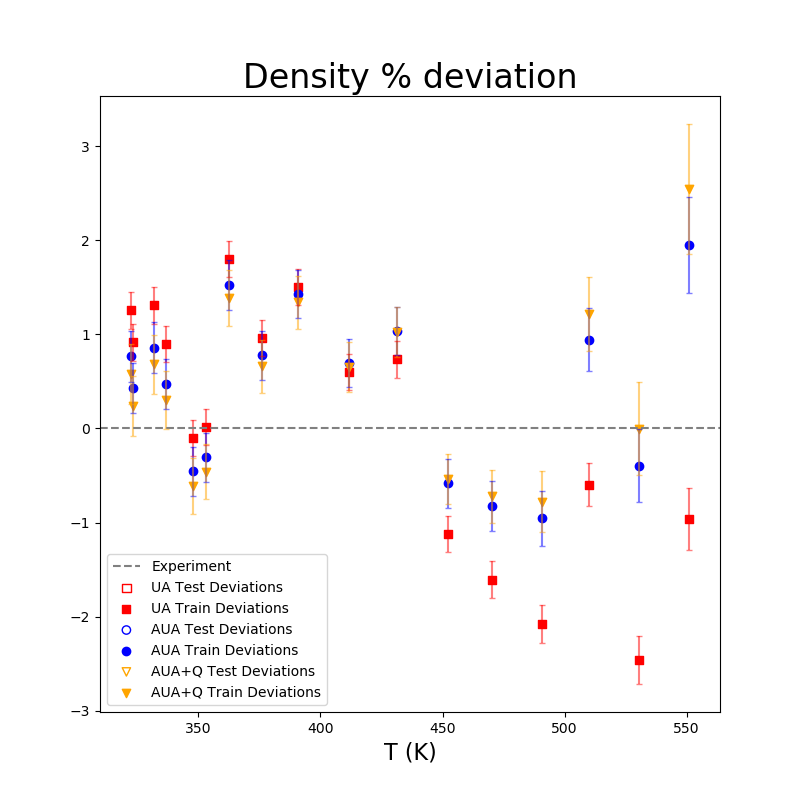}
    \includegraphics[width=0.48\textwidth]{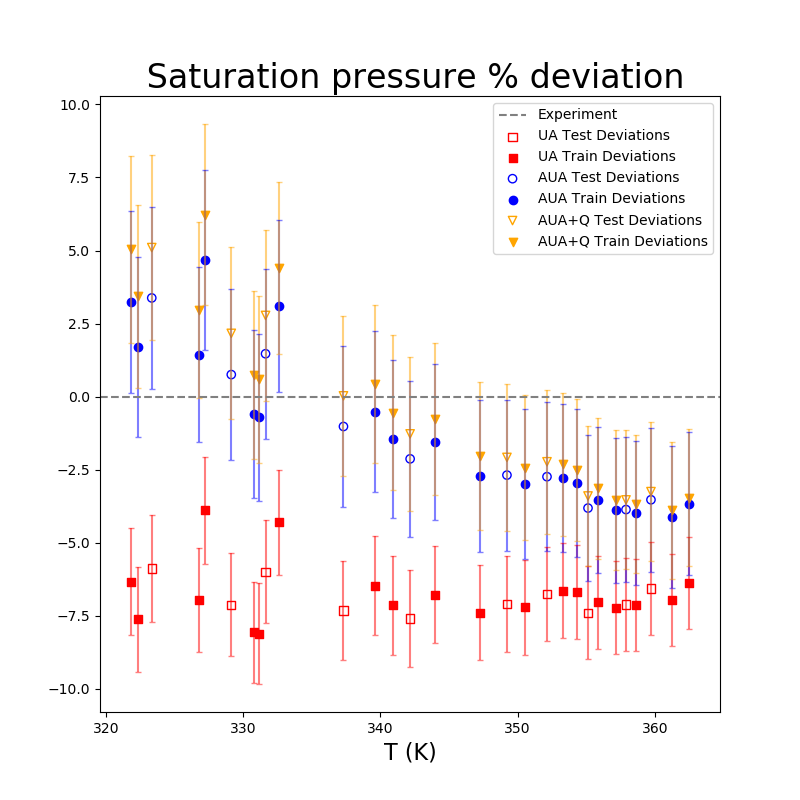}
    \caption{ Average $\rho_l$ (left panel), $P_{sat}$ (right panel) \% deviation plots for Br$_2$. Parameter sets drawn from posterior probability distribution, evaluated against separate benchmark data points (open points) as well as points used in calculated Bayes factor (filled points).}
\end{figure}
\newpage
\subsubsection{N$_2$}
\begin{figure}[h]
\centering
    \includegraphics[width=0.48\textwidth]{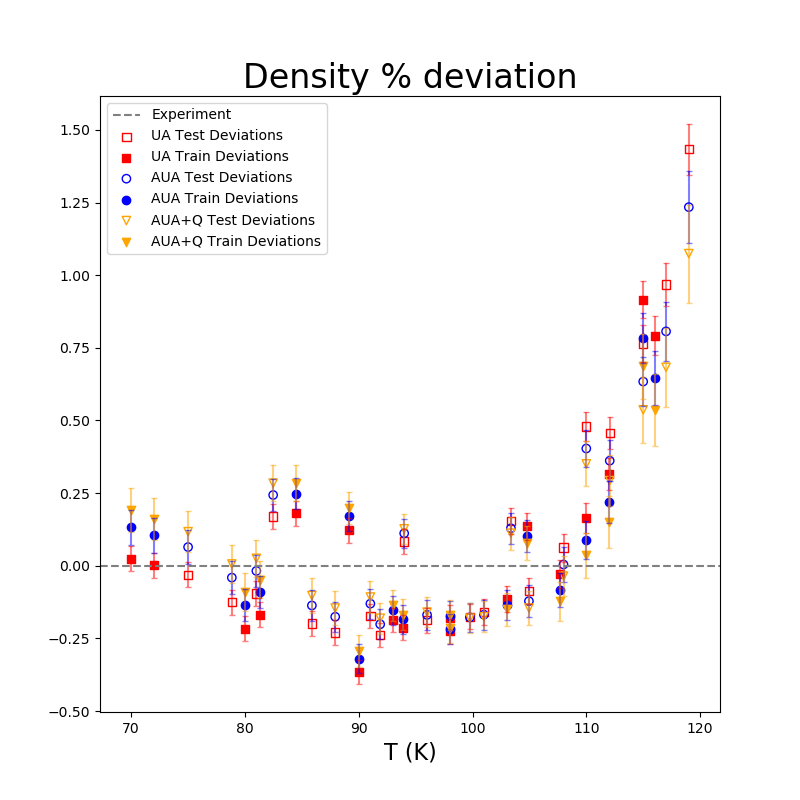}
    \includegraphics[width=0.48\textwidth]{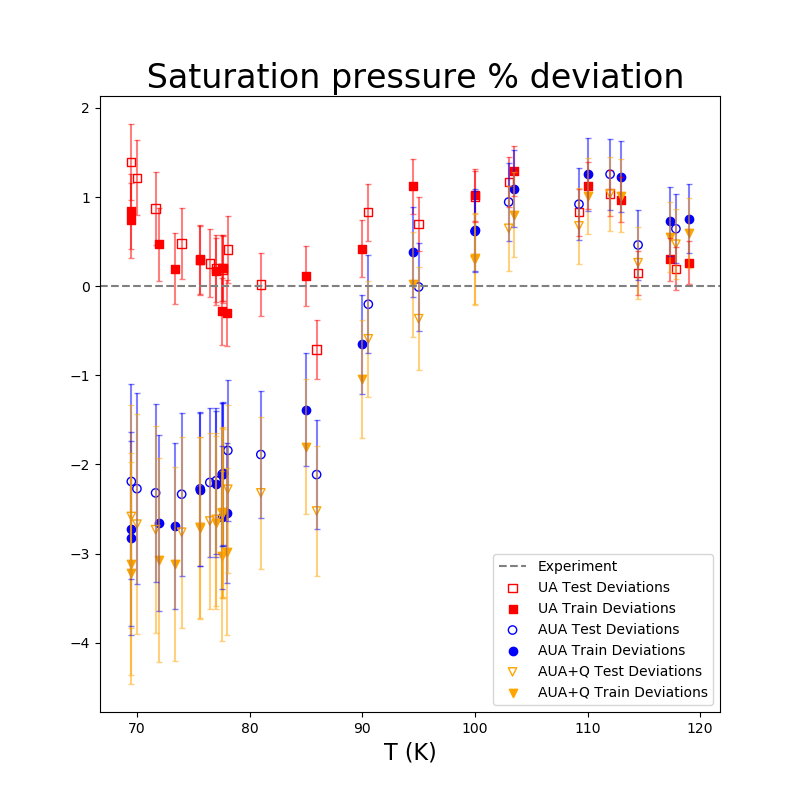}
    \caption{ Average $\rho_l$ (left panel), $P_{sat}$ (right panel) \% deviation plots for N$_2$. Parameter sets drawn from posterior probability distribution, evaluated against separate benchmark data points (open points) as well as points used in calculated Bayes factor (filled points).}
\end{figure}
\newpage
\subsubsection{O$_2$}
\begin{figure}[h]
\centering
    \includegraphics[width=0.48\textwidth]{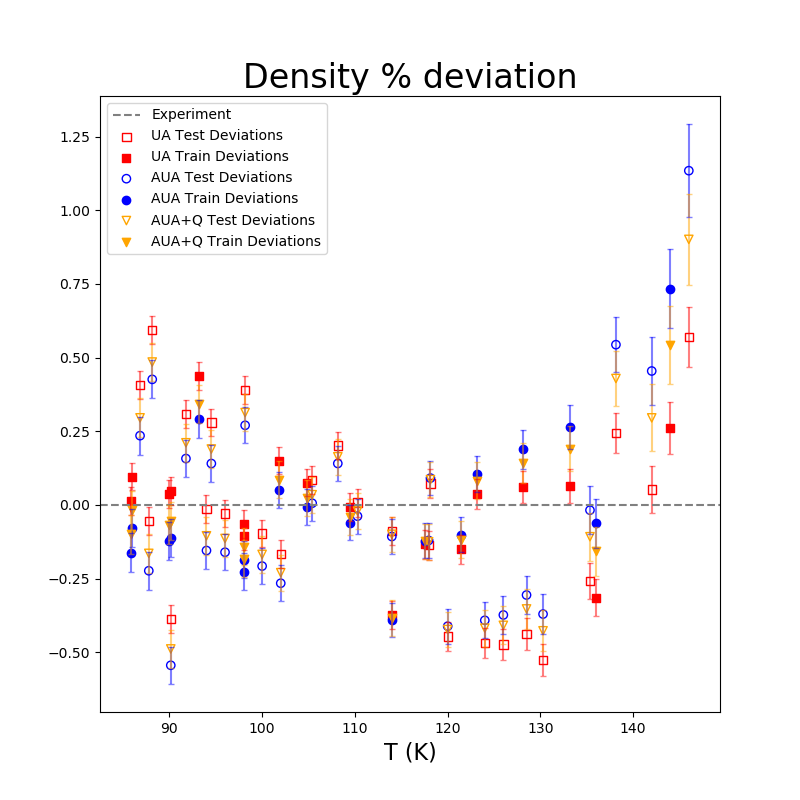}
    \includegraphics[width=0.48\textwidth]{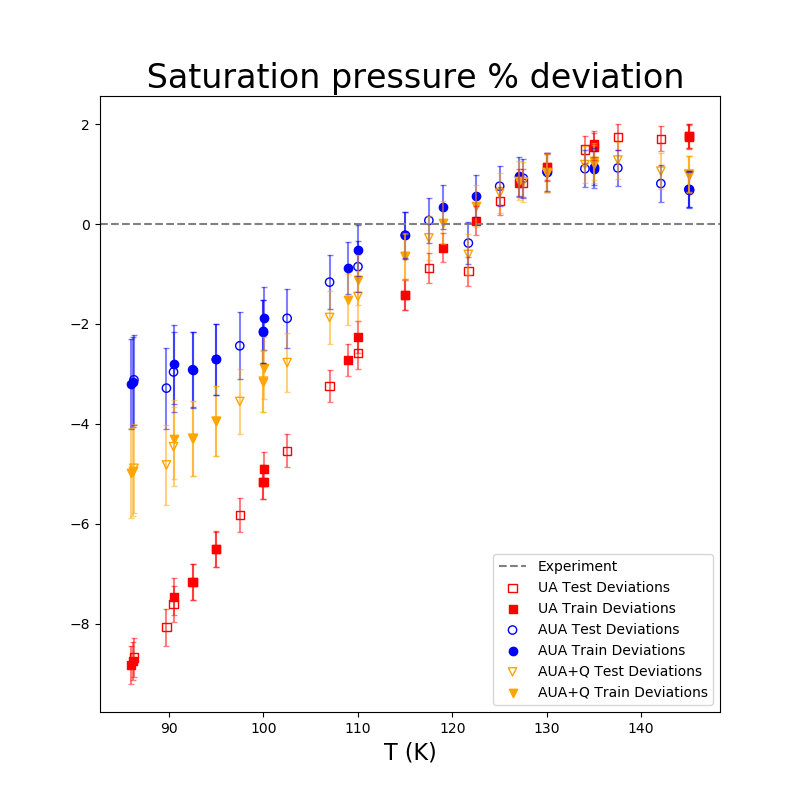}
    \caption{ Average $\rho_l$ (left panel), $P_{sat}$ (right panel) \% deviation plots for O$_2$. Parameter sets drawn from posterior probability distribution, evaluated against separate benchmark data points (open points) as well as points used in calculated Bayes factor (filled points).}
\end{figure}
\newpage
\subsubsection{C$_2$H$_2$}
\begin{figure}[h]
\centering
    \includegraphics[width=0.48\textwidth]{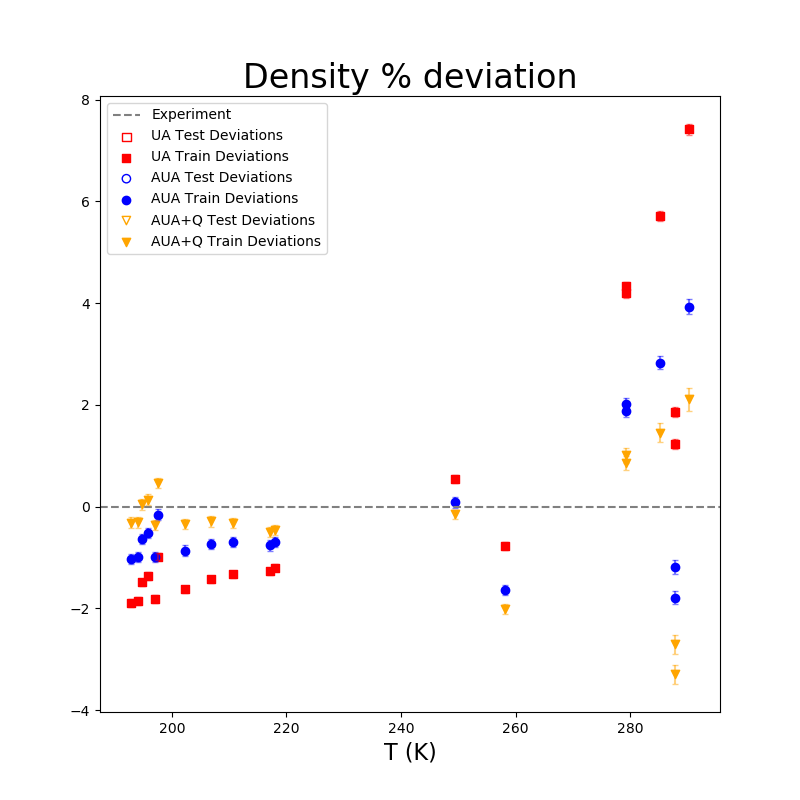}
    \includegraphics[width=0.48\textwidth]{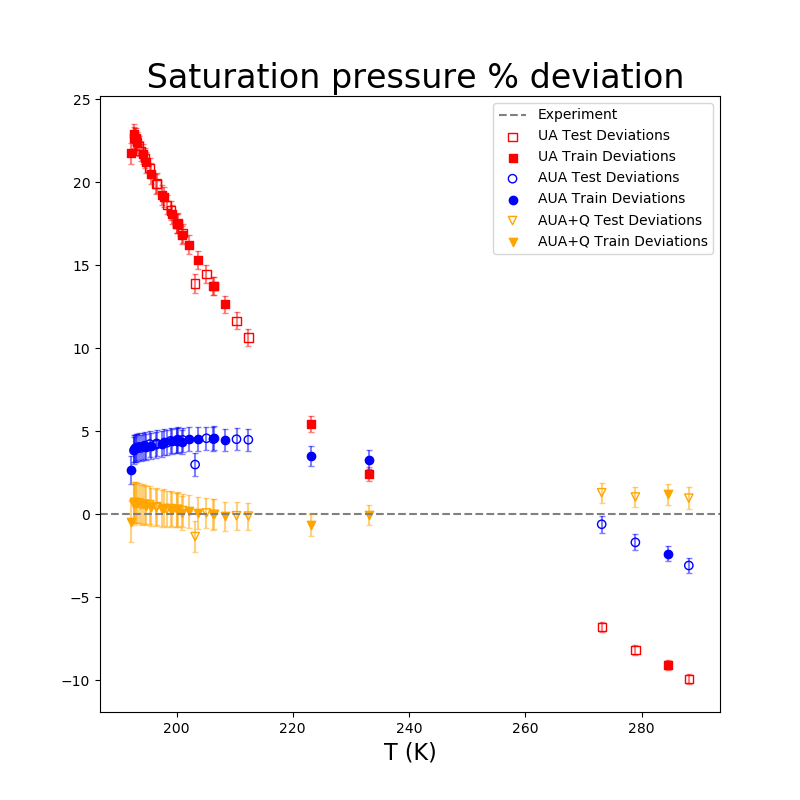}
    \caption{ Average $\rho_l$ (left panel), $P_{sat}$ (right panel) \% deviation plots for C$_2$H$_2$. Parameter sets drawn from posterior probability distribution, evaluated against separate benchmark data points (open points) as well as points used in calculated Bayes factor (filled points).}
\end{figure}
\newpage
\subsubsection{C$_2$H$_4$}
\begin{figure}[h]
\centering
    \includegraphics[width=0.48\textwidth]{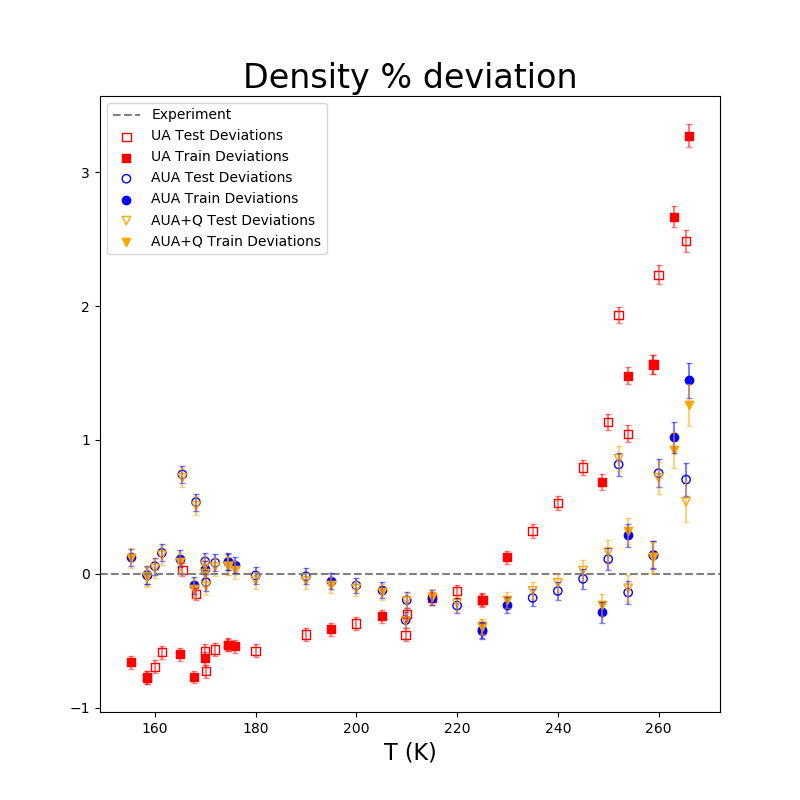}
    \includegraphics[width=0.48\textwidth]{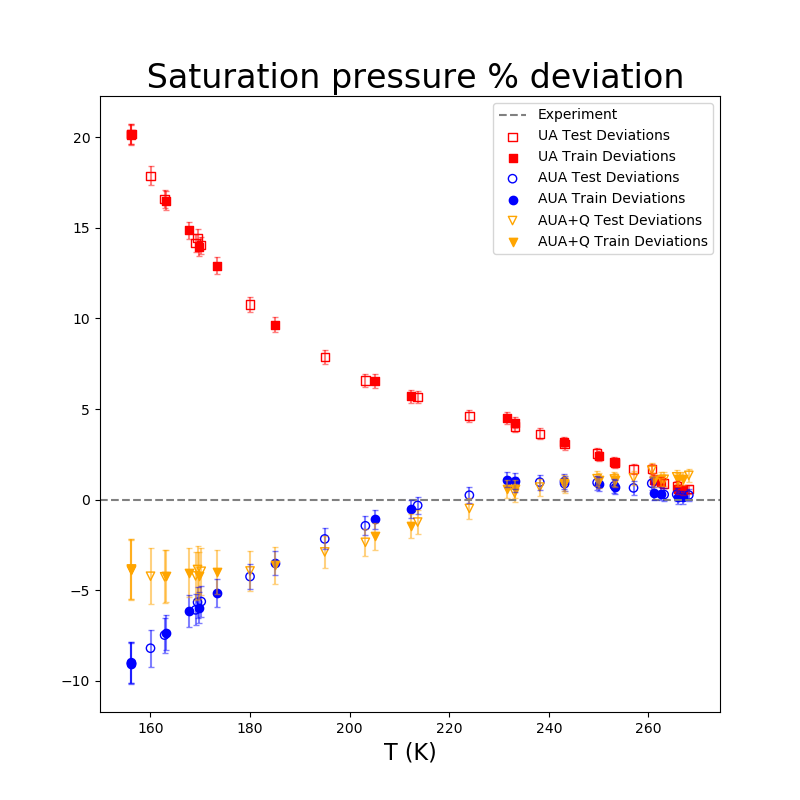}
    \caption{ Average $\rho_l$ (left panel), $P_{sat}$ (right panel) \% deviation plots for C$_2$H$_4$. Parameter sets drawn from posterior probability distribution, evaluated against separate benchmark data points (open points) as well as points used in calculated Bayes factor (filled points).}
\end{figure}
\newpage
\subsubsection{C$_2$H$_6$}
\begin{figure}[h]
\centering
    \includegraphics[width=0.48\textwidth]{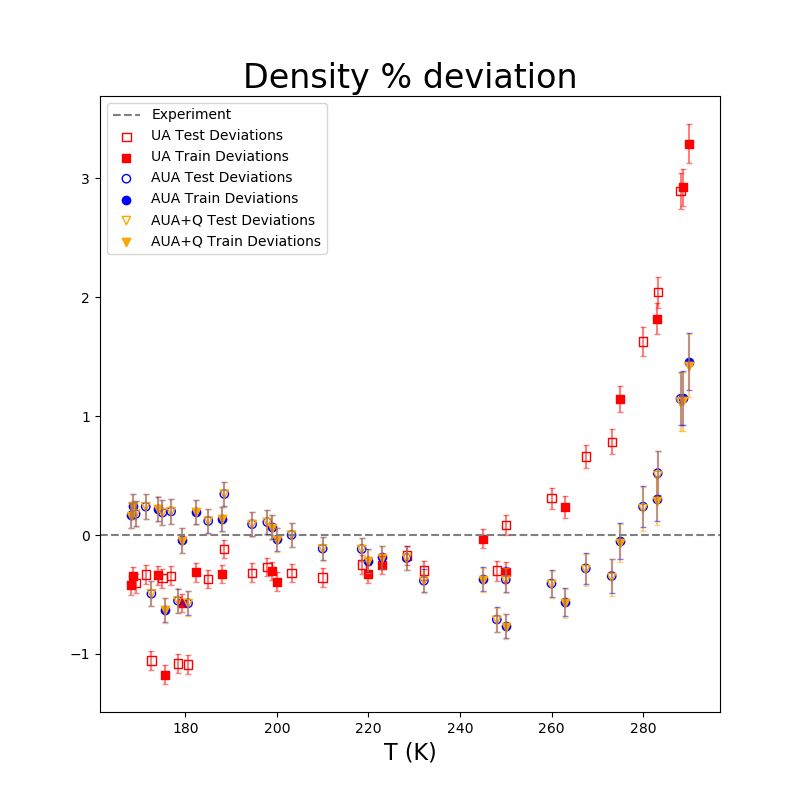}
    \includegraphics[width=0.48\textwidth]{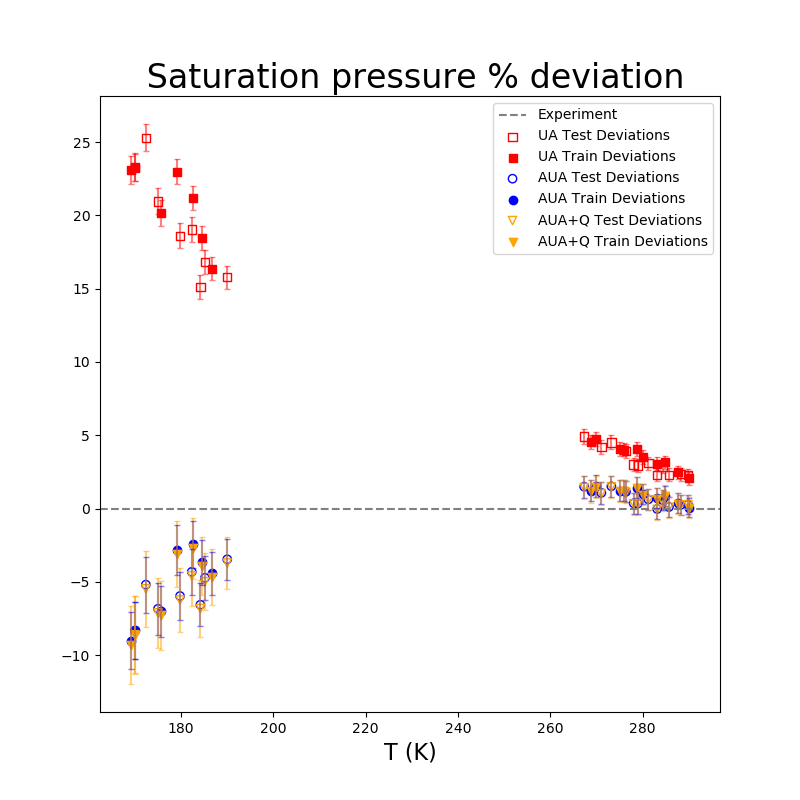}
    \caption{ Average $\rho_l$ (left panel), $P_{sat}$ (right panel) \% deviation plots for C$_2$H$_6$. Parameter sets drawn from posterior probability distribution, evaluated against separate benchmark data points (open points) as well as points used in calculated Bayes factor (filled points).}
\end{figure}
\newpage
\subsubsection{C$_2$F$_4$}
\begin{figure}[h]
\centering
    \includegraphics[width=0.48\textwidth]{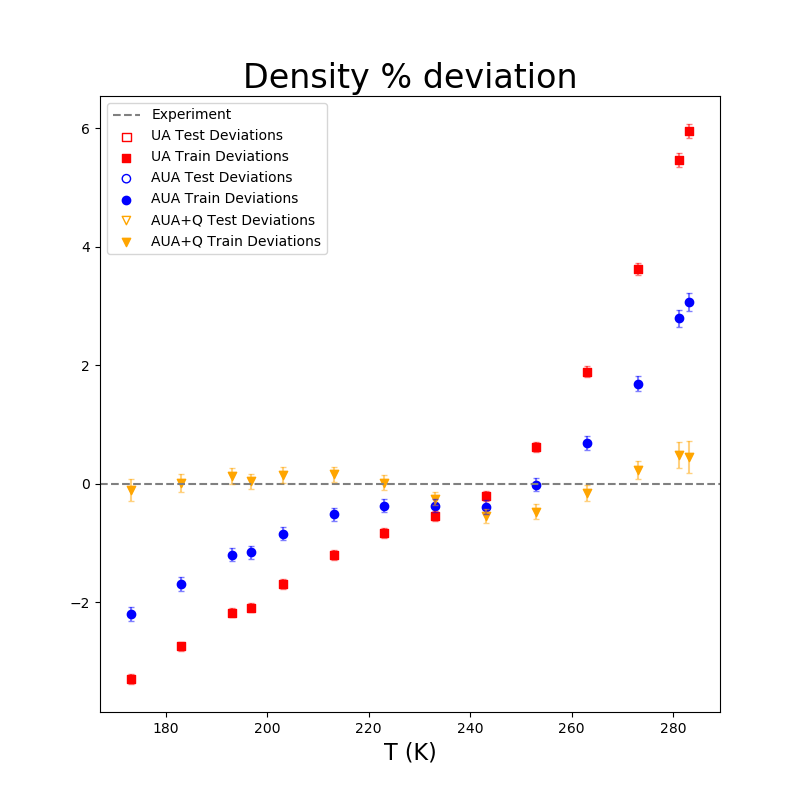}
    \includegraphics[width=0.48\textwidth]{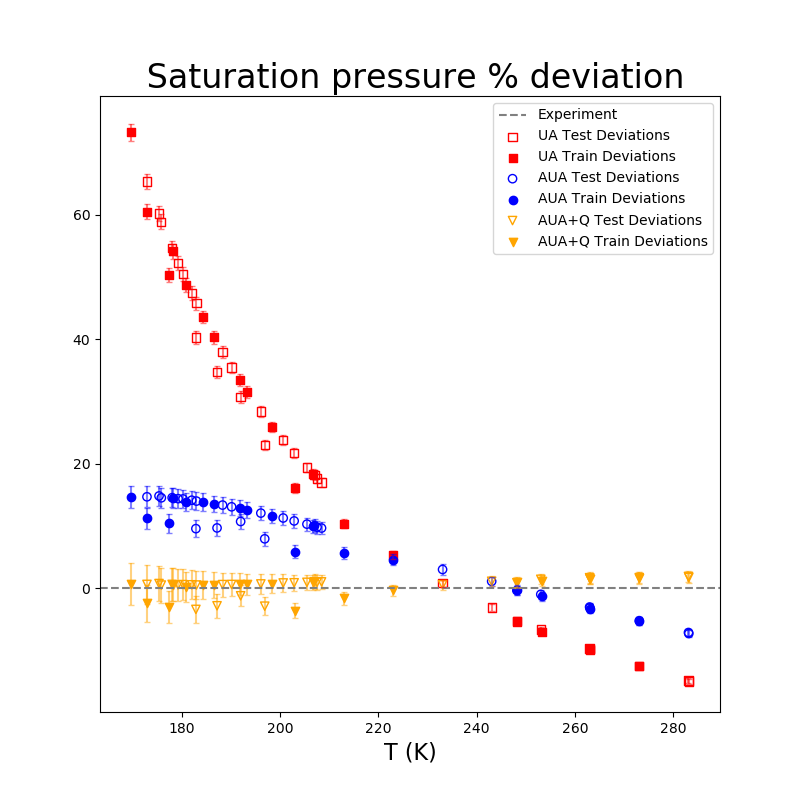}
    \caption{ Average $\rho_l$ (left panel), $P_{sat}$ (right panel) \% deviation plots for C$_2$F$_4$. Parameter sets drawn from posterior probability distribution, evaluated against separate benchmark data points (open points) as well as points used in calculated Bayes factor (filled points).}
\end{figure}
\newpage
\subsection{$\rho_l, P_{sat}, \gamma$ target}
\subsubsection{F$_2$}
\begin{figure}[h]
\centering
    \includegraphics[width=0.4\textwidth]{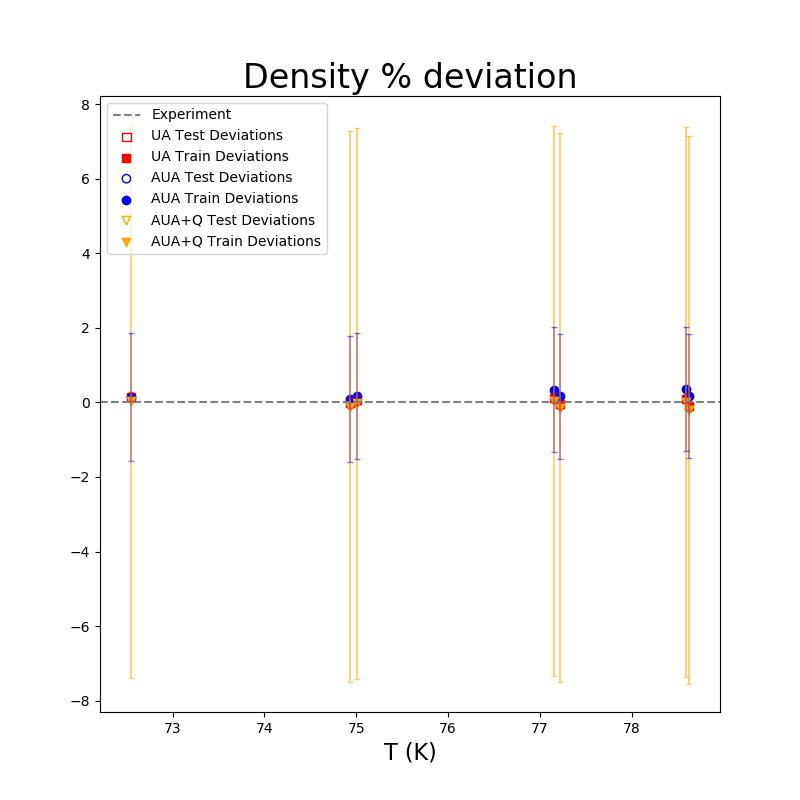}
    \includegraphics[width=0.4\textwidth]{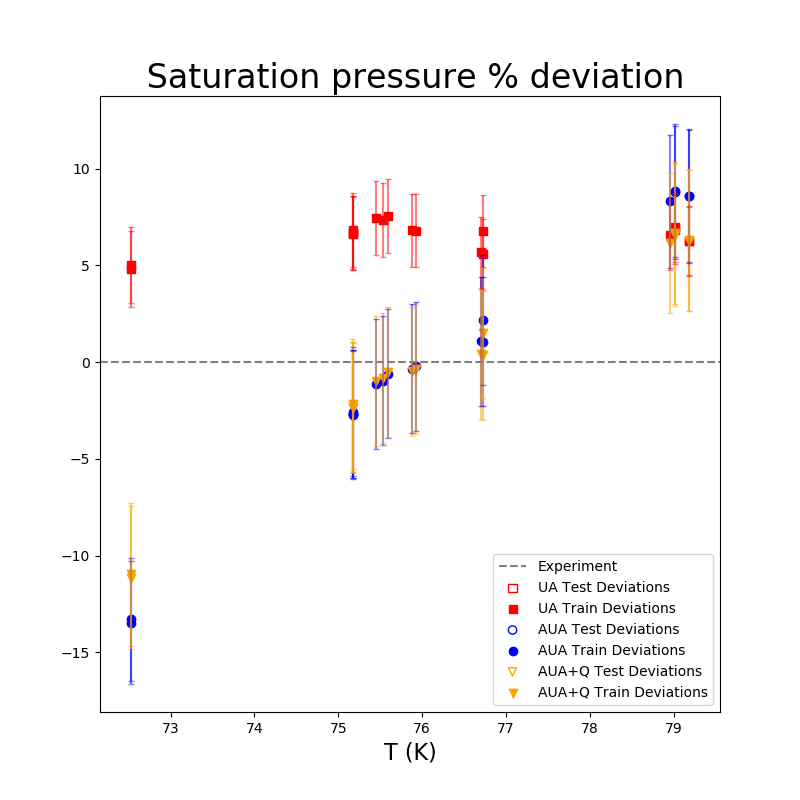}
    \includegraphics[width=0.4\textwidth]{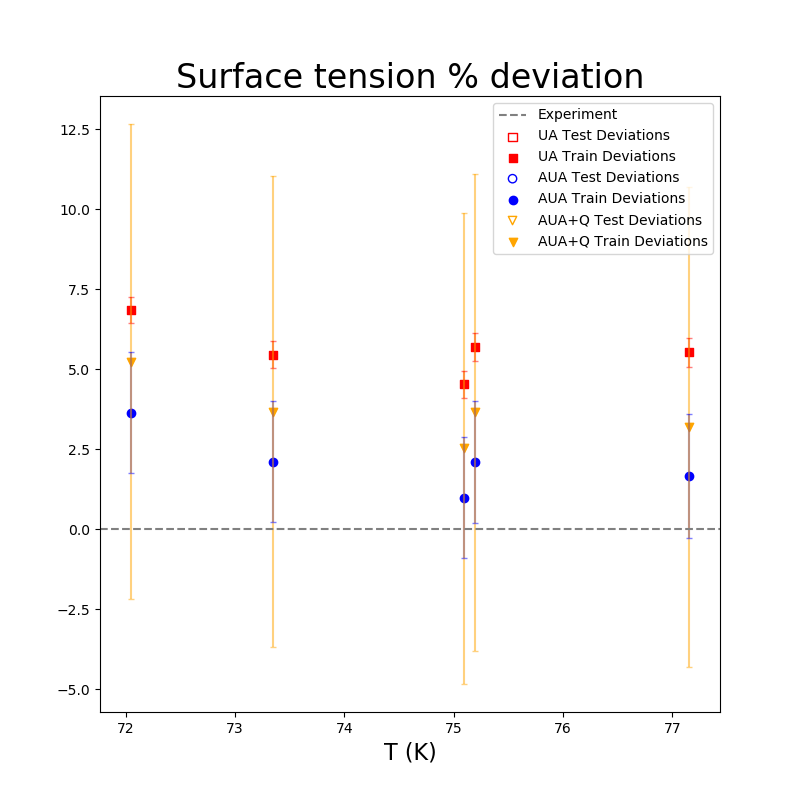}
    \caption{ Average $\rho_l$ (top left panel), $P_{sat}$ (top right panel), $\gamma$ (bottom panel) \% deviation plots for F$_2$. Parameter sets drawn from posterior probability distribution, evaluated against separate benchmark data points (open points) as well as points used in calculated Bayes factor (filled points).}
\end{figure}
\newpage
\subsubsection{Br$_2$}
\begin{figure}[h]
\centering
    \includegraphics[width=0.4\textwidth]{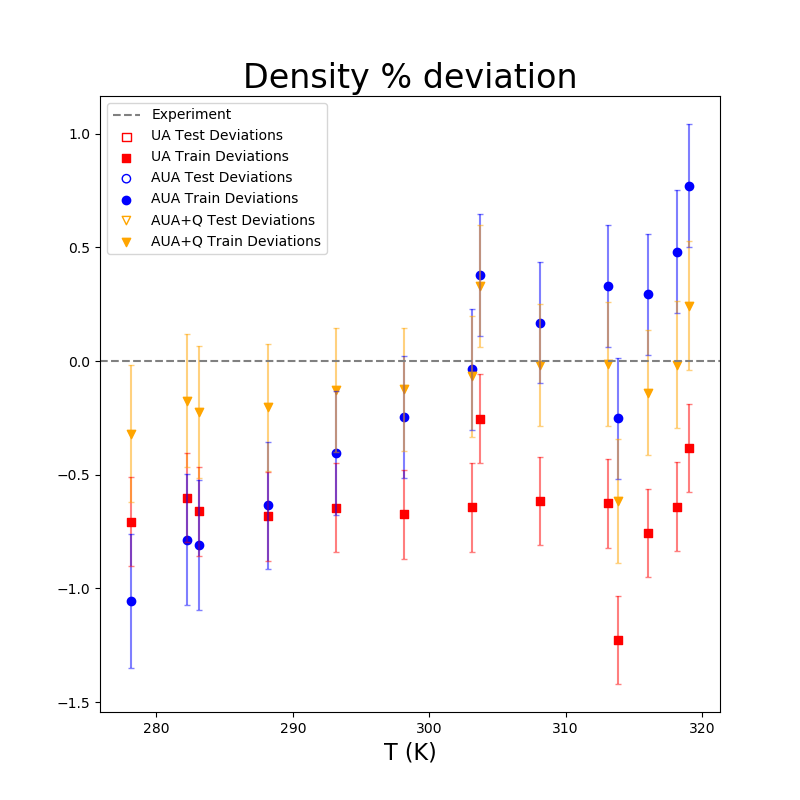}
    \includegraphics[width=0.4\textwidth]{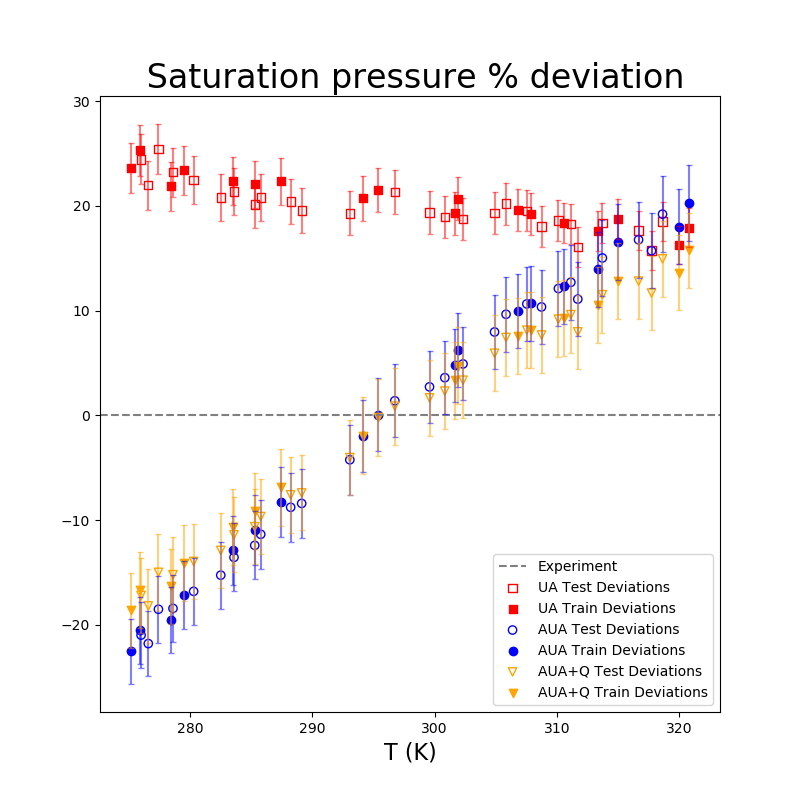}
    \includegraphics[width=0.4\textwidth]{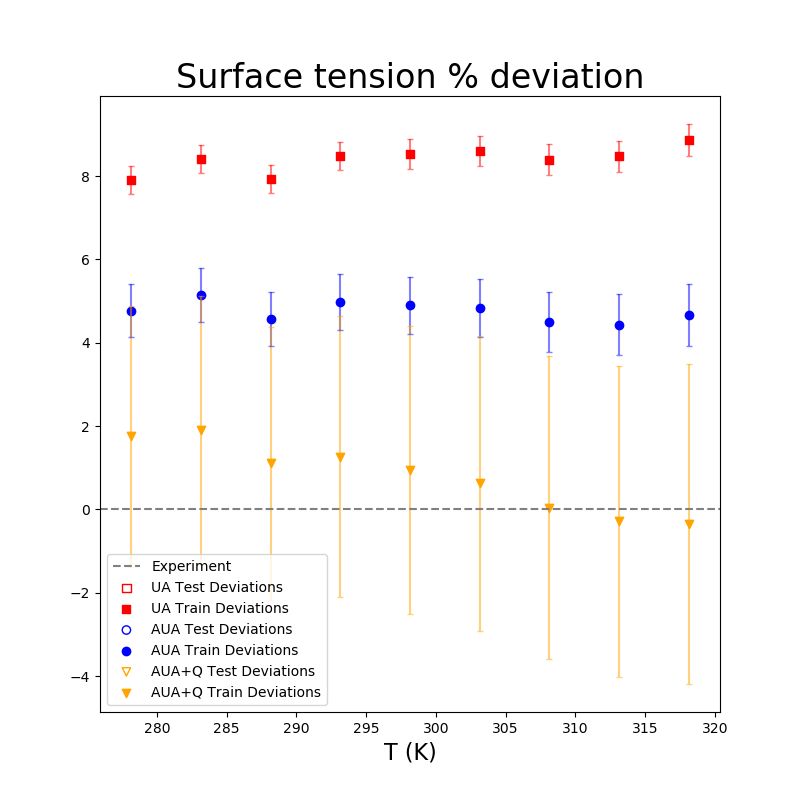}
    \caption{ Average $\rho_l$ (top left panel), $P_{sat}$ (top right panel), $\gamma$ (bottom panel) \% deviation plots for Br$_2$. Parameter sets drawn from posterior probability distribution, evaluated against separate benchmark data points (open points) as well as points used in calculated Bayes factor (filled points).}
\end{figure}
\newpage
\subsubsection{N$_2$}
\begin{figure}[h]
\centering
    \includegraphics[width=0.4\textwidth]{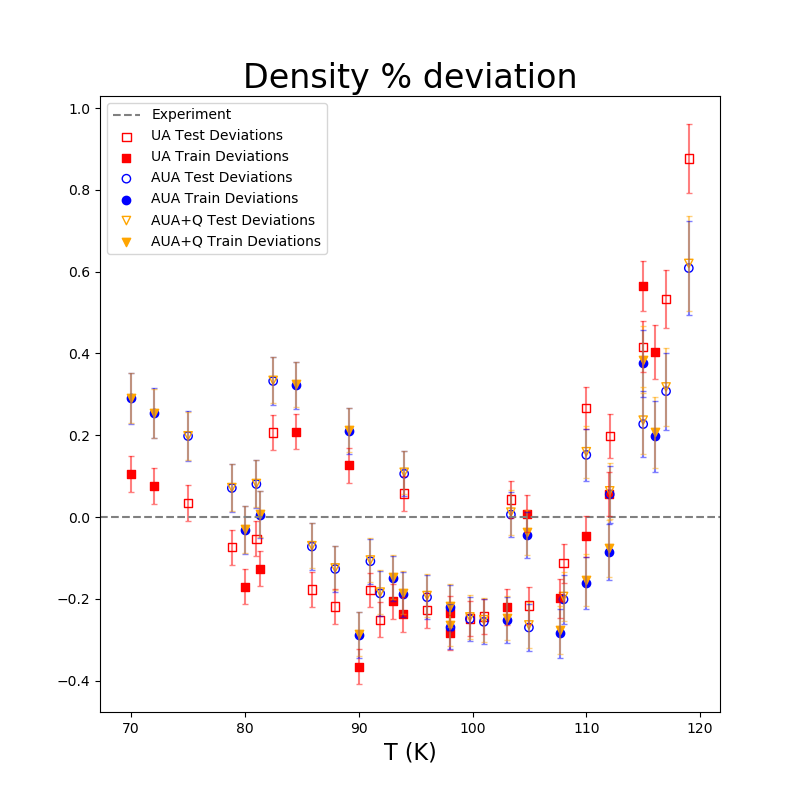}
    \includegraphics[width=0.4\textwidth]{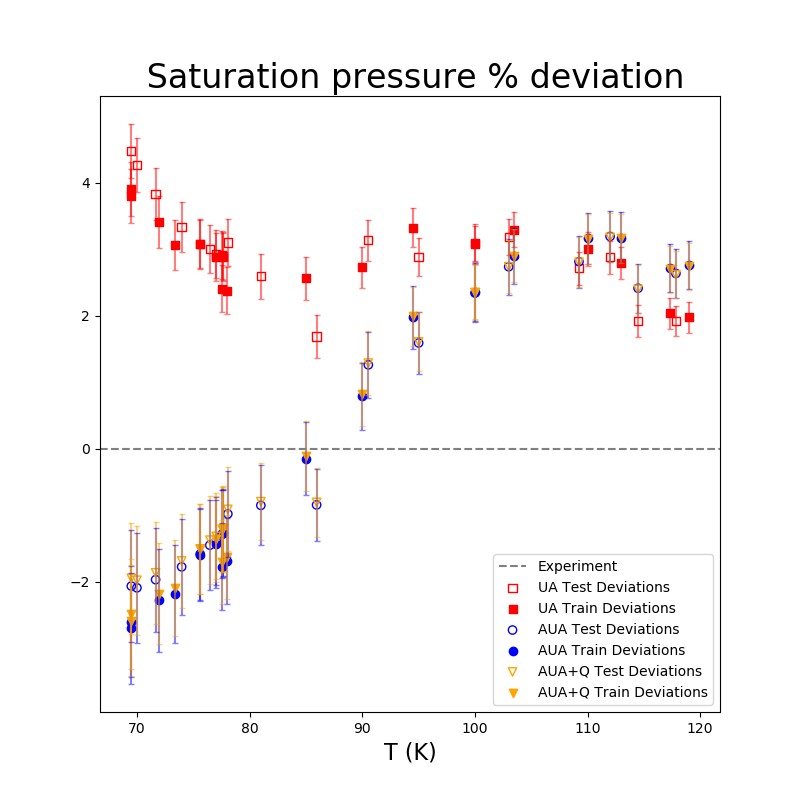}
    \includegraphics[width=0.4\textwidth]{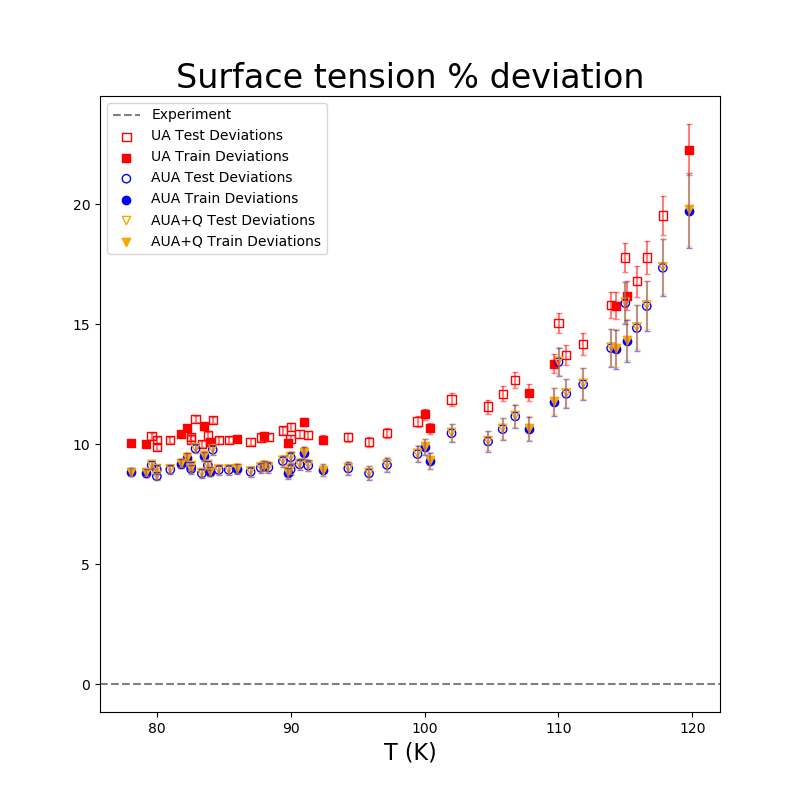}
    \caption{ Average $\rho_l$ (top left panel), $P_{sat}$ (top right panel), $\gamma$ (bottom panel) \% deviation plots for N$_2$. Parameter sets drawn from posterior probability distribution, evaluated against separate benchmark data points (open points) as well as points used in calculated Bayes factor (filled points).}
\end{figure}
\newpage
\subsubsection{O$_2$}
\begin{figure}[h]
\centering
    \includegraphics[width=0.4\textwidth]{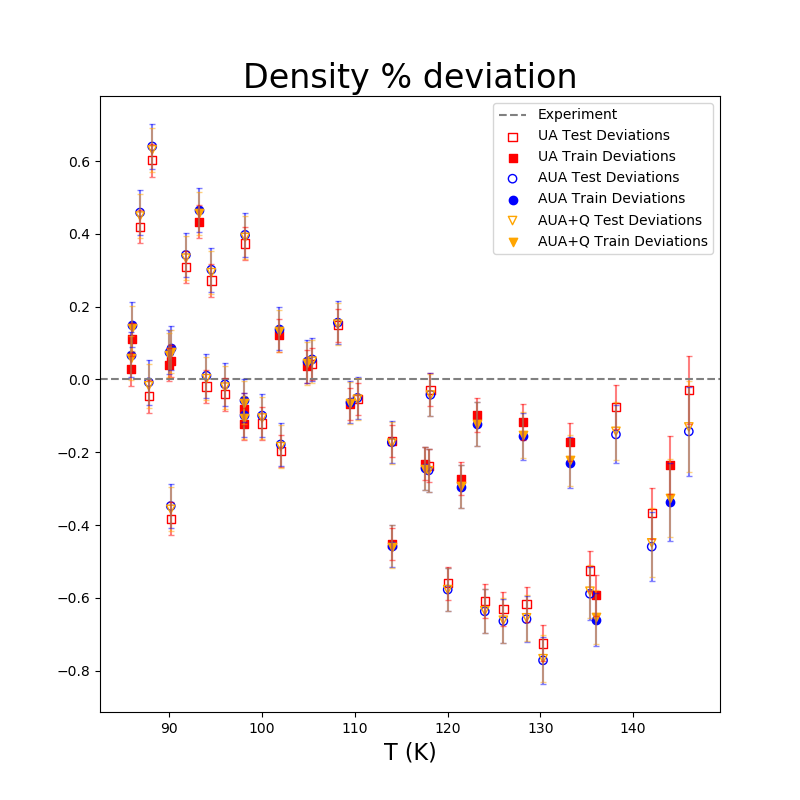}
    \includegraphics[width=0.4\textwidth]{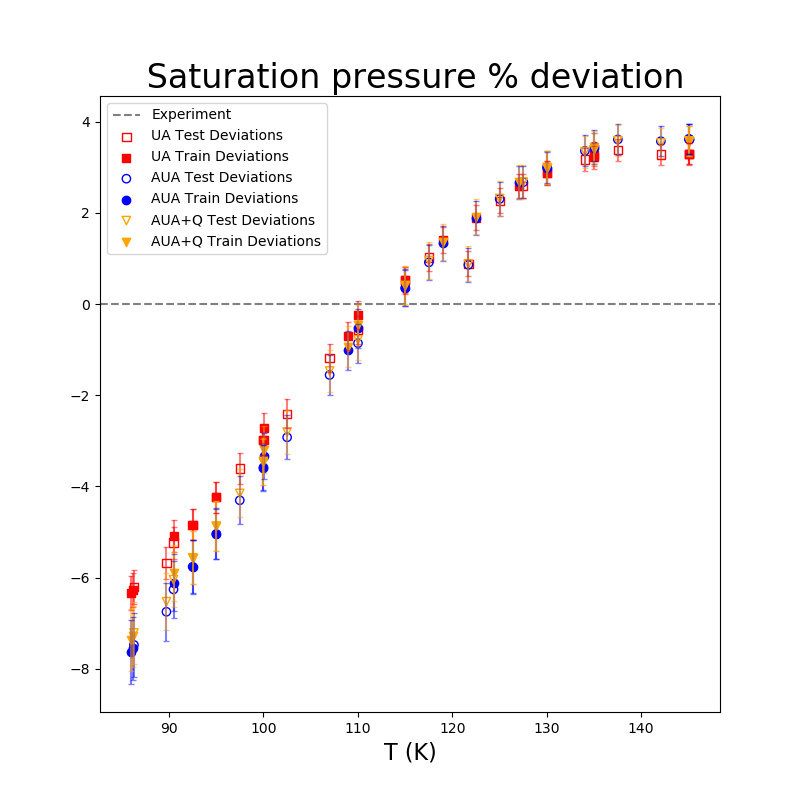}
    \includegraphics[width=0.4\textwidth]{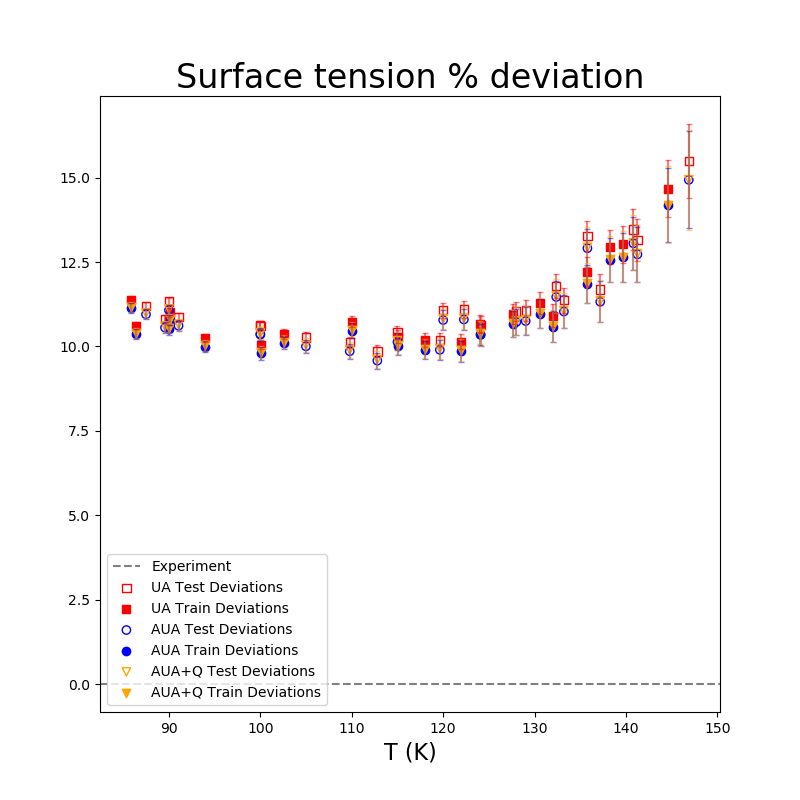}
    \caption{ Average $\rho_l$ (top left panel), $P_{sat}$ (top right panel), $\gamma$ (bottom panel) \% deviation plots for O$_2$. Parameter sets drawn from posterior probability distribution, evaluated against separate benchmark data points (open points) as well as points used in calculated Bayes factor (filled points).}
\end{figure}
\newpage
\subsubsection{C$_2$H$_2$}
\begin{figure}[h]
\centering
    \includegraphics[width=0.4\textwidth]{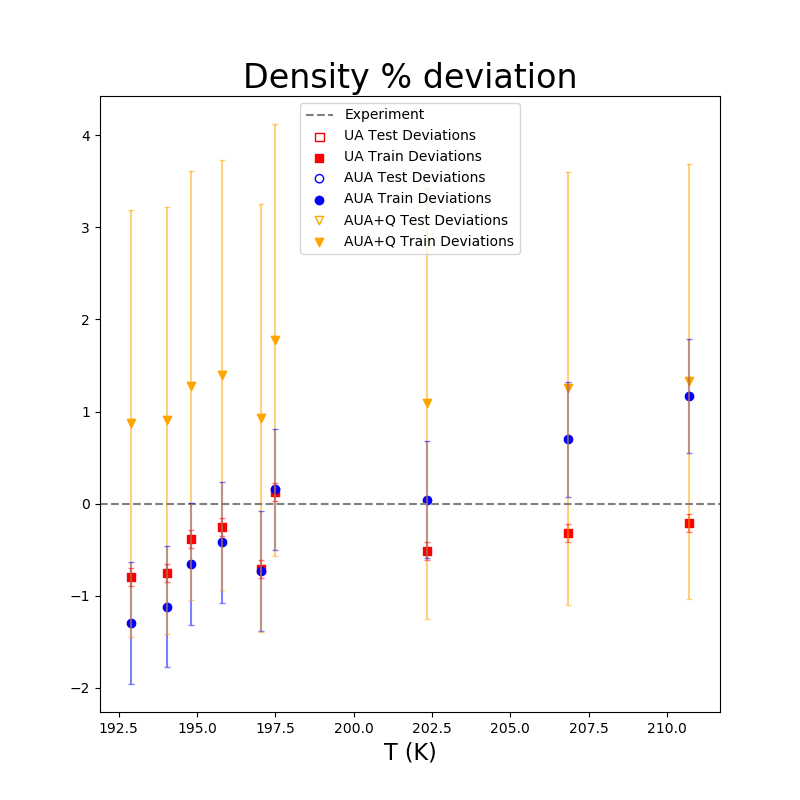}
    \includegraphics[width=0.4\textwidth]{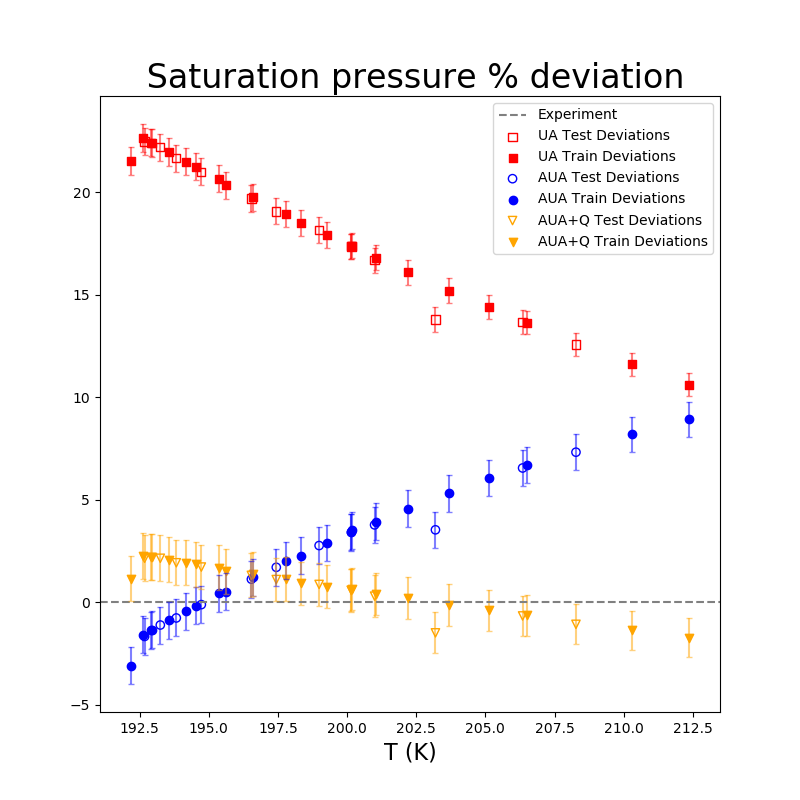}
    \includegraphics[width=0.4\textwidth]{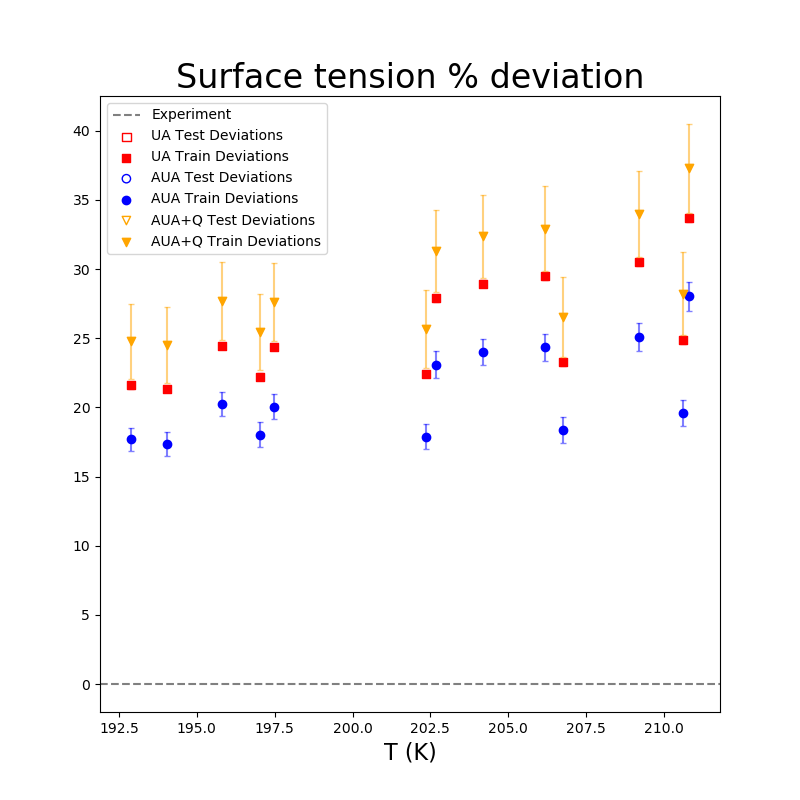}
    \caption{ Average $\rho_l$ (top left panel), $P_{sat}$ (top right panel), $\gamma$ (bottom panel) \% deviation plots for C$_2$H$_2$. Parameter sets drawn from posterior probability distribution, evaluated against separate benchmark data points (open points) as well as points used in calculated Bayes factor (filled points).}
\end{figure}
\newpage
\subsubsection{C$_2$H$_4$}
\begin{figure}[h]
\centering
    \includegraphics[width=0.4\textwidth]{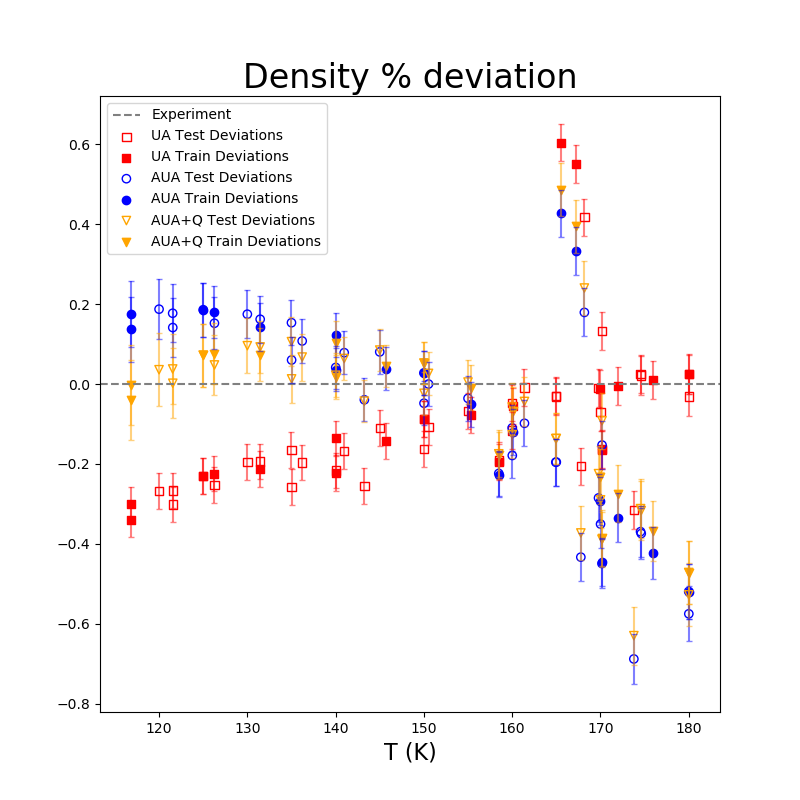}
    \includegraphics[width=0.4\textwidth]{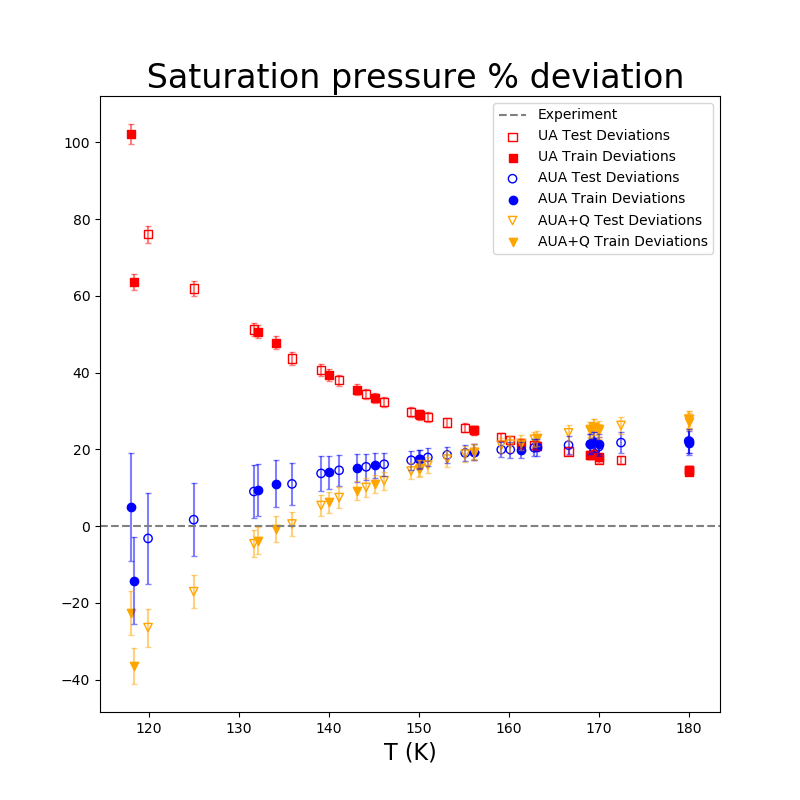}
    \includegraphics[width=0.4\textwidth]{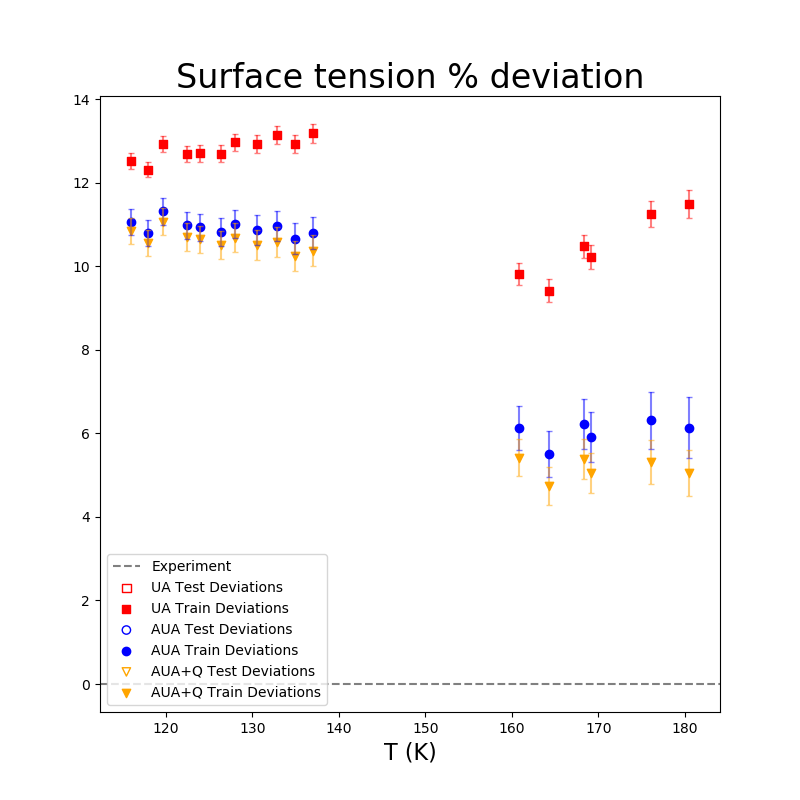}
    \caption{ Average $\rho_l$ (top left panel), $P_{sat}$ (top right panel), $\gamma$ (bottom panel) \% deviation plots for C$_2$H$_4$. Parameter sets drawn from posterior probability distribution, evaluated against separate benchmark data points (open points) as well as points used in calculated Bayes factor (filled points).}
\end{figure}
\newpage
\subsubsection{C$_2$H$_6$}
\begin{figure}[h]
\centering
    \includegraphics[width=0.4\textwidth]{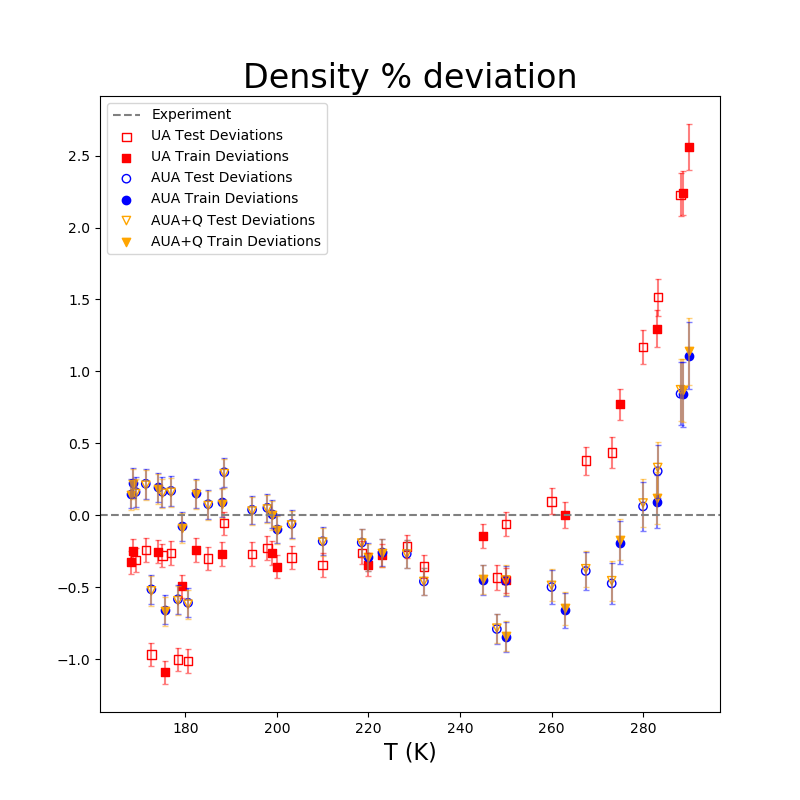}
    \includegraphics[width=0.4\textwidth]{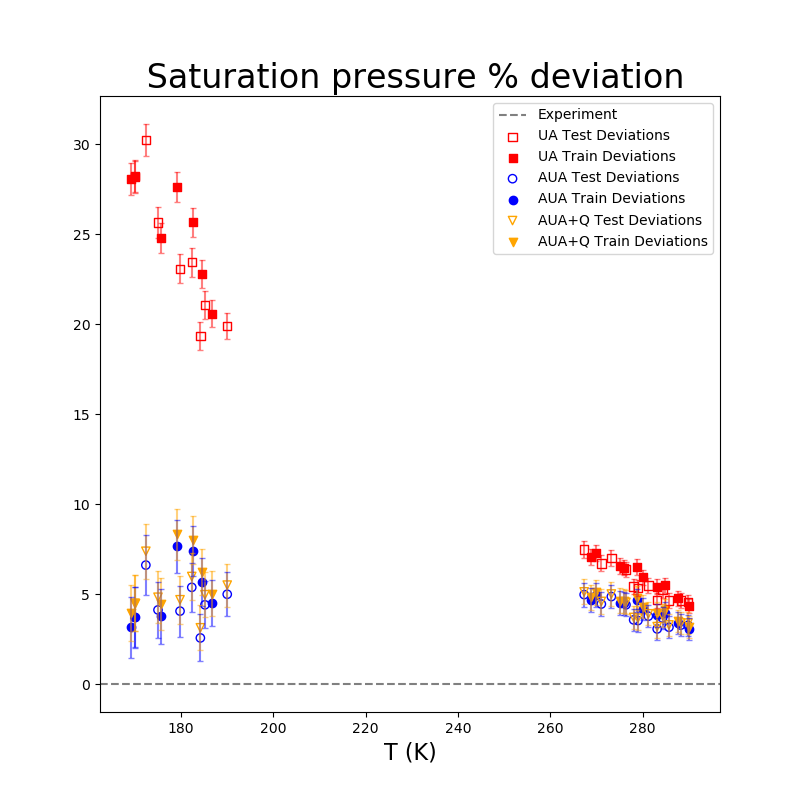}
    \includegraphics[width=0.4\textwidth]{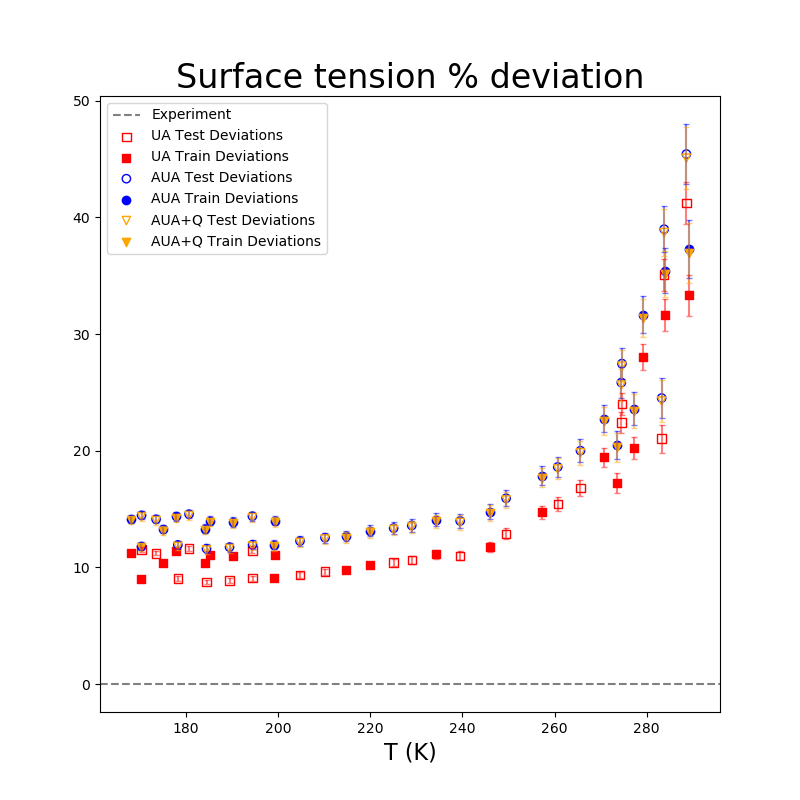}
    \caption{ Average $\rho_l$ (top left panel), $P_{sat}$ (top right panel), $\gamma$ (bottom panel) \% deviation plots for C$_2$H$_6$. Parameter sets drawn from posterior probability distribution, evaluated against separate benchmark data points (open points) as well as points used in calculated Bayes factor (filled points).}
\end{figure}
\newpage
\section{Parameter Distributions from Bayes factor MCMC samples}
These triangle plots are taken from the MCMC sampling of the model posteriors from the MBAR Bayes factor calculations for all 3 models (UA, AUA, AUA+Q), with priors set from the high information training samples (n=8 data points per property).

All measurements are in nm ($\sigma$), K ($\epsilon $), nm ($L$), $\mathrm{D}\cdot \mathrm{nm}$ ($Q$).
\subsection{$\rho_l, P_{sat}$ target}
\subsubsection{F$_2$}
\begin{figure}[h]
    \includegraphics[width=0.3\textwidth]{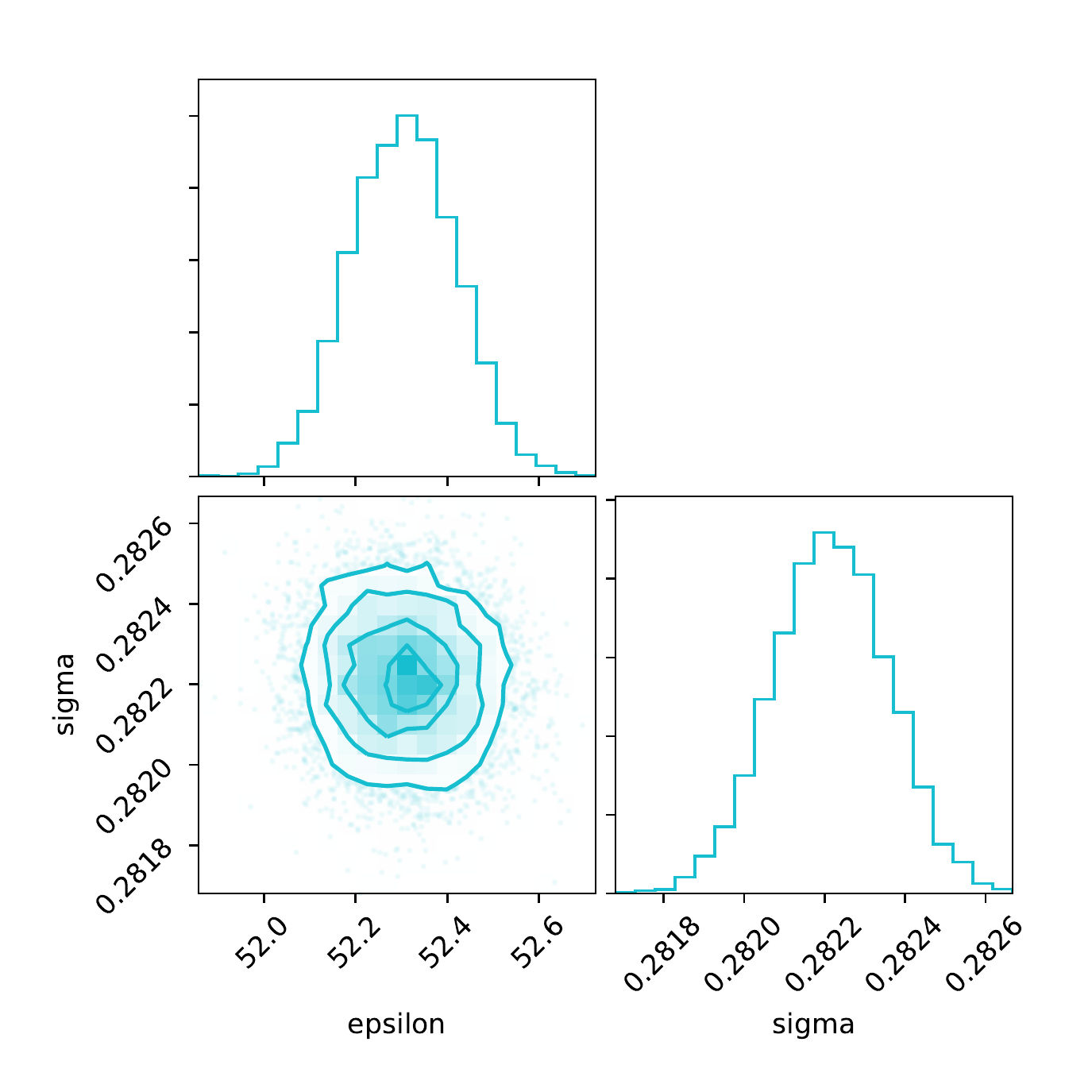}
    \includegraphics[width=0.3\textwidth]{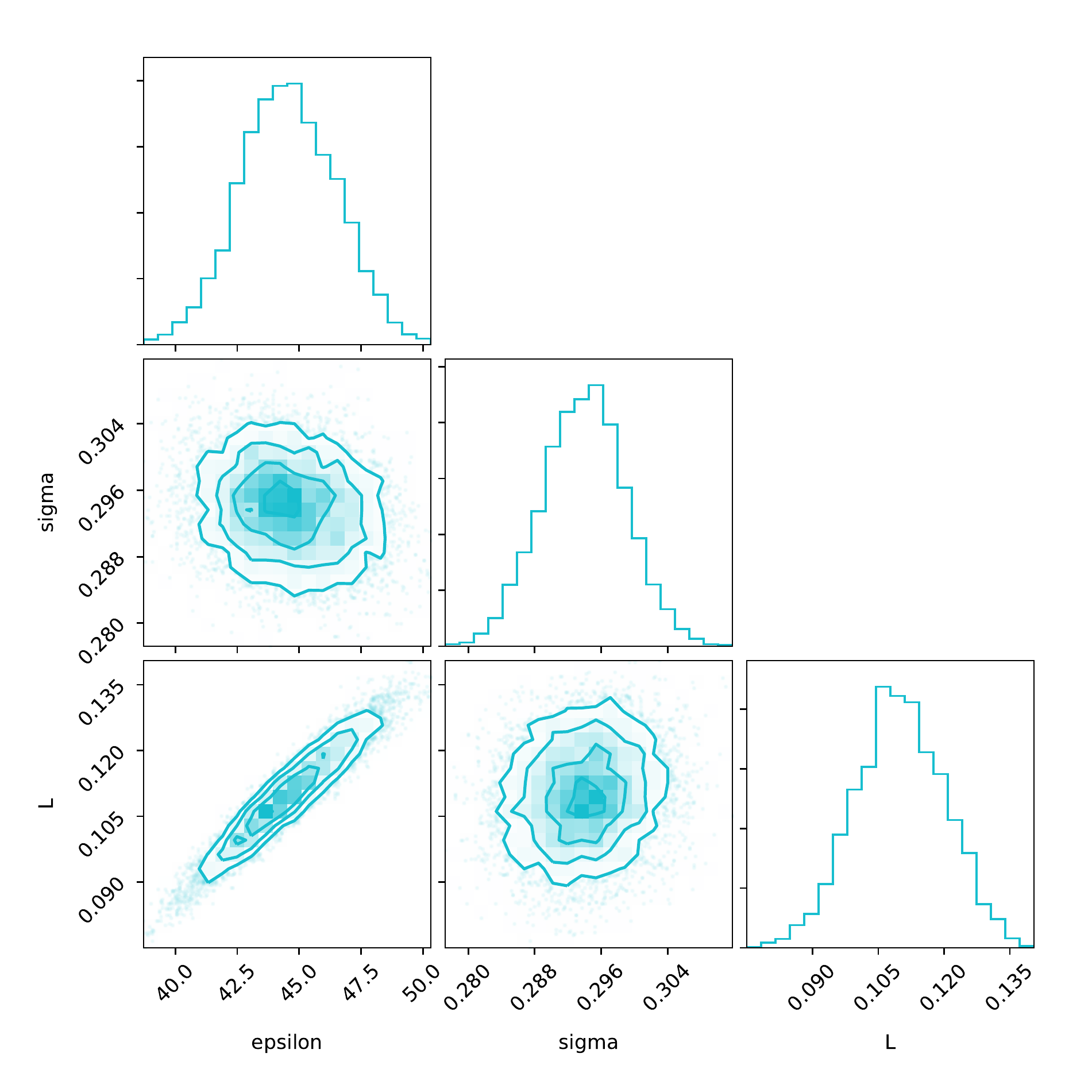}
        \includegraphics[width=0.3\textwidth]{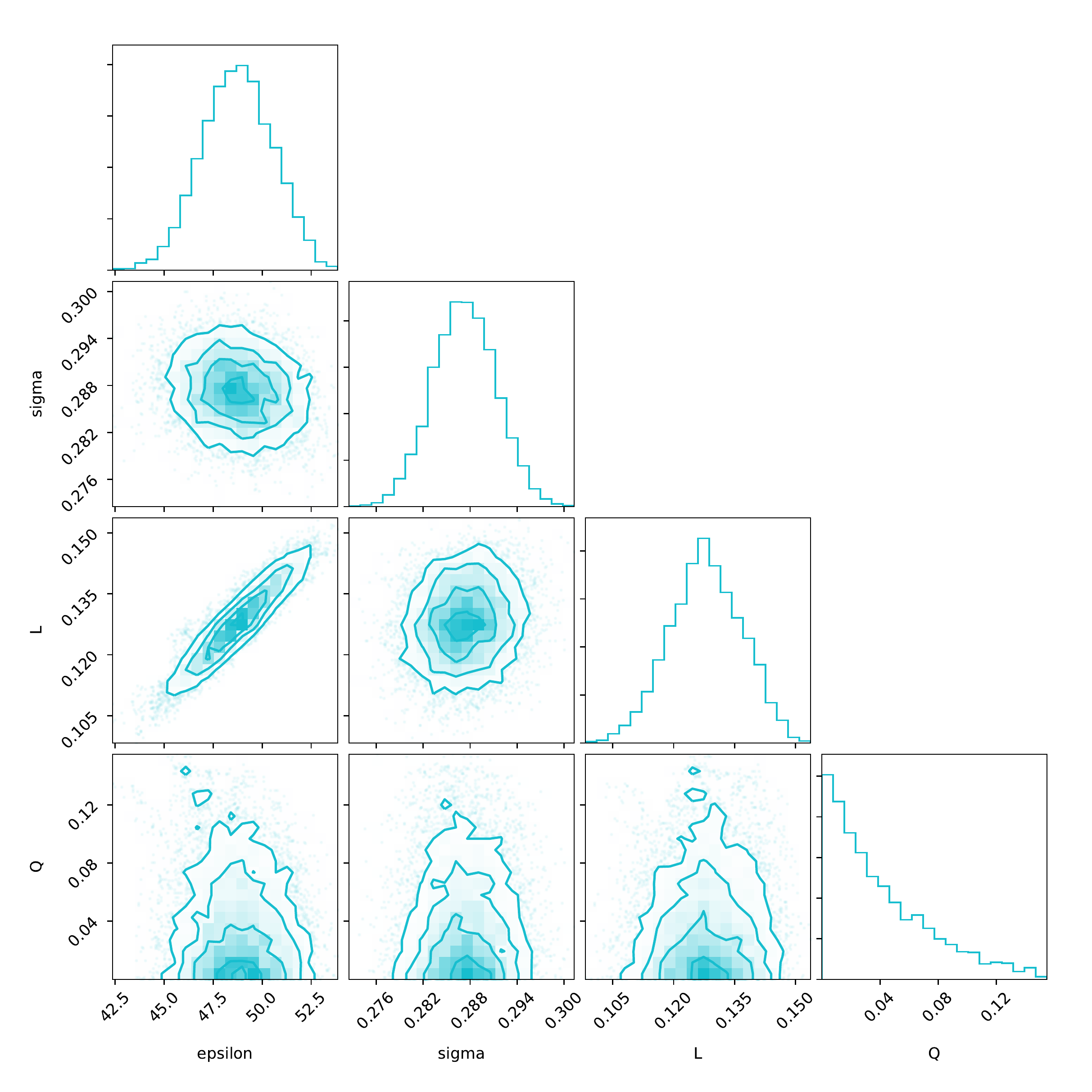}
    \caption{Parameter distributions for F$_2$, $\rho_l, P_{sat}$ target. From left to right: UA, AUA, AUA+Q}
    \label{fig:2crit_F2_triangle}
\end{figure}
\subsubsection{Br$_2$}
\begin{figure}[h]
    \includegraphics[width=0.3\textwidth]{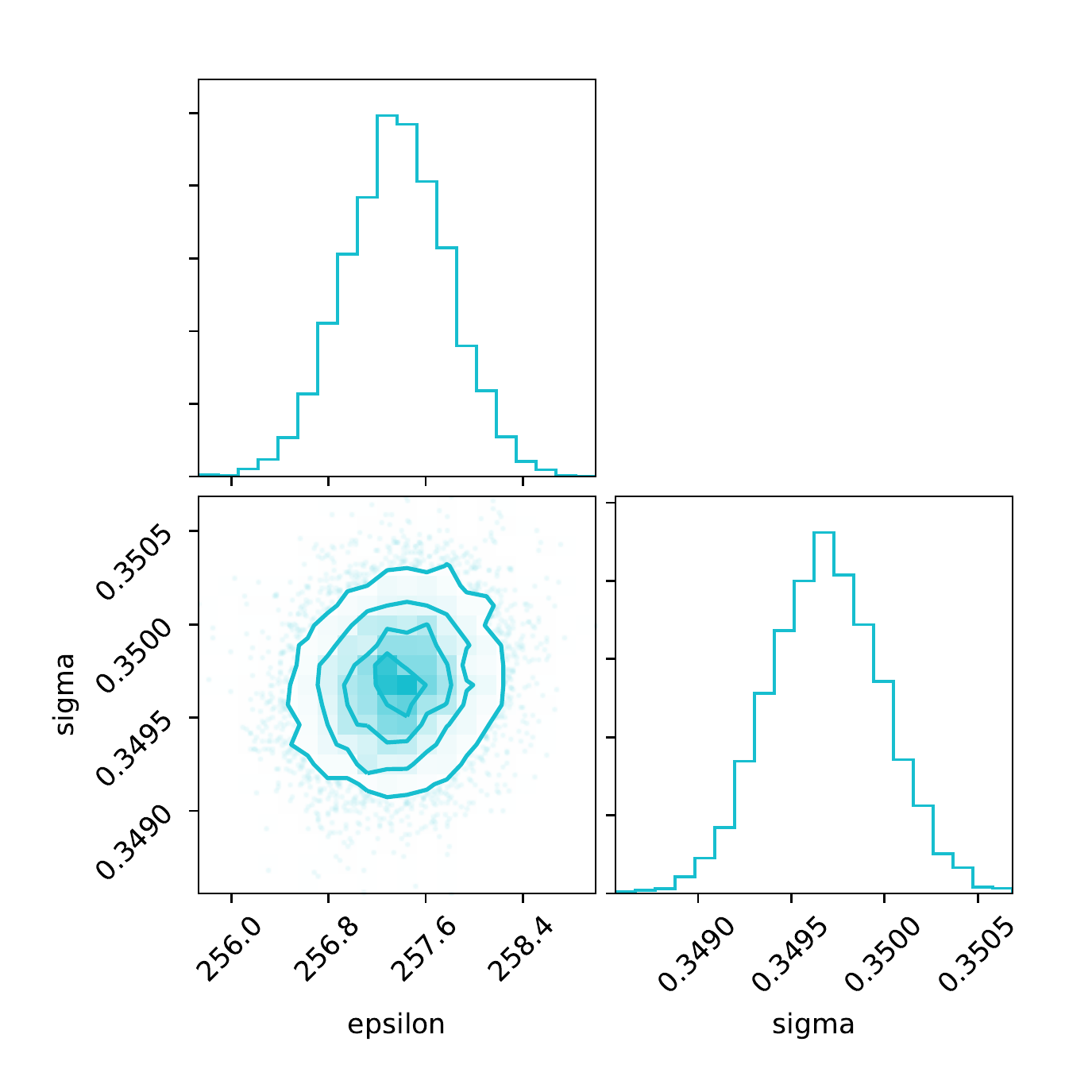}
    \includegraphics[width=0.3\textwidth]{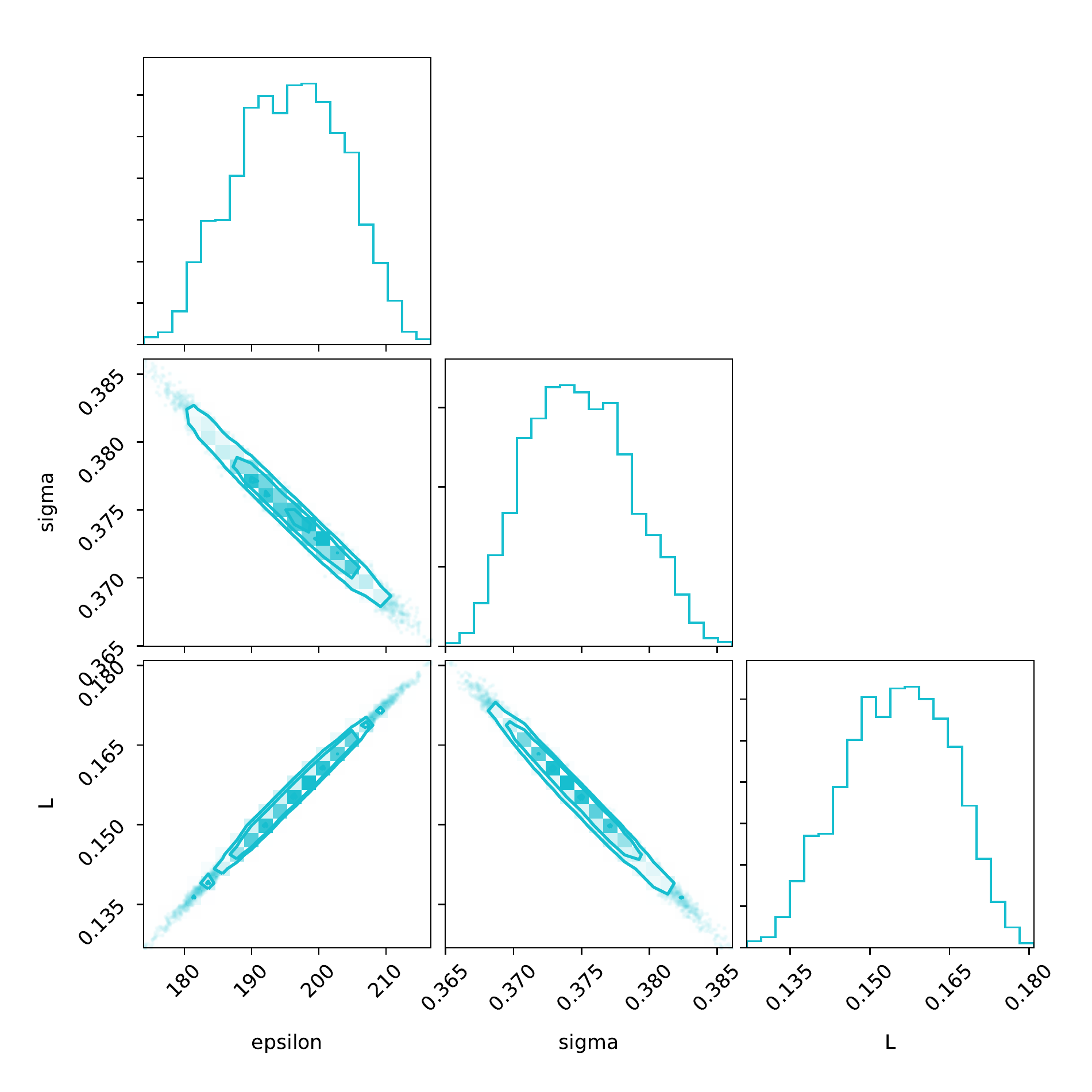}
        \includegraphics[width=0.3\textwidth]{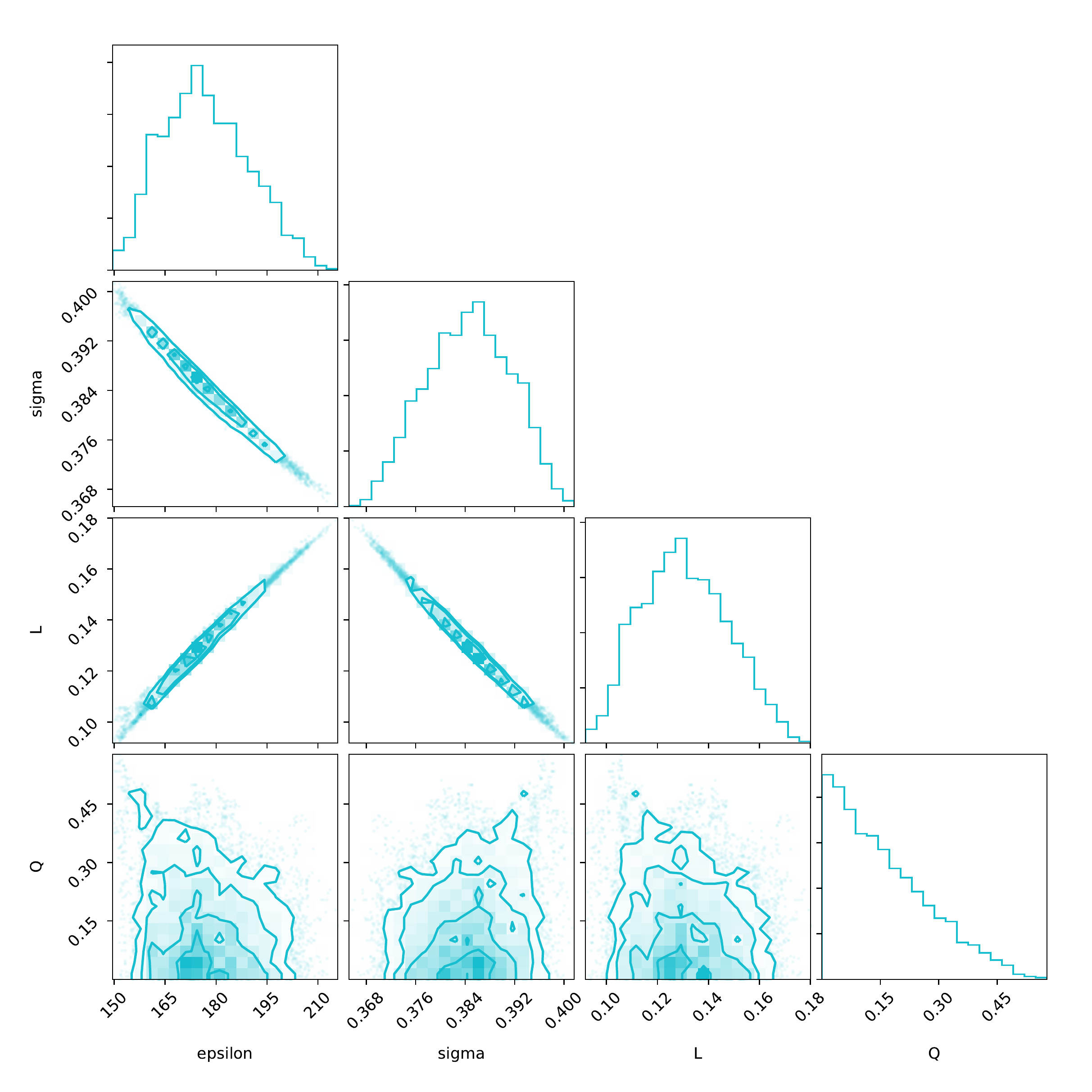}
    \caption{Parameter distributions for Br$_2$, $\rho_l, P_{sat}$ target. From left to right: UA, AUA, AUA+Q}
    \label{fig:2crit_Br2_triangle}
\end{figure}
\newpage
\subsubsection{N$_2$}
\begin{figure}[h]
    \includegraphics[width=0.3\textwidth]{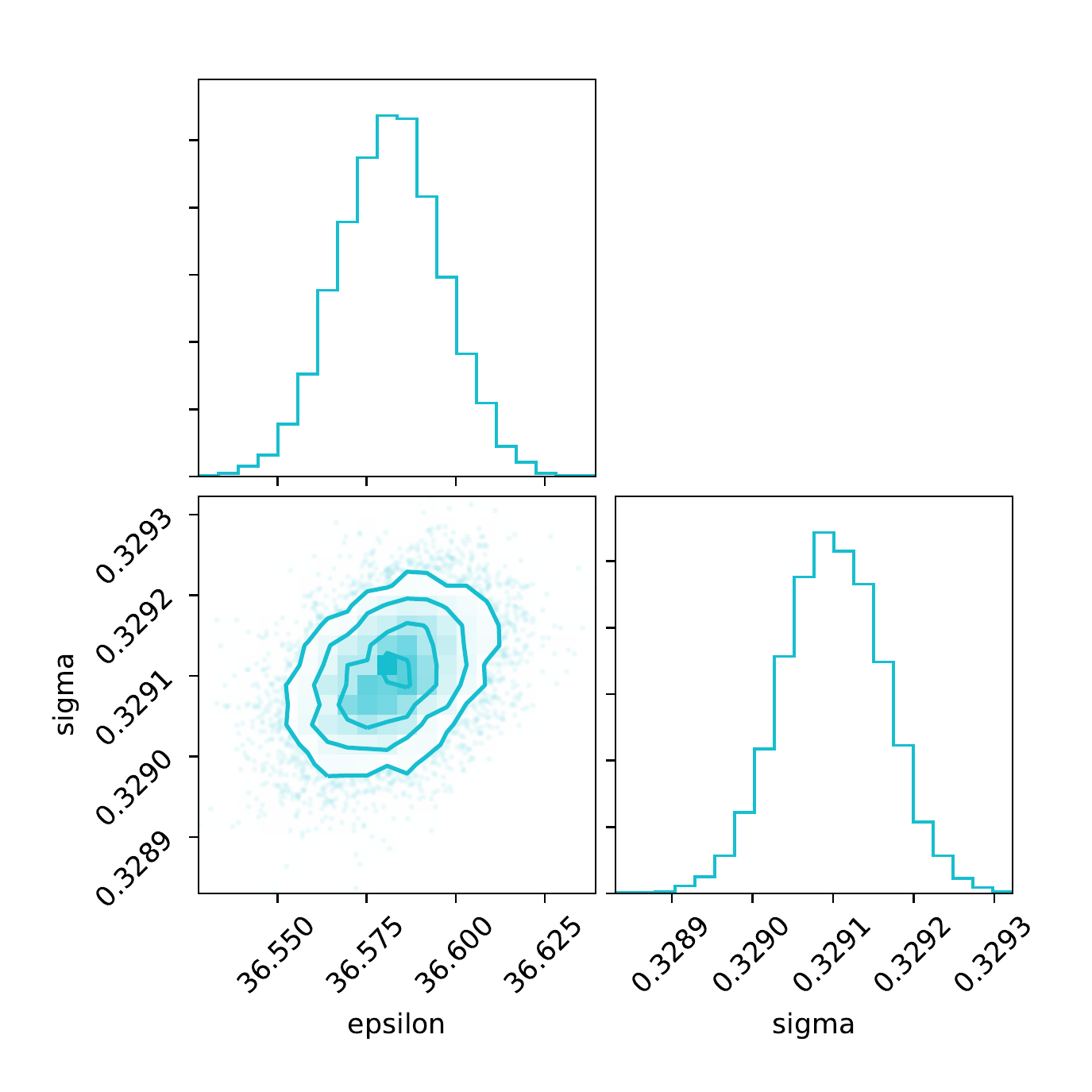}
    \includegraphics[width=0.3\textwidth]{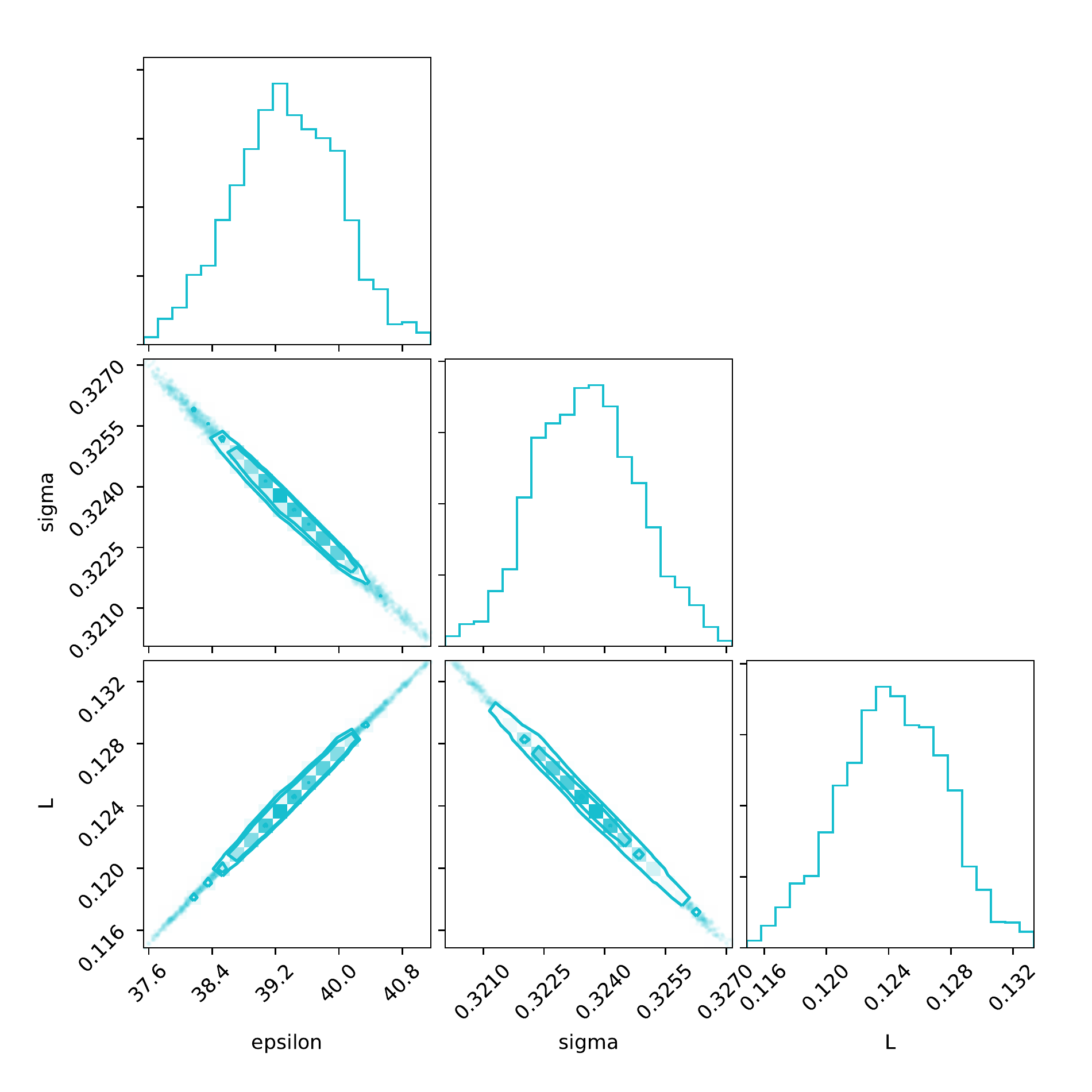}
        \includegraphics[width=0.3\textwidth]{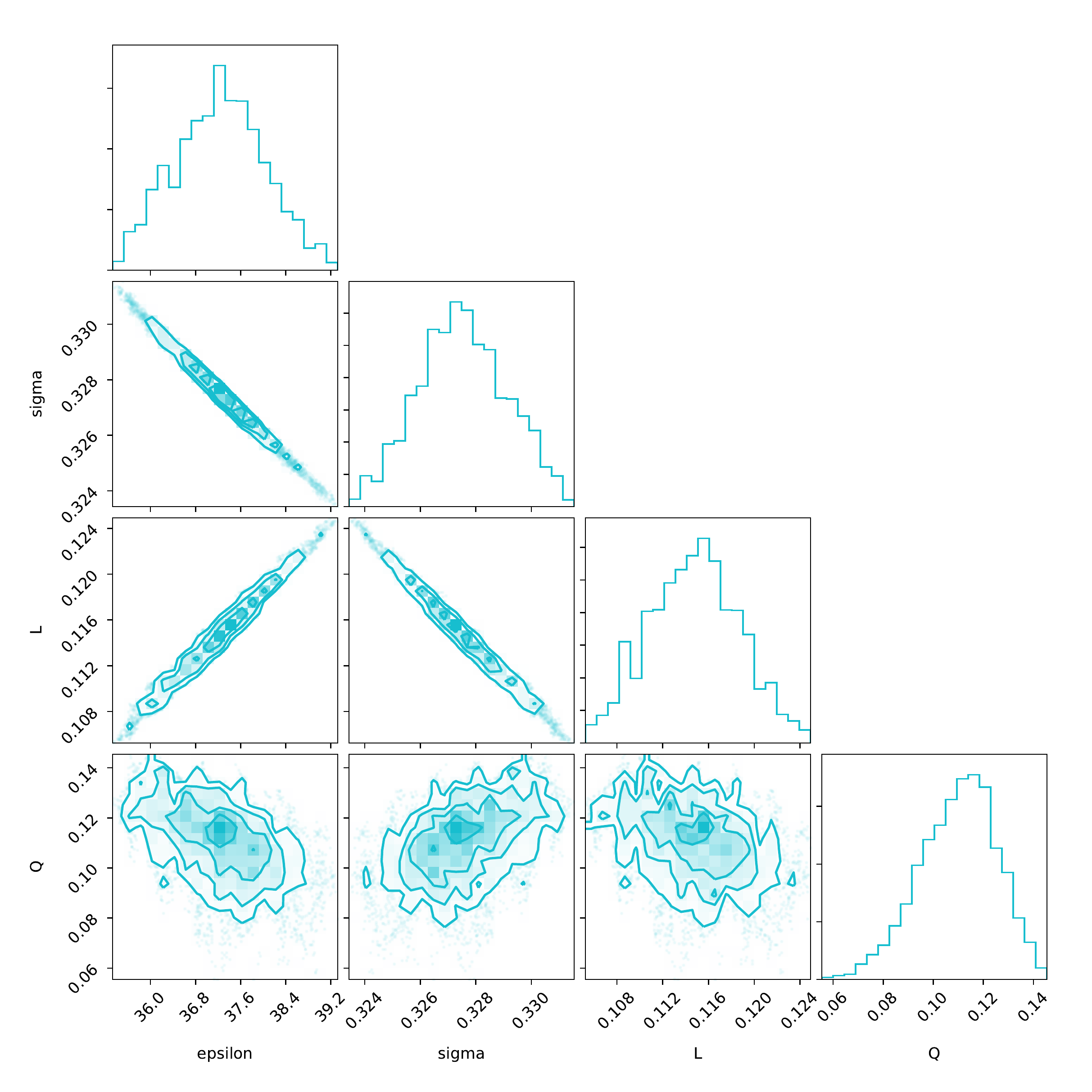}
    \caption{Parameter distributions for N$_2$, $\rho_l, P_{sat}$ target. From left to right: UA, AUA, AUA+Q}
    \label{fig:2crit_N2_triangle}
\end{figure}
\subsubsection{O$_2$}
\begin{figure}[h]
    \includegraphics[width=0.3\textwidth]{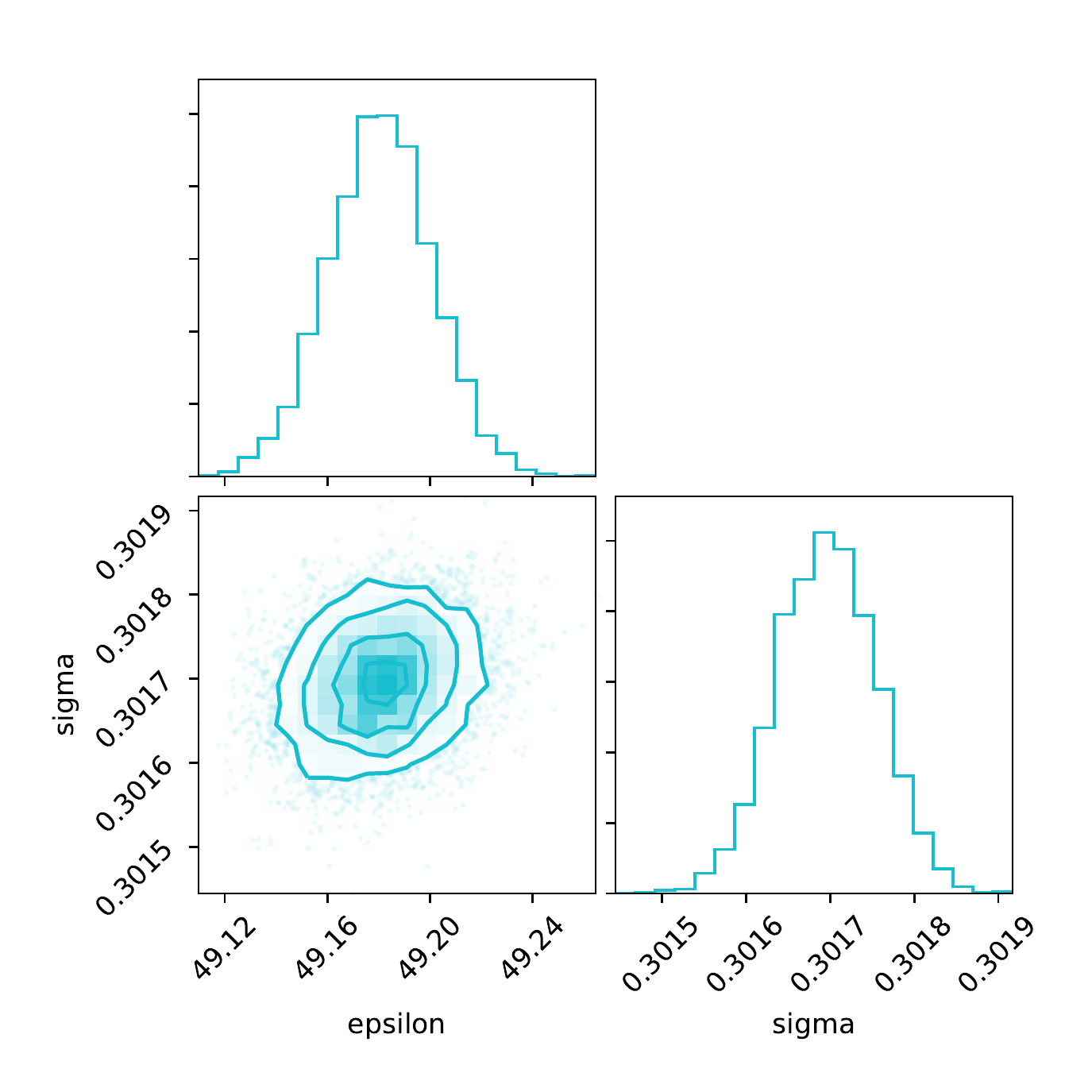}
    \includegraphics[width=0.3\textwidth]{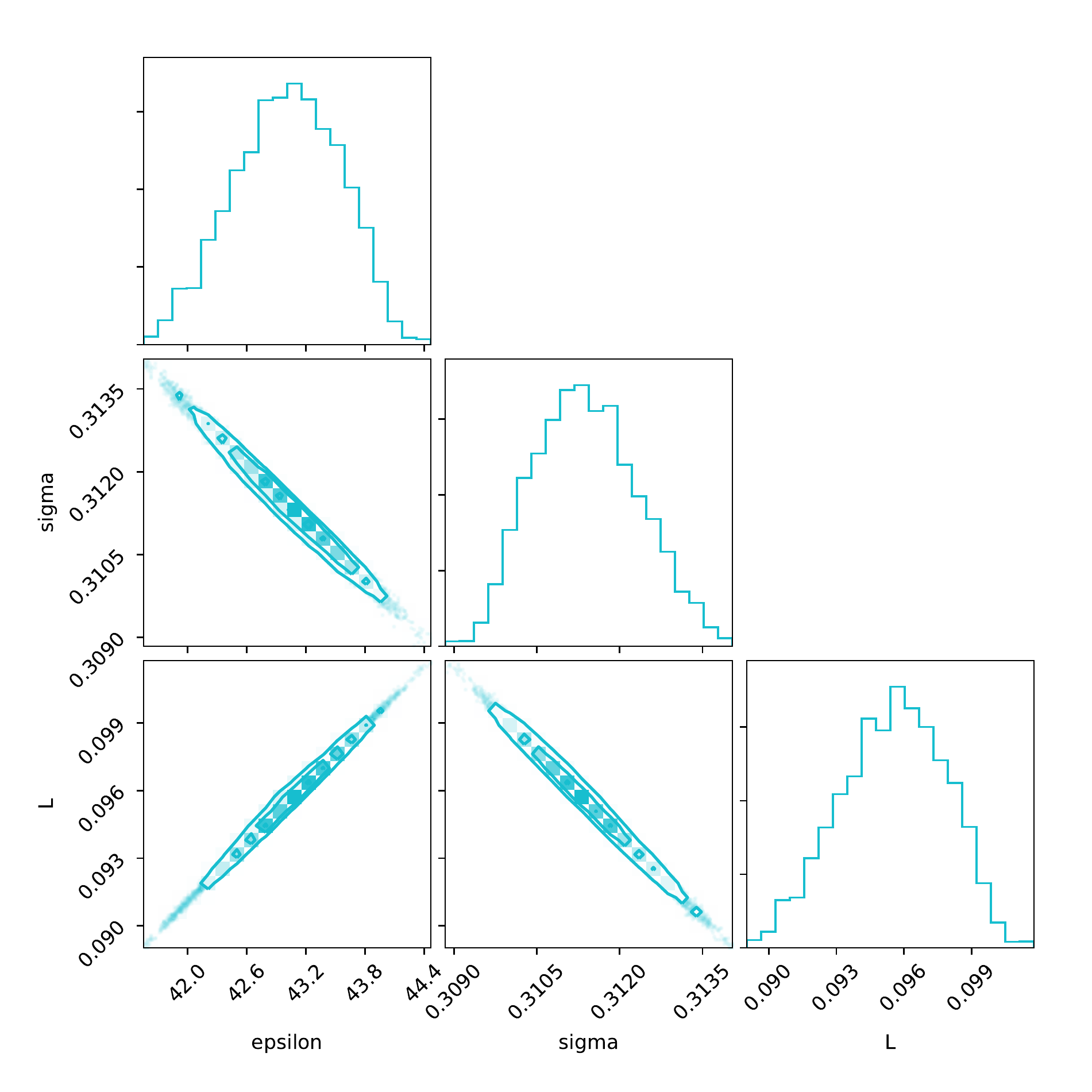}
        \includegraphics[width=0.3\textwidth]{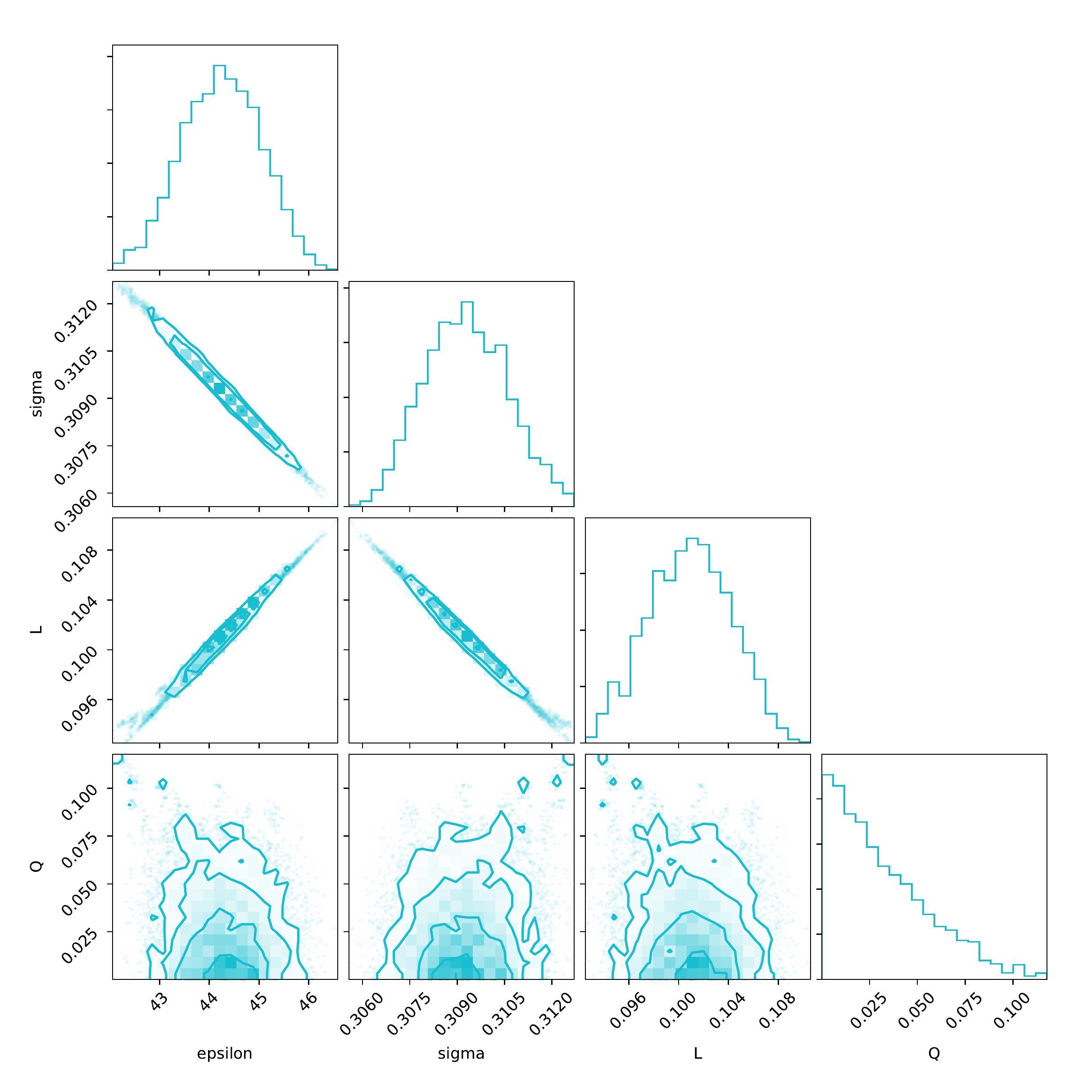}
    \caption{Parameter distributions for O$_2$, $\rho_l, P_{sat}$ target. From left to right: UA, AUA, AUA+Q}
    \label{fig:2crit_N2_triangle}
\end{figure}
\newpage
\subsubsection{C$_2$H$_2$}
\begin{figure}[h]
    \includegraphics[width=0.3\textwidth]{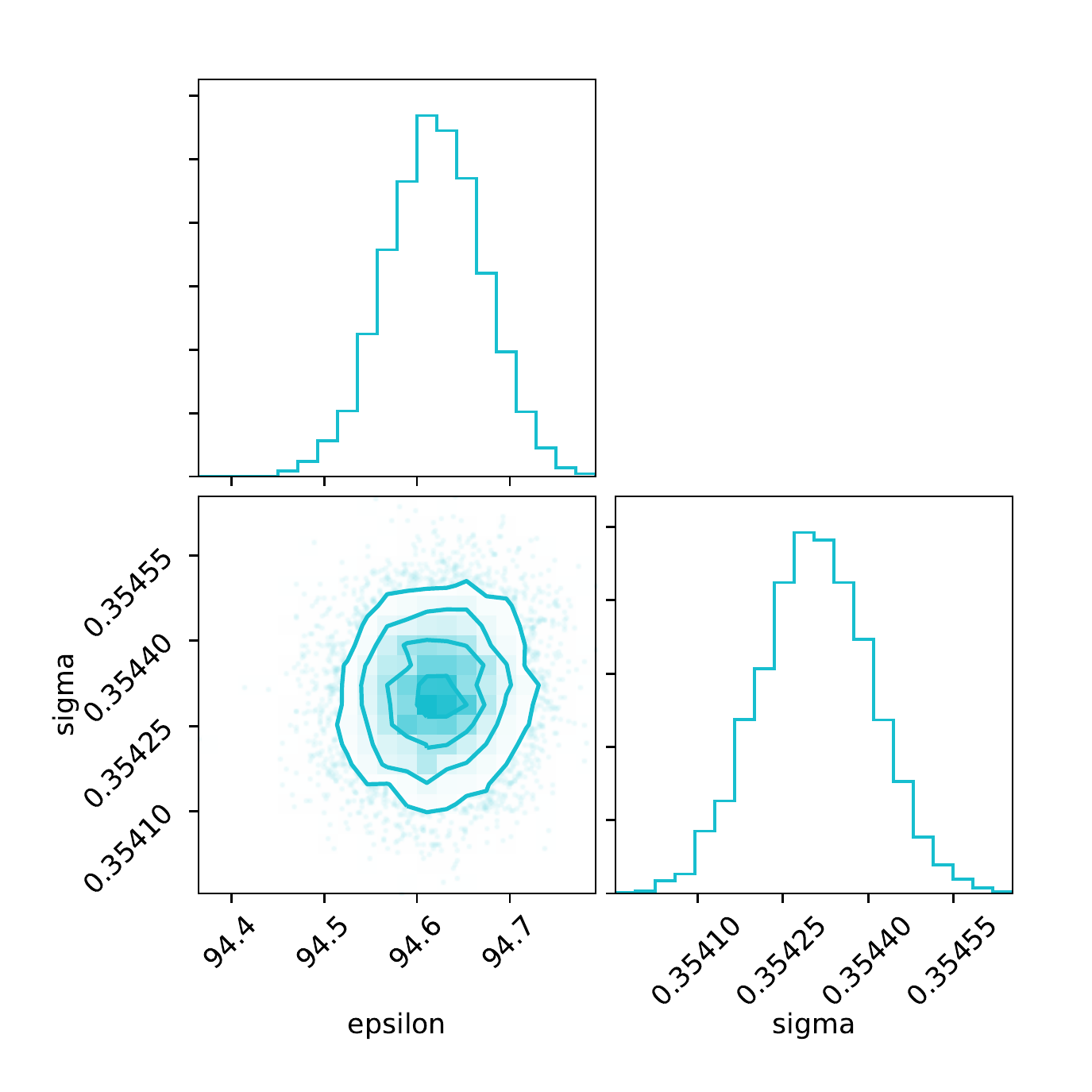}
    \includegraphics[width=0.3\textwidth]{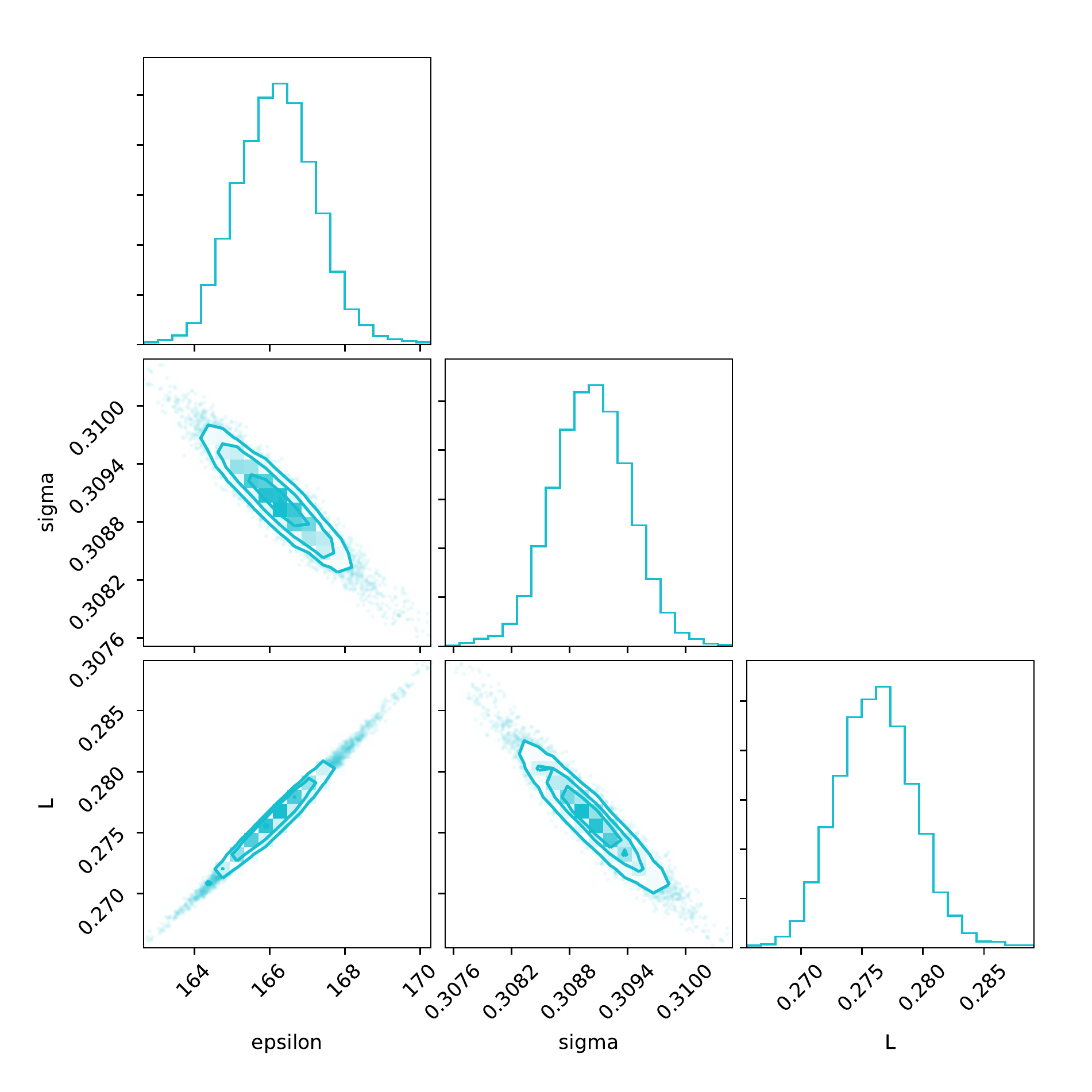}
        \includegraphics[width=0.3\textwidth]{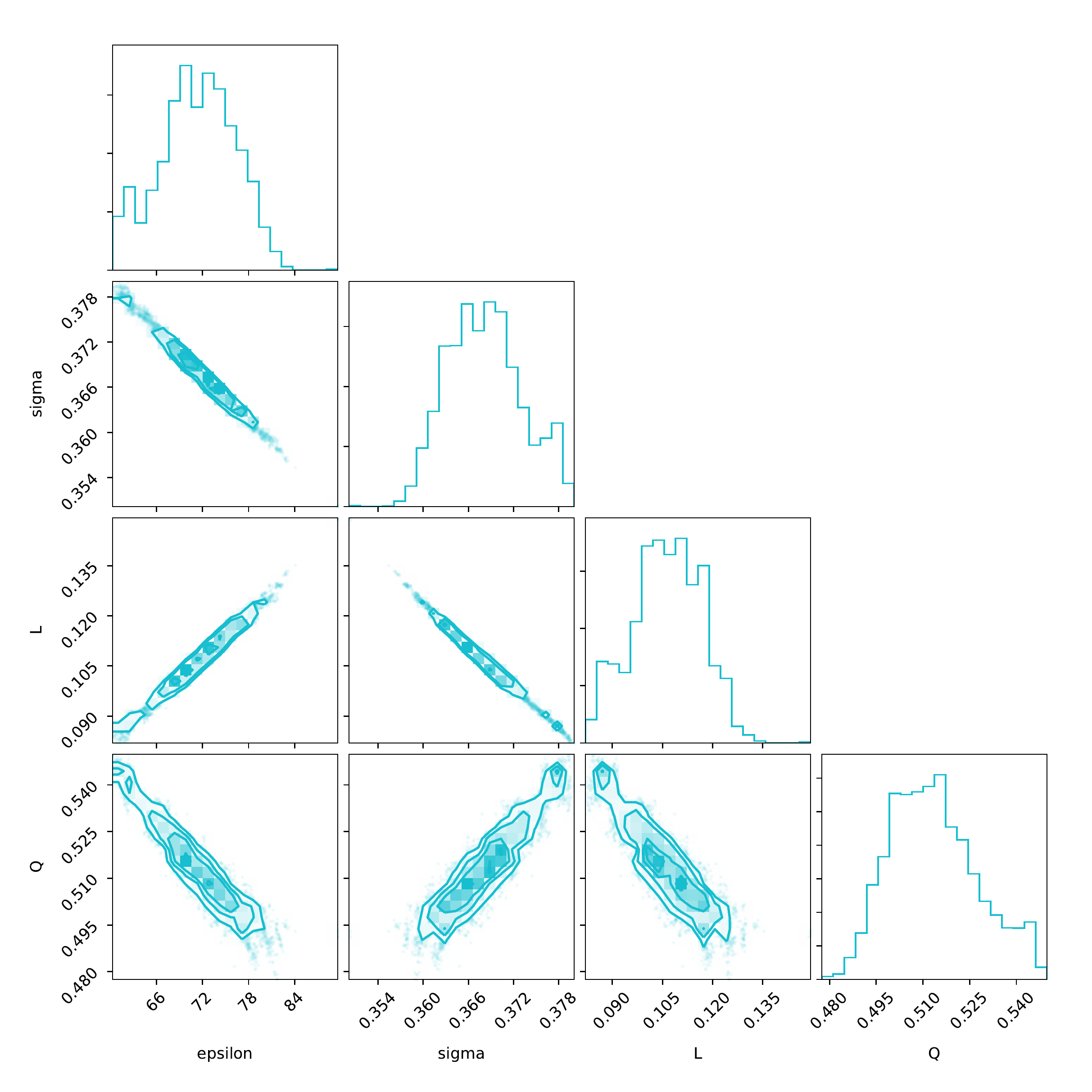}
    \caption{Parameter distributions for C$_2$H$_2$, $\rho_l, P_{sat}$ target. From left to right: UA, AUA, AUA+Q}
    \label{fig:2crit_C2H2_triangle}
\end{figure}
\subsubsection{C$_2$H$_4$}
\begin{figure}[h]
    \includegraphics[width=0.3\textwidth]{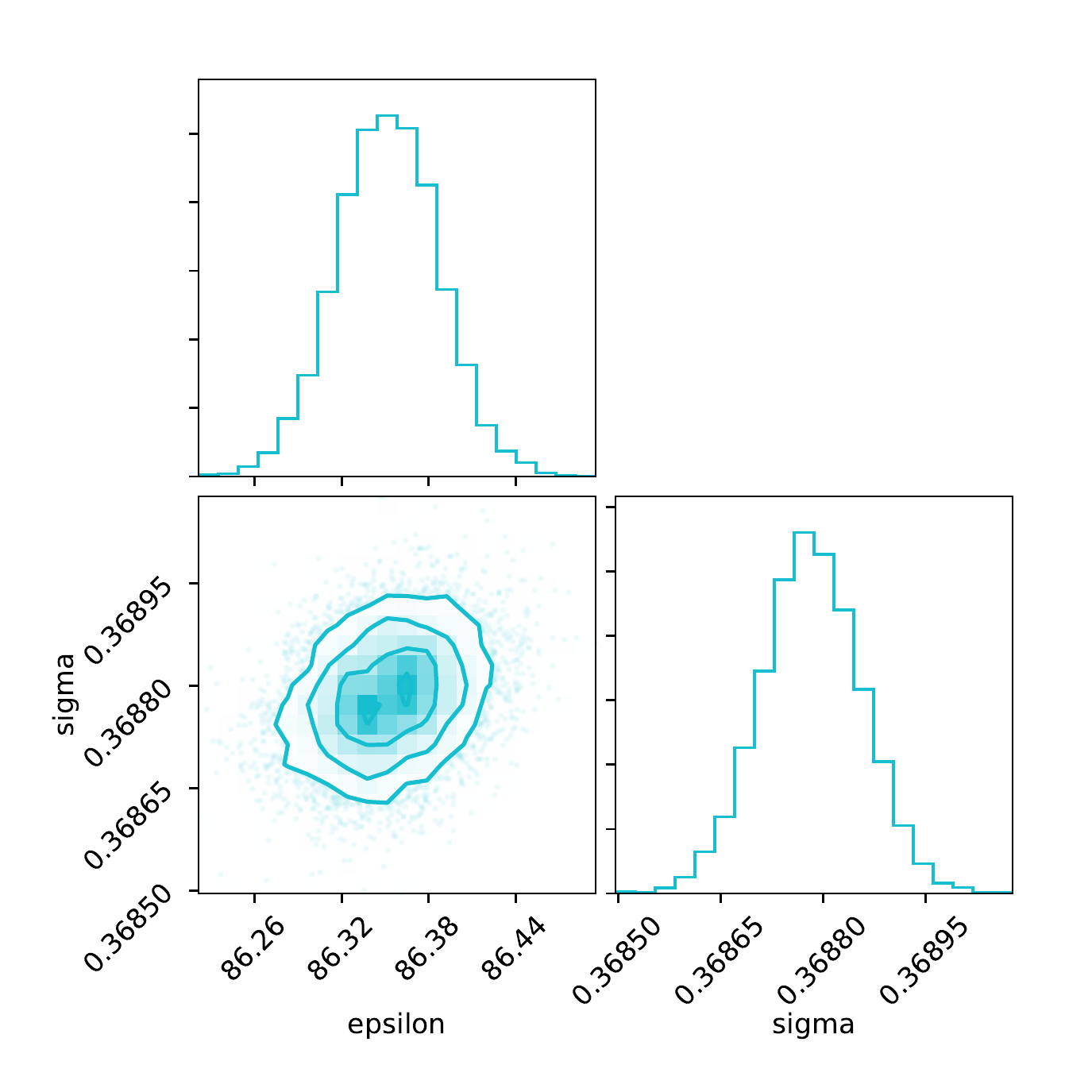}
    \includegraphics[width=0.3\textwidth]{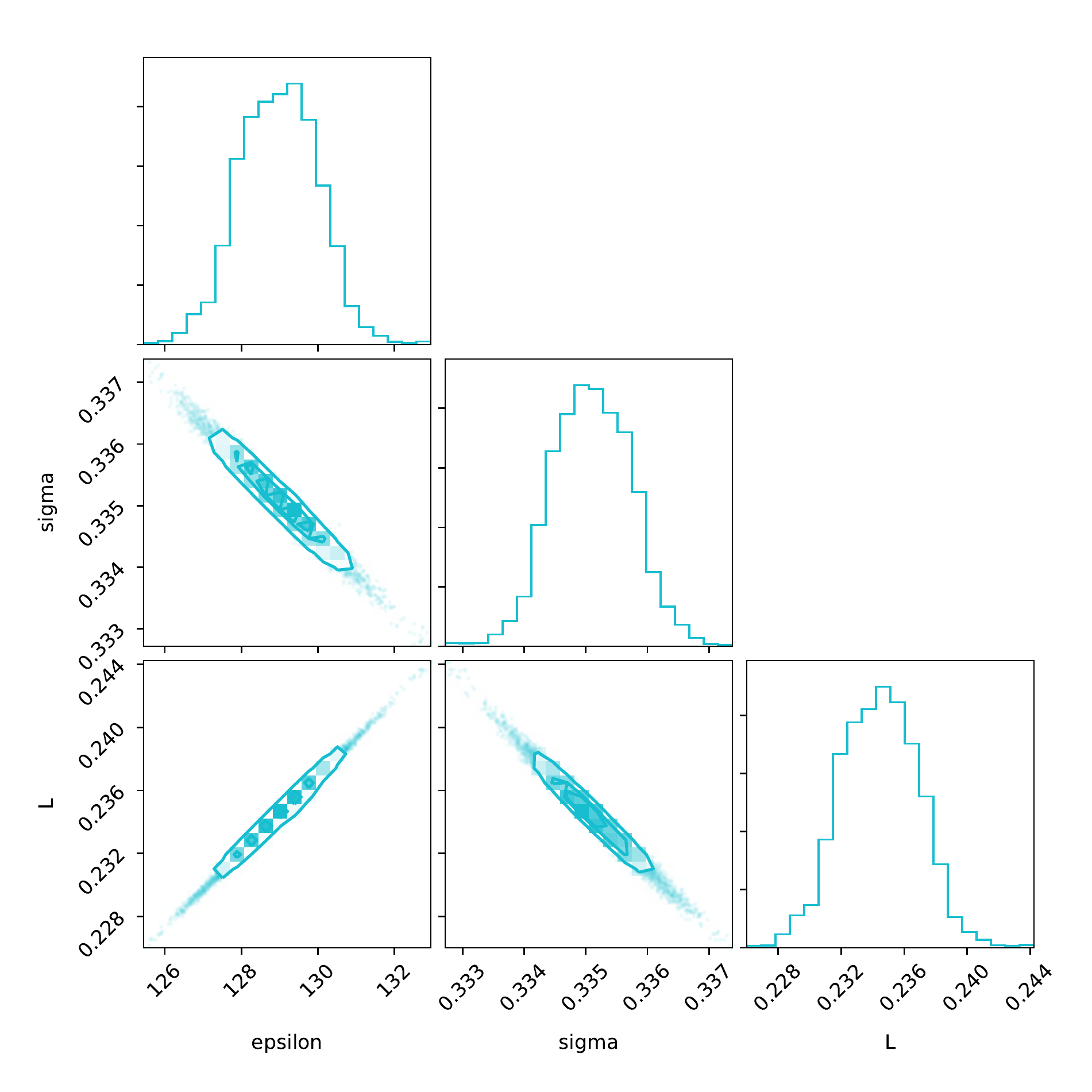}
        \includegraphics[width=0.3\textwidth]{figures/supplementary_figures/triangle/2crit_C2H4_AUA+Q_corner.pdf}
    \caption{Parameter distributions for C$_2$H$_4$, $\rho_l, P_{sat}$ target. From left to right: UA, AUA, AUA+Q}
    \label{fig:2crit_C2H4_triangle}
\end{figure}
\newpage
\subsubsection{C$_2$H$_6$}
\begin{figure}[h]
    \includegraphics[width=0.3\textwidth]{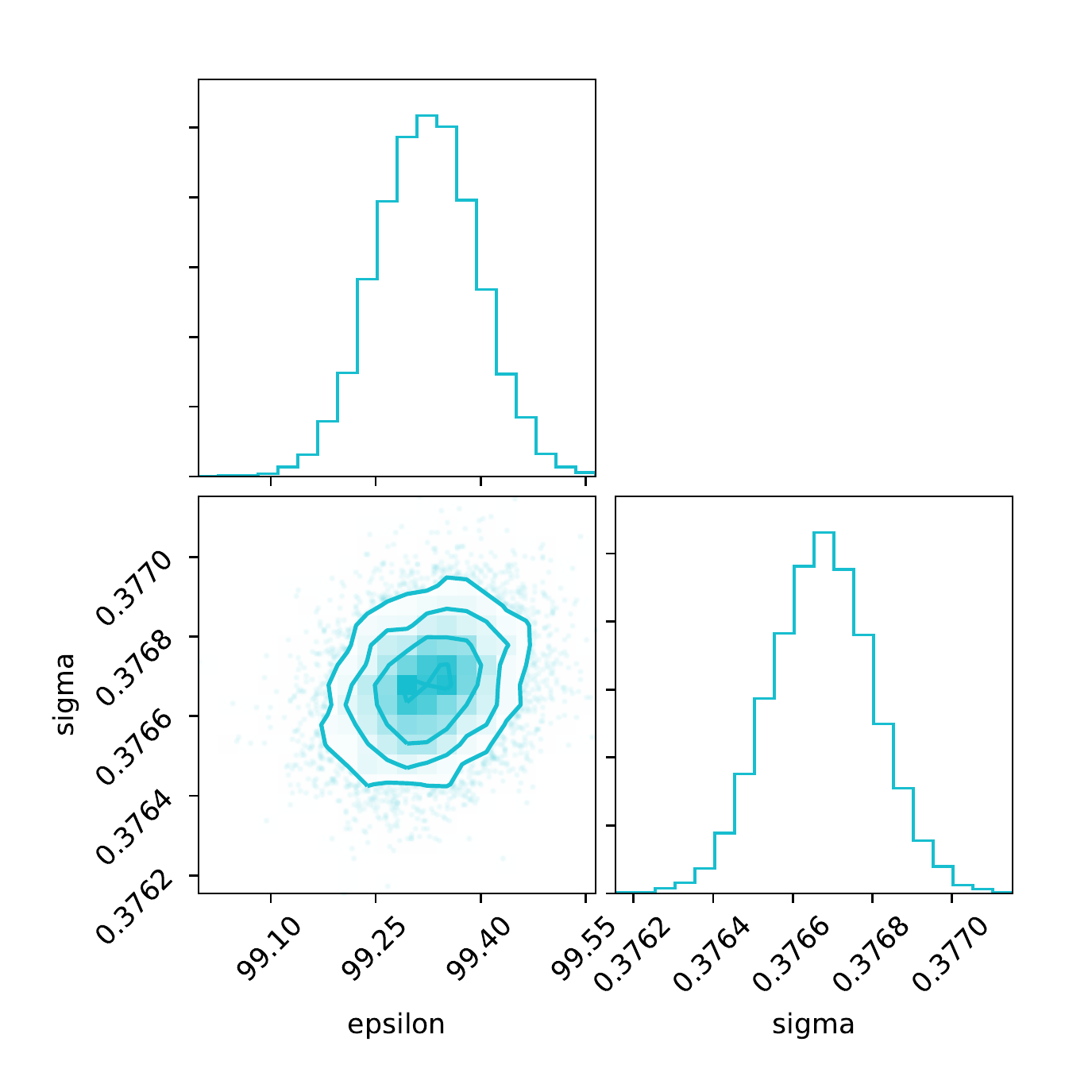}
    \includegraphics[width=0.3\textwidth]{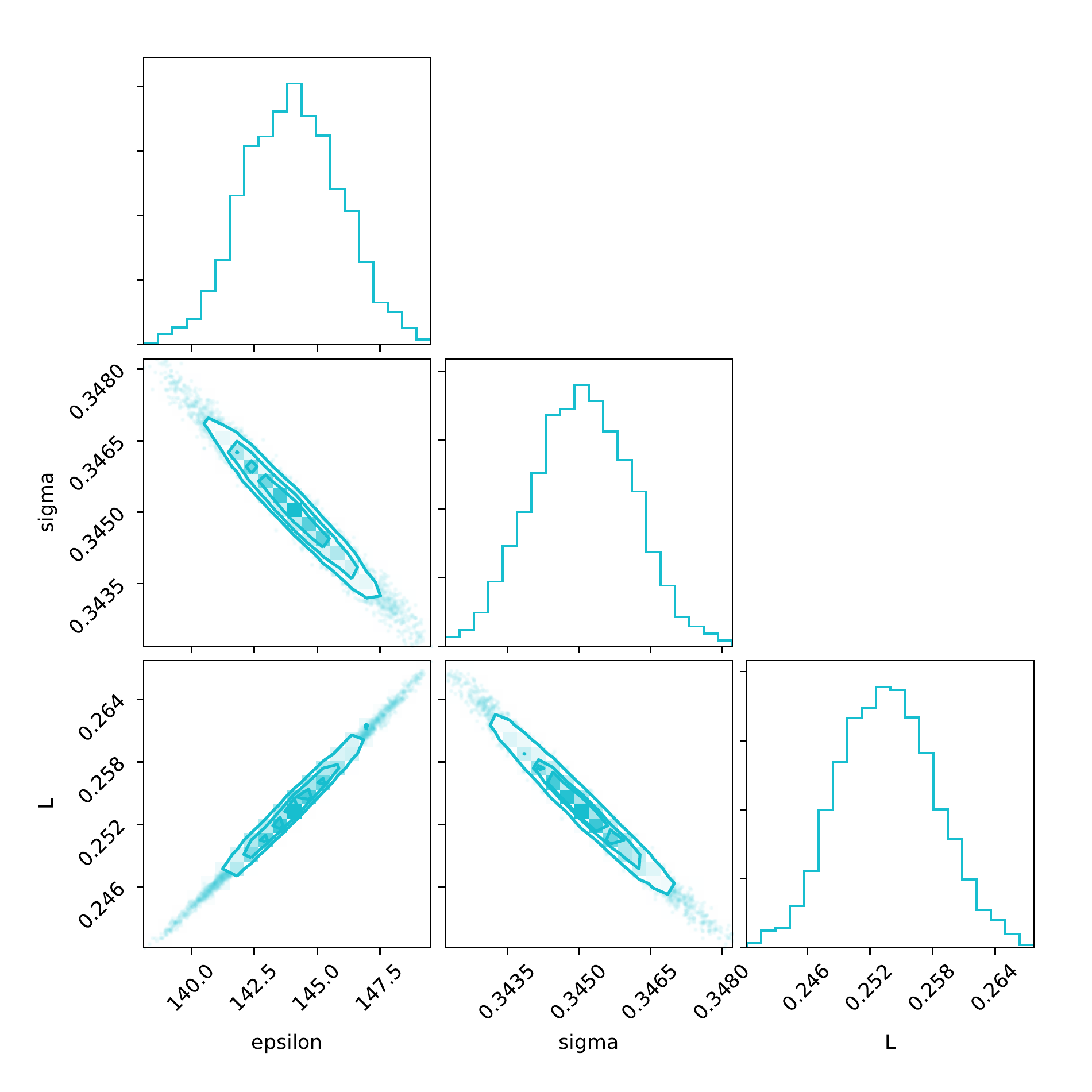}
        \includegraphics[width=0.3\textwidth]{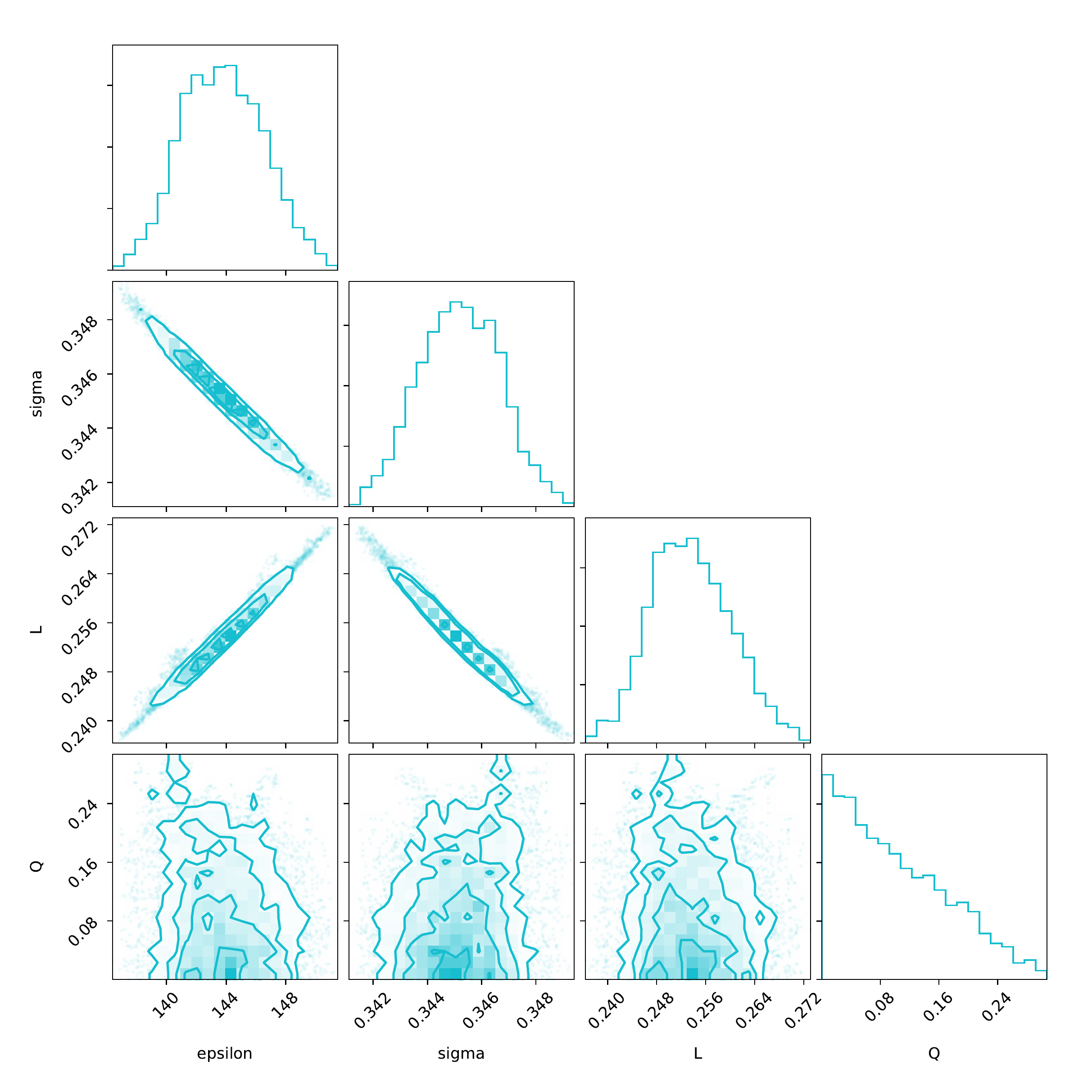}
    \caption{Parameter distributions for C$_2$H$_6$, $\rho_l, P_{sat}$ target. From left to right: UA, AUA, AUA+Q}
    \label{fig:2crit_C2H6_triangle}
\end{figure}
\subsubsection{C$_2$F$_4$}
\begin{figure}[h]
    \includegraphics[width=0.3\textwidth]{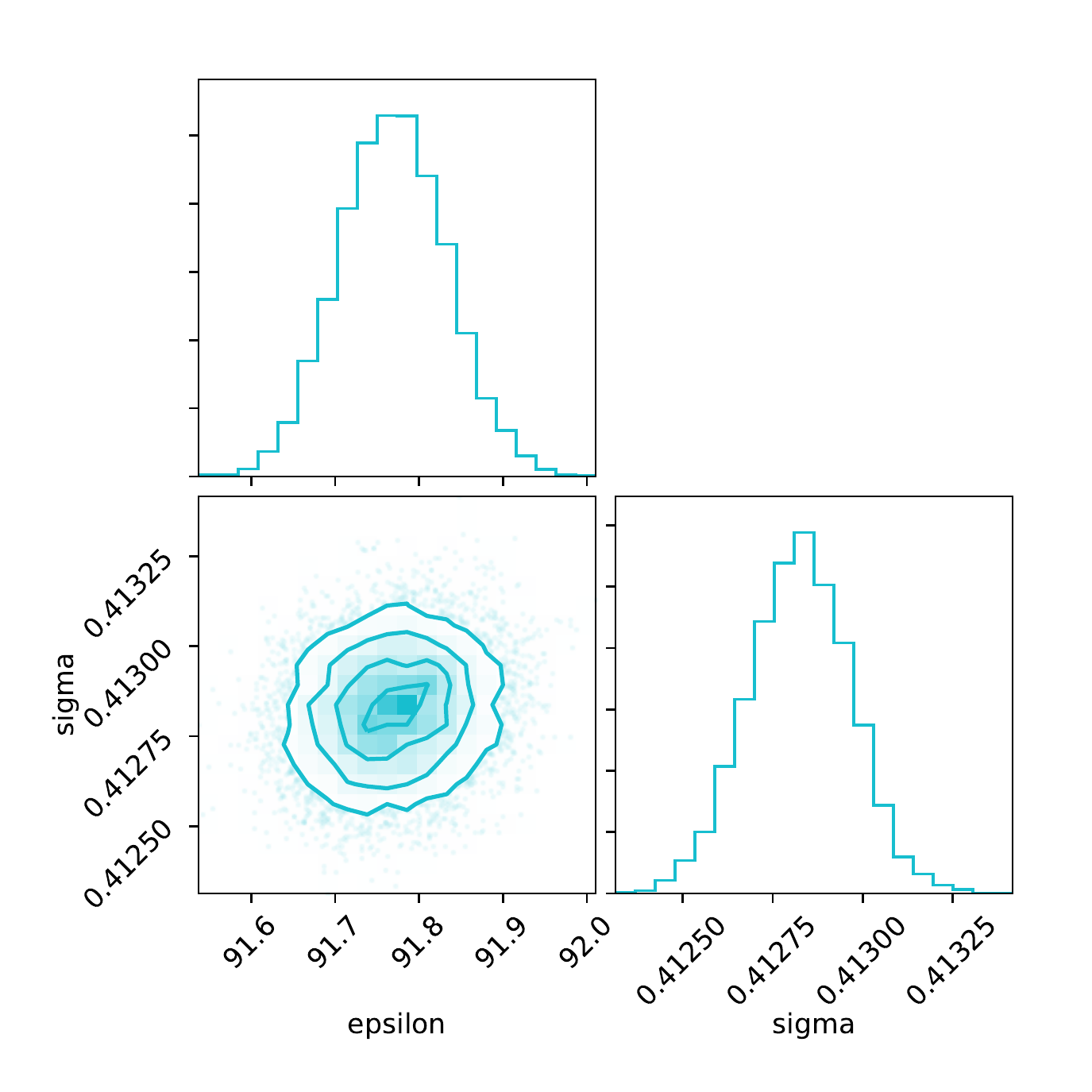}
    \includegraphics[width=0.3\textwidth]{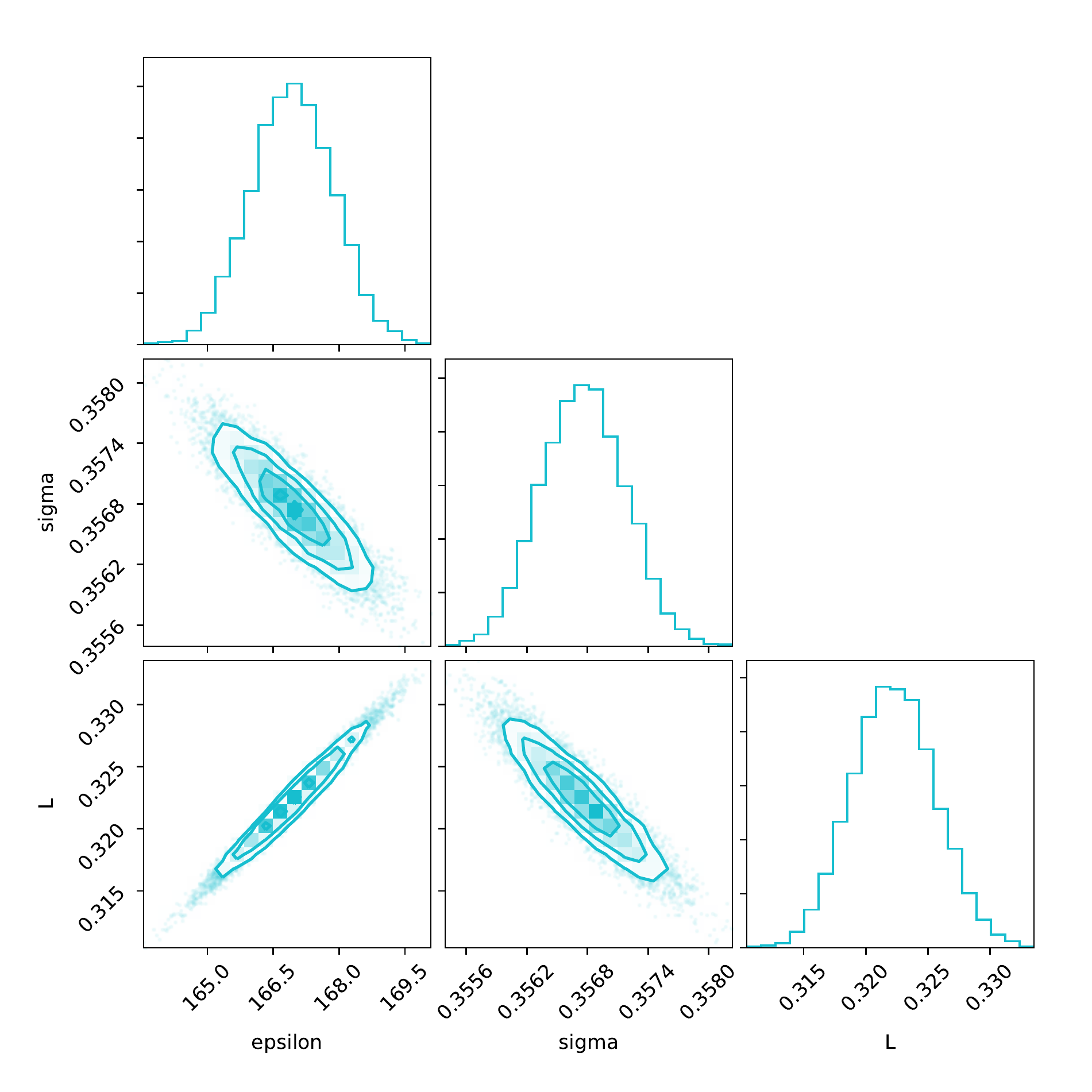}
        \includegraphics[width=0.3\textwidth]{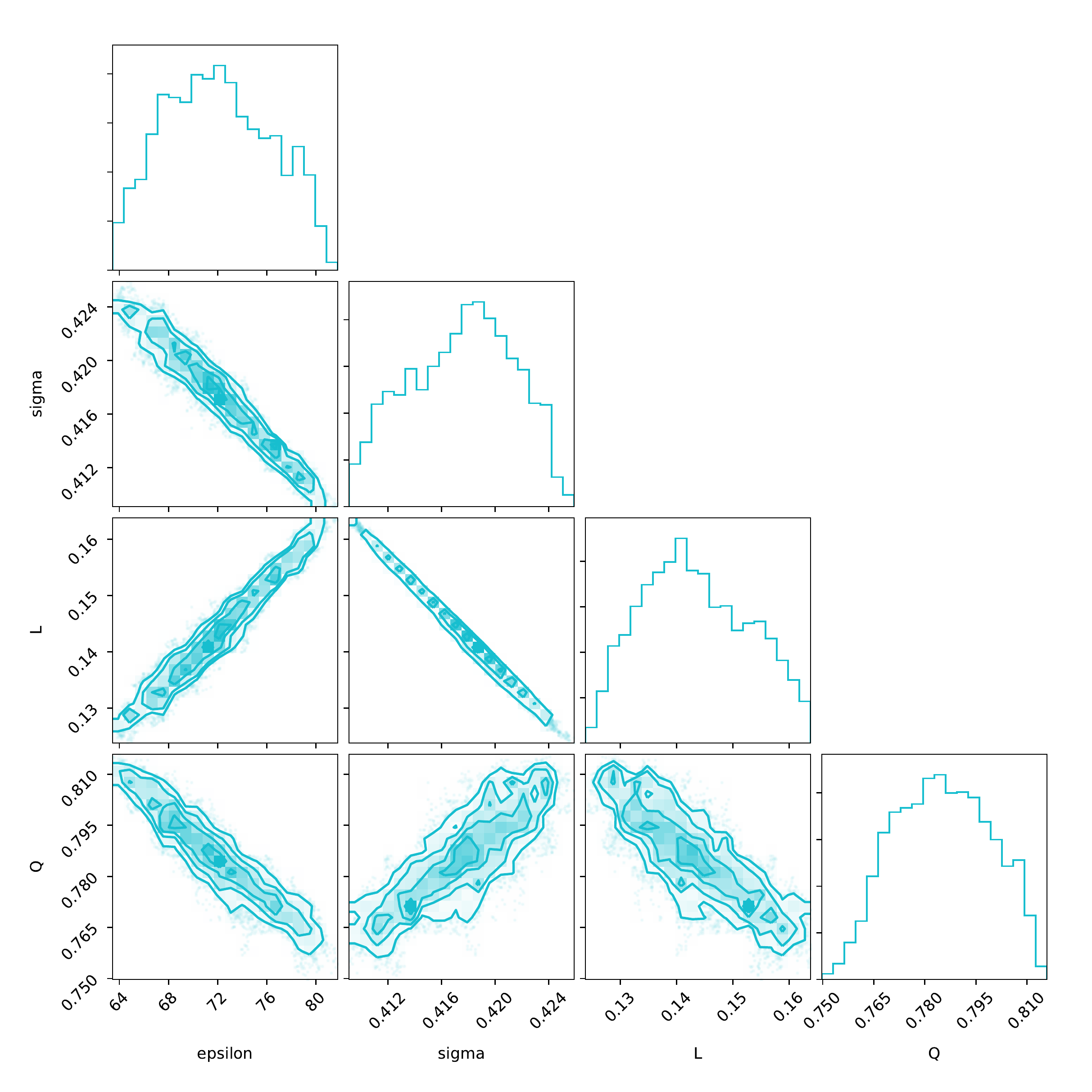}
    \caption{Parameter distributions for C$_2$F$_4$, $\rho_l, P_{sat}$ target. From left to right: UA, AUA, AUA+Q}
    \label{fig:2crit_C2F4_triangle}
\end{figure}
\newpage
\subsection{$\rho_l, P_{sat}, \gamma$ target}
\subsubsection{F$_2$}
\begin{figure}[h]
    \includegraphics[width=0.3\textwidth]{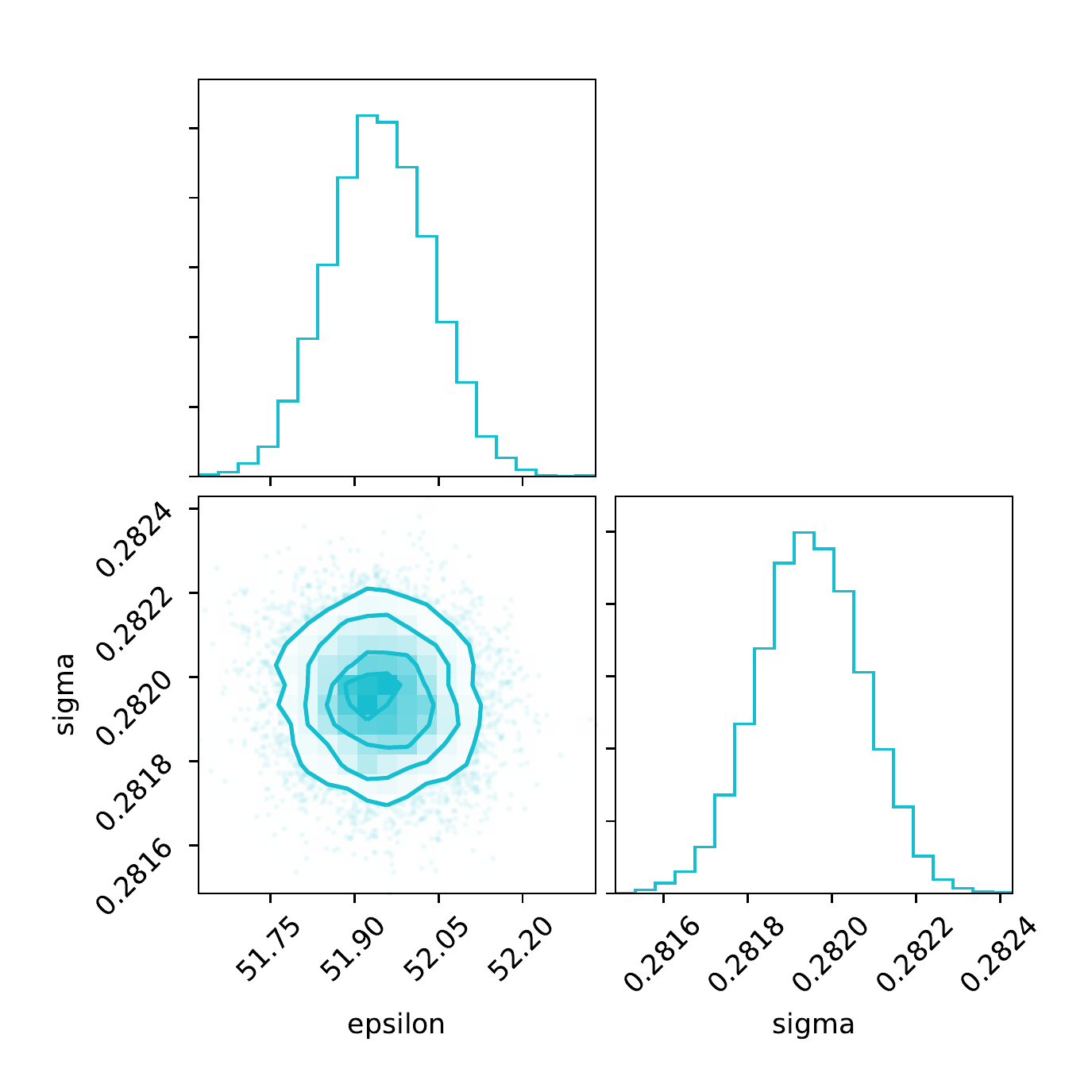}
    \includegraphics[width=0.3\textwidth]{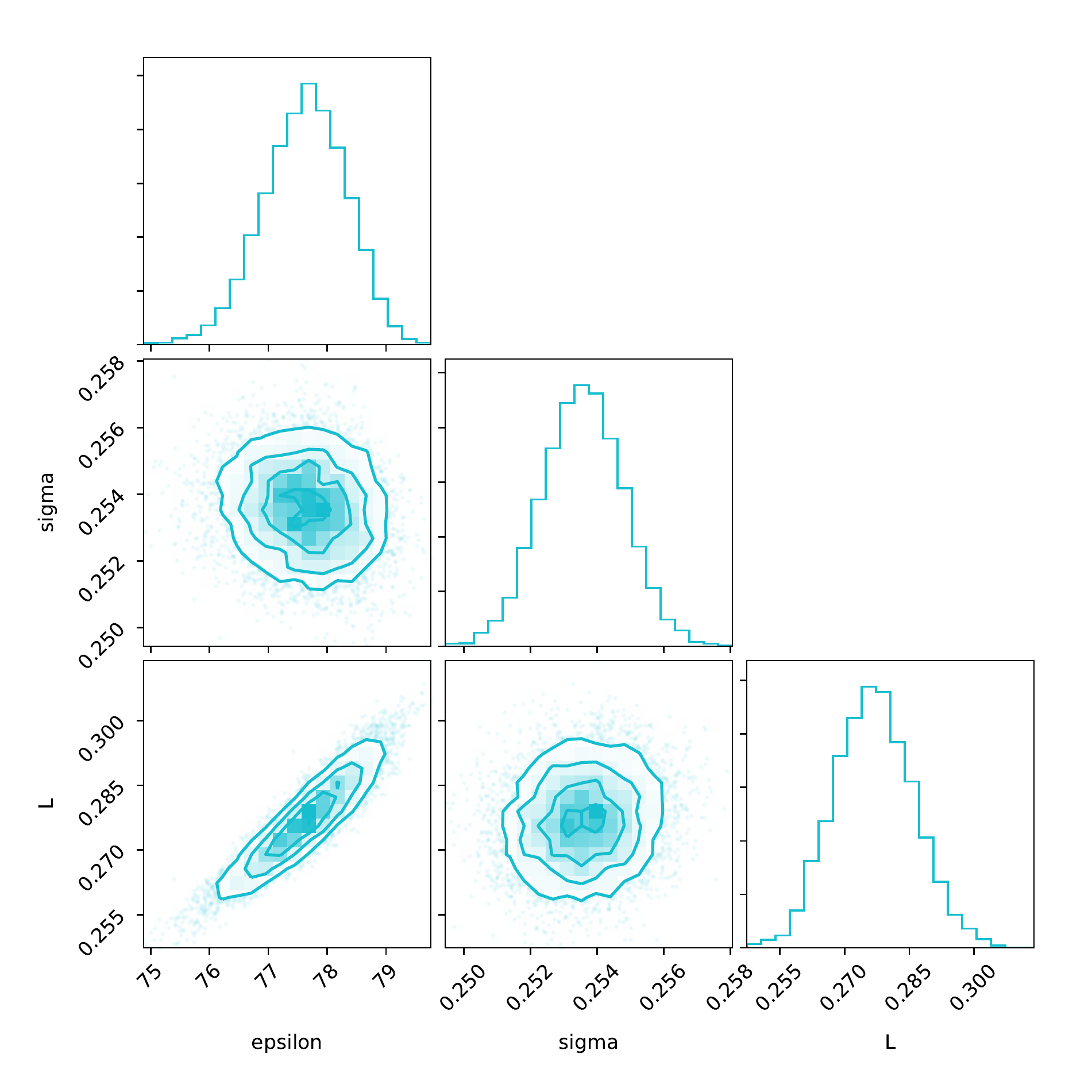}
        \includegraphics[width=0.3\textwidth]{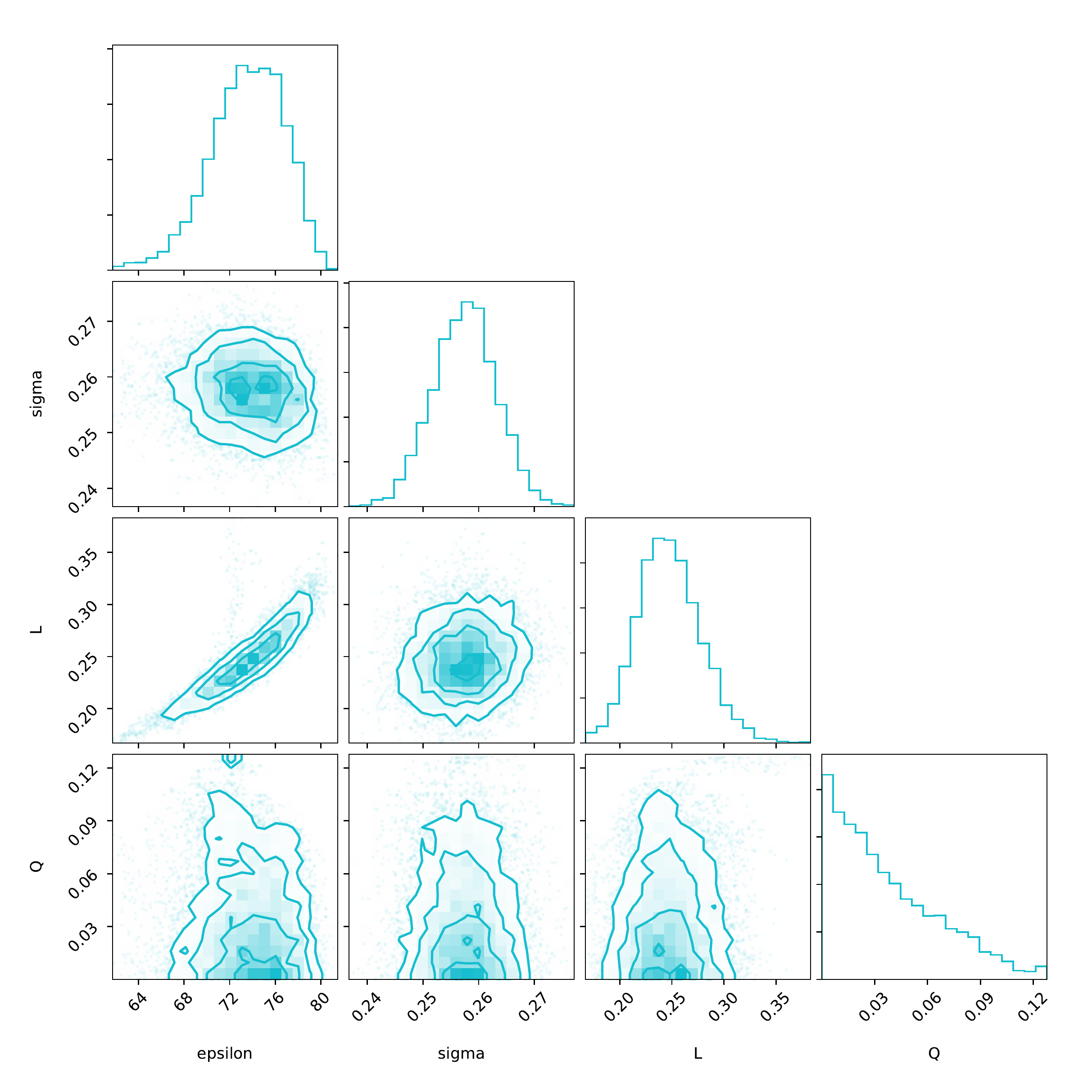}
    \caption{Parameter distributions for F$_2$, $\rho_l, P_{sat}, \gamma$ target. From left to right: UA, AUA, AUA+Q}
    \label{fig:3crit_F2_triangle}
\end{figure}
\subsubsection{Br$_2$}
\begin{figure}[h]
    \includegraphics[width=0.3\textwidth]{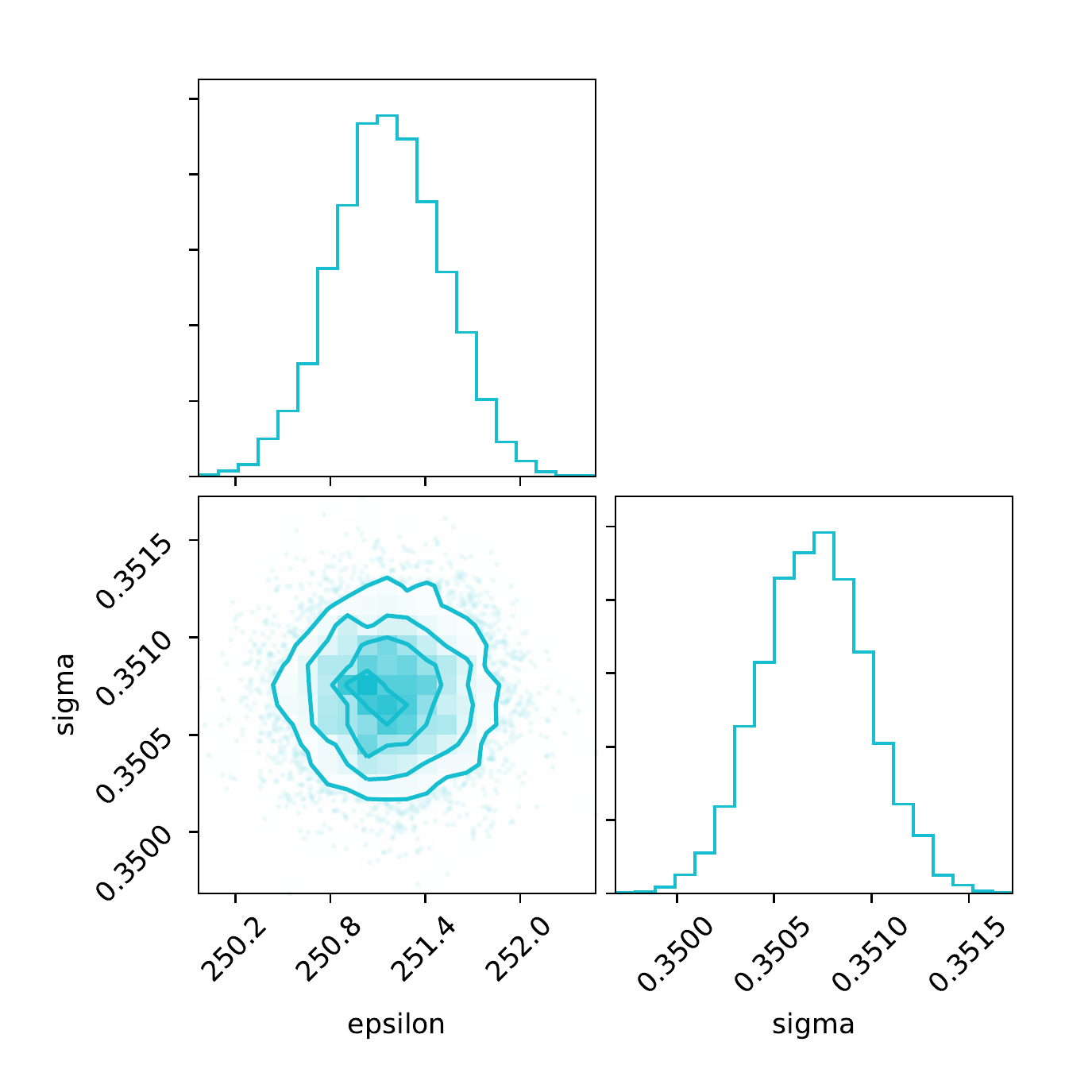}
    \includegraphics[width=0.3\textwidth]{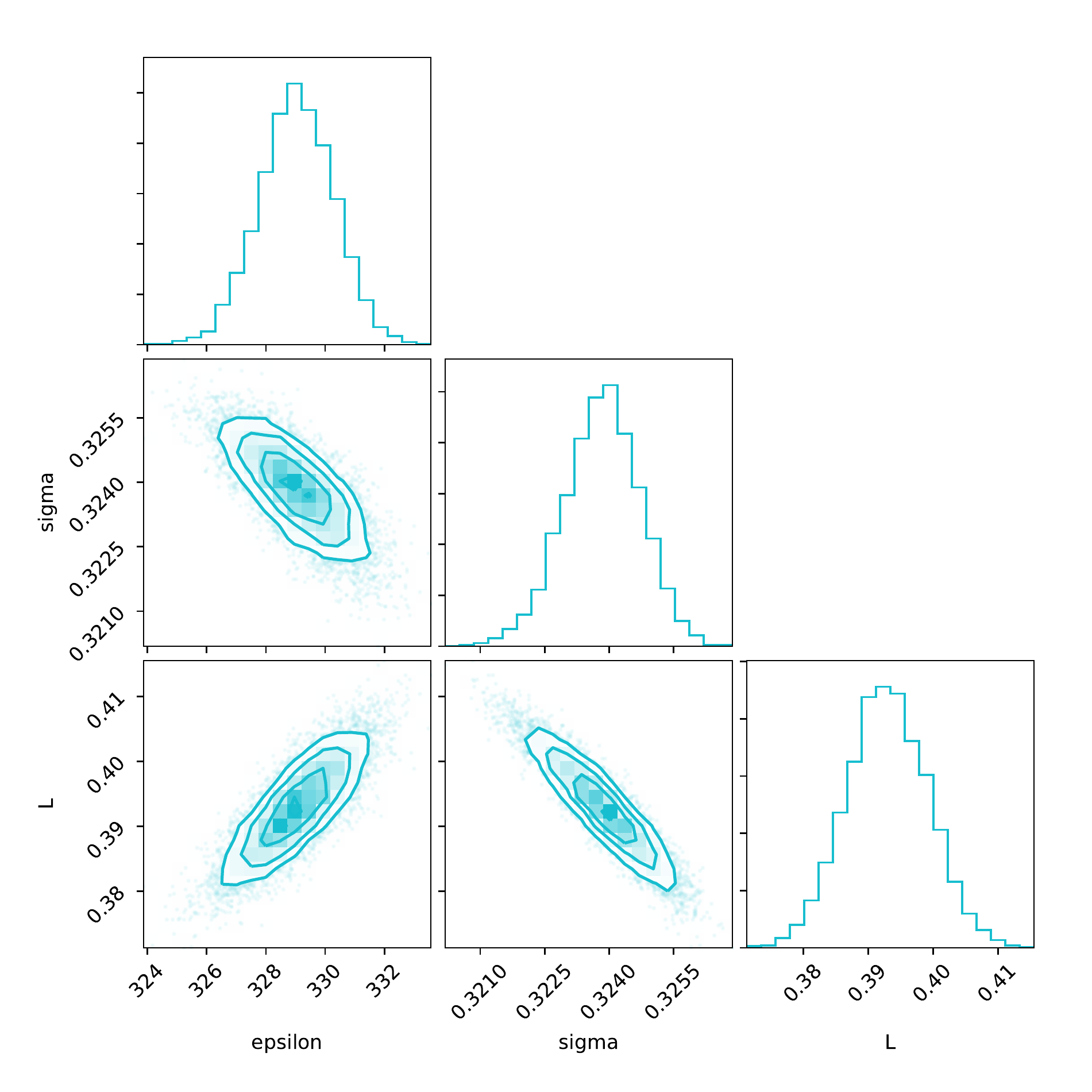}
        \includegraphics[width=0.3\textwidth]{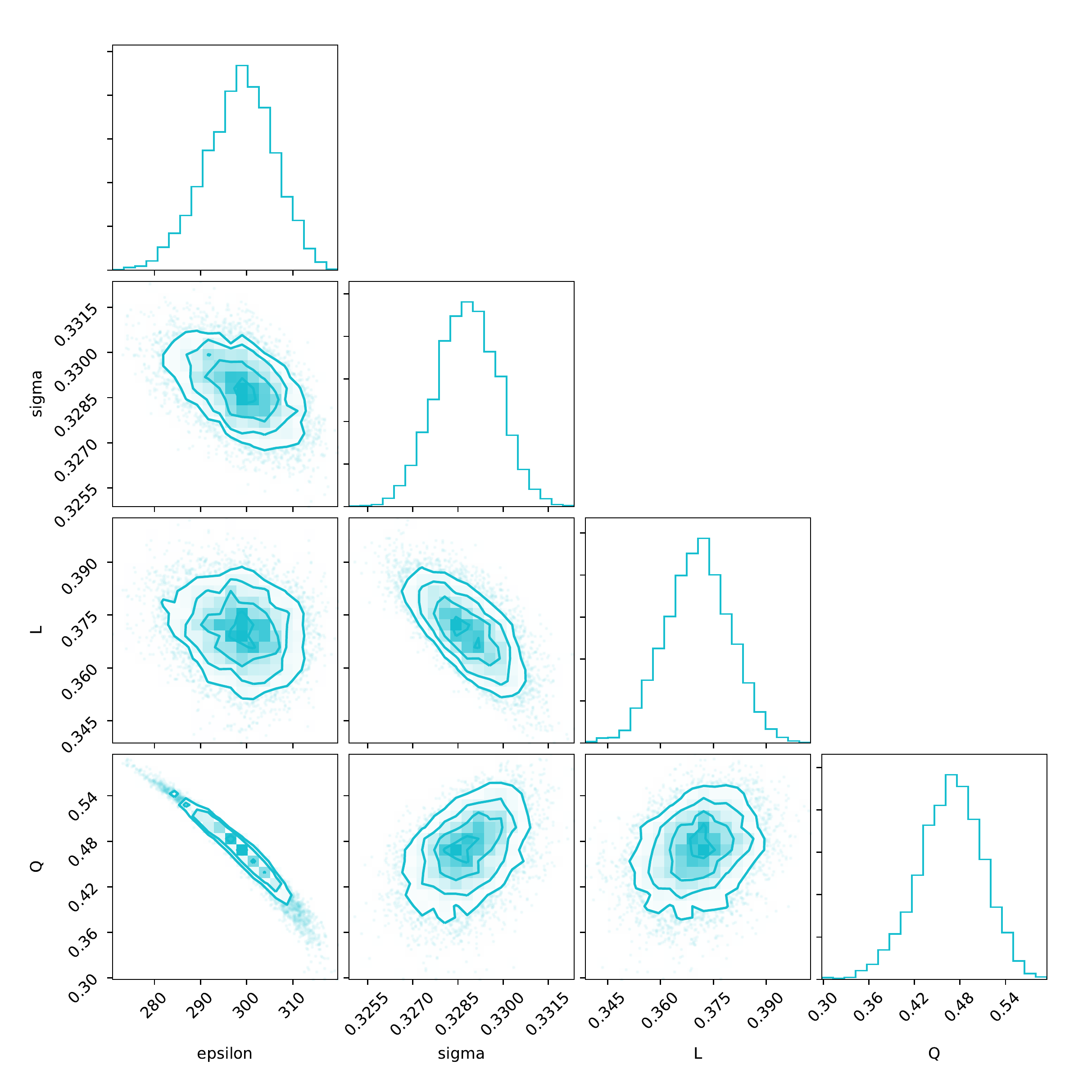}
    \caption{Parameter distributions for Br$_2$, $\rho_l, P_{sat}, \gamma$ target. From left to right: UA, AUA, AUA+Q}
    \label{fig:3crit_Br2_triangle}
\end{figure}
\newpage
\subsubsection{N$_2$}
\begin{figure}[h]
    \includegraphics[width=0.3\textwidth]{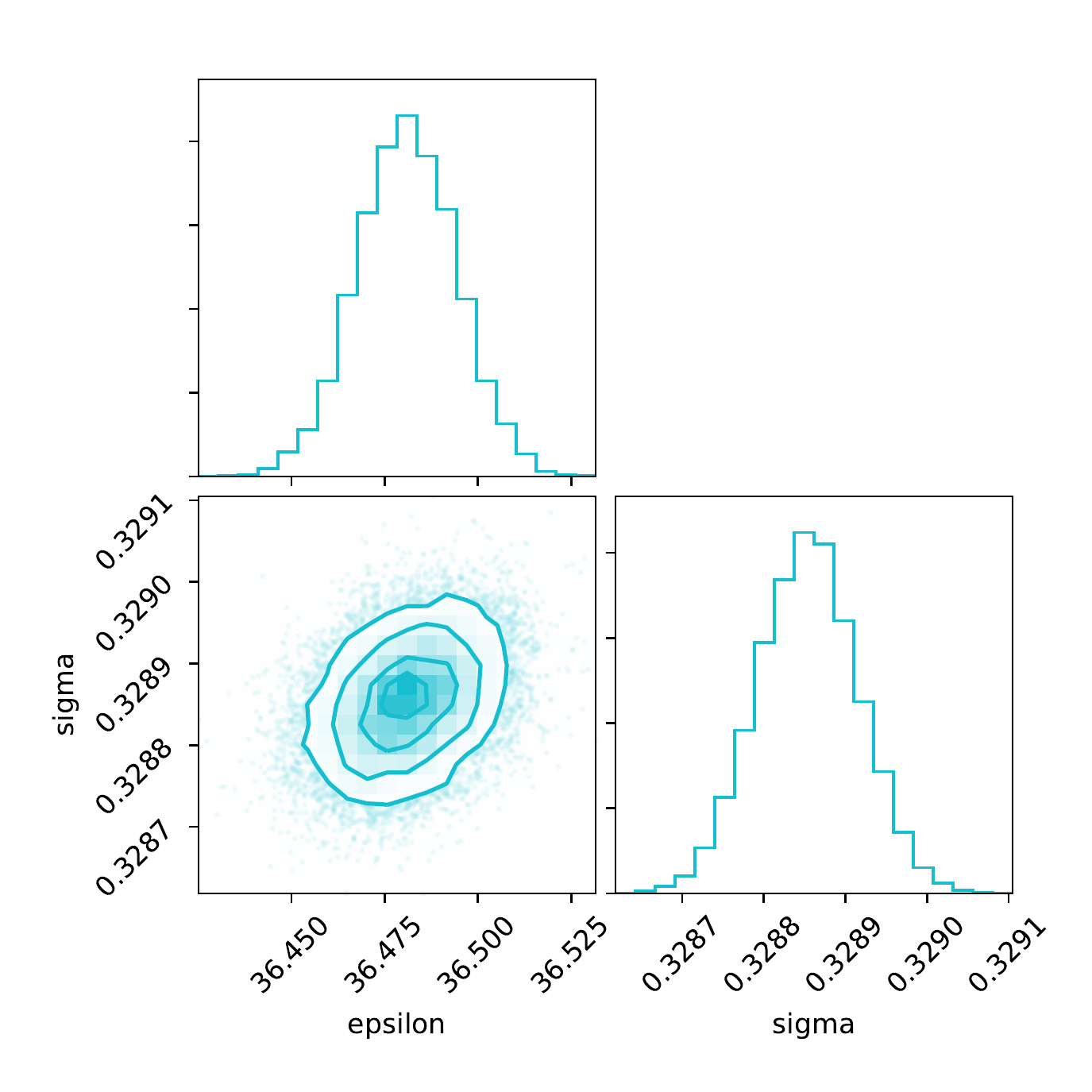}
    \includegraphics[width=0.3\textwidth]{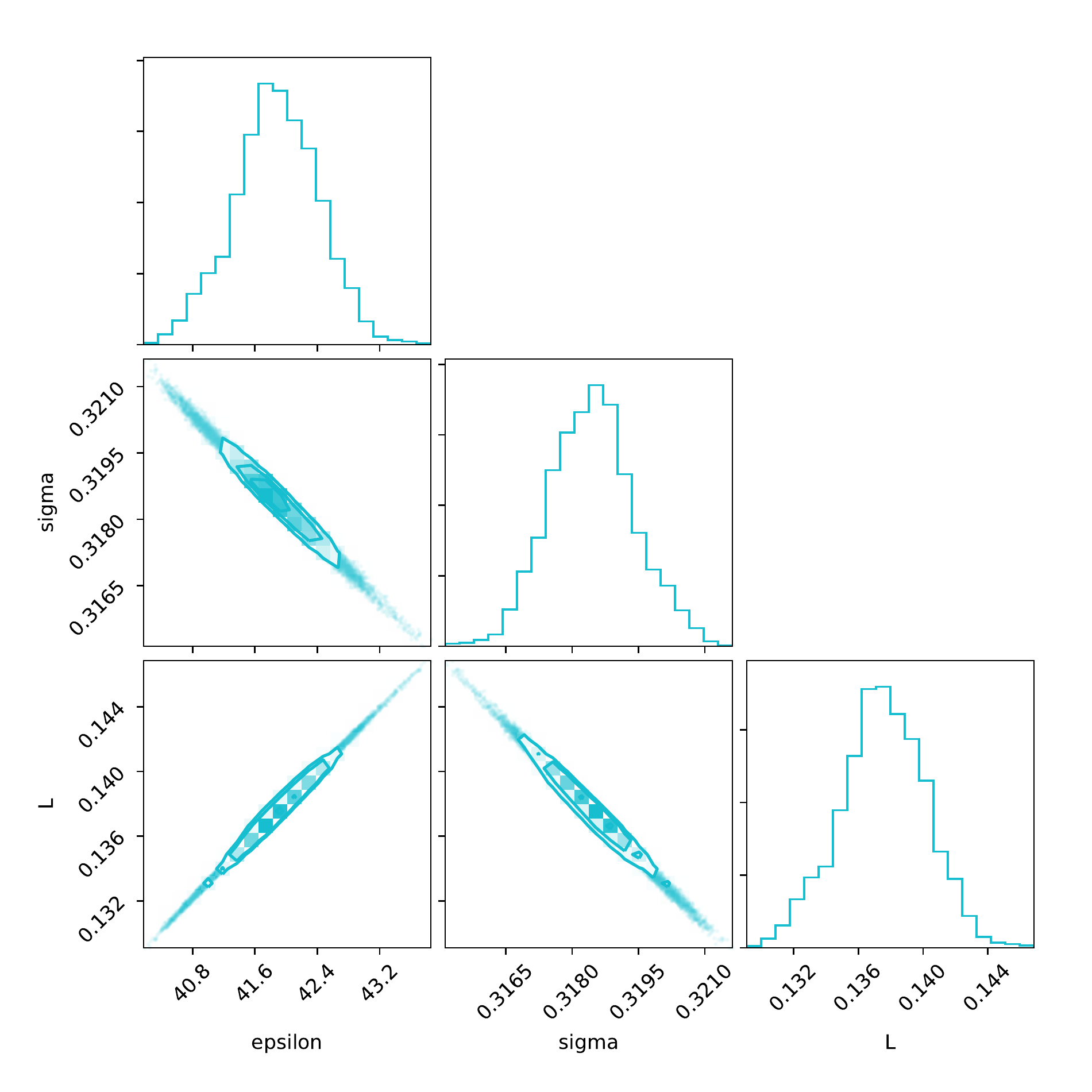}
        \includegraphics[width=0.3\textwidth]{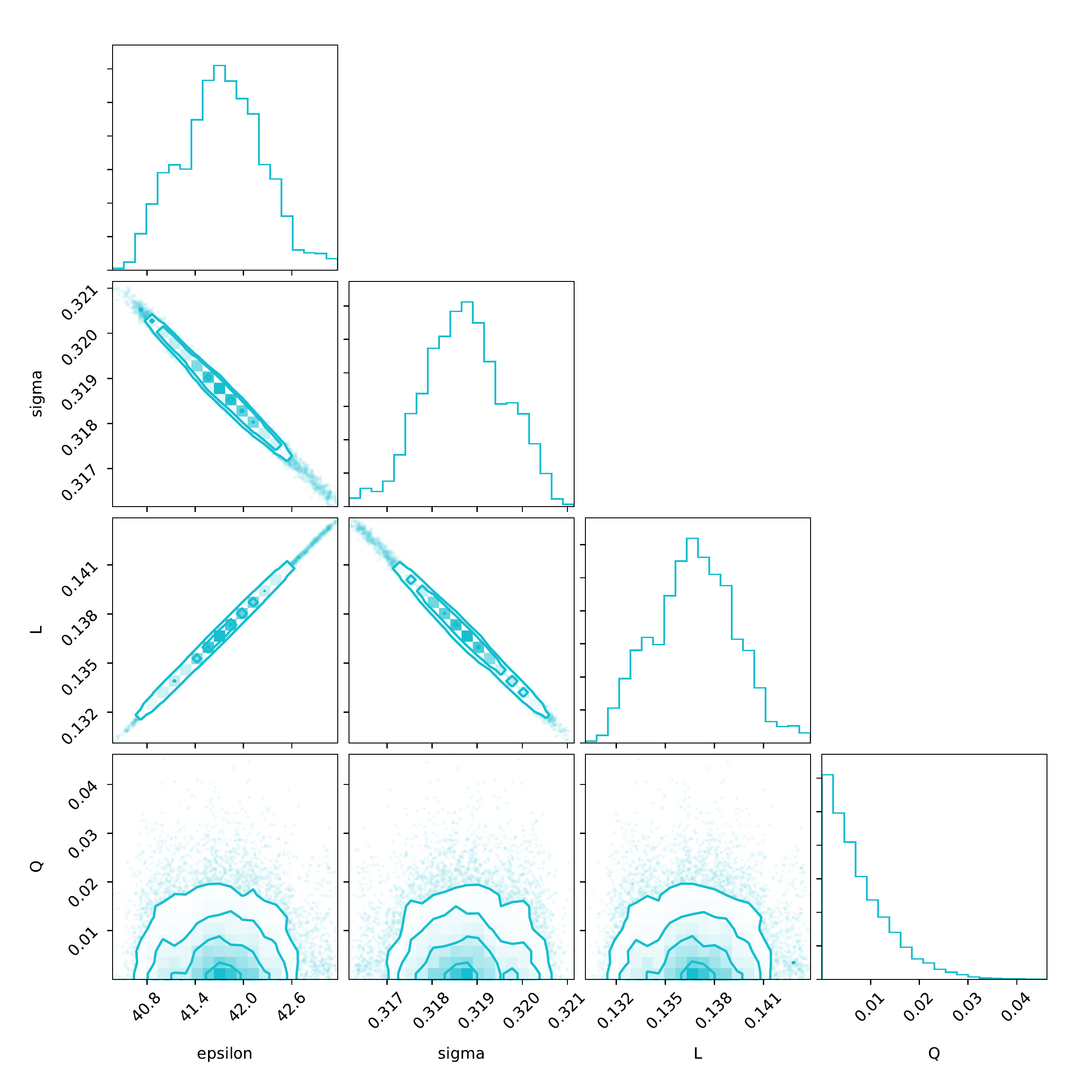}
    \caption{Parameter distributions for N$_2$, $\rho_l, P_{sat}, \gamma$ target. From left to right: UA, AUA, AUA+Q}
    \label{fig:3crit_N2_triangle}
\end{figure}

\subsubsection{O$_2$}
\begin{figure}[h]
    \includegraphics[width=0.3\textwidth]{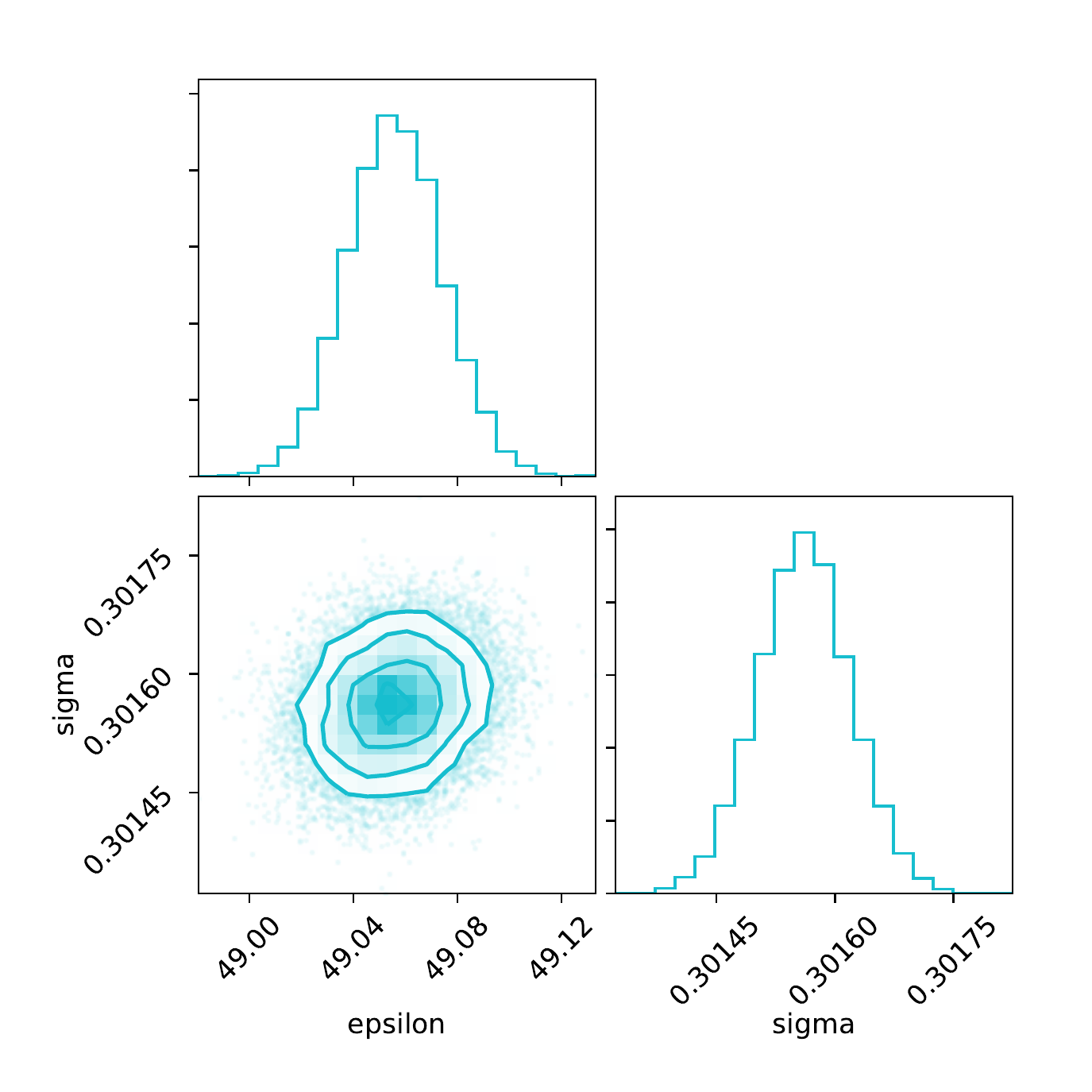}
    \includegraphics[width=0.3\textwidth]{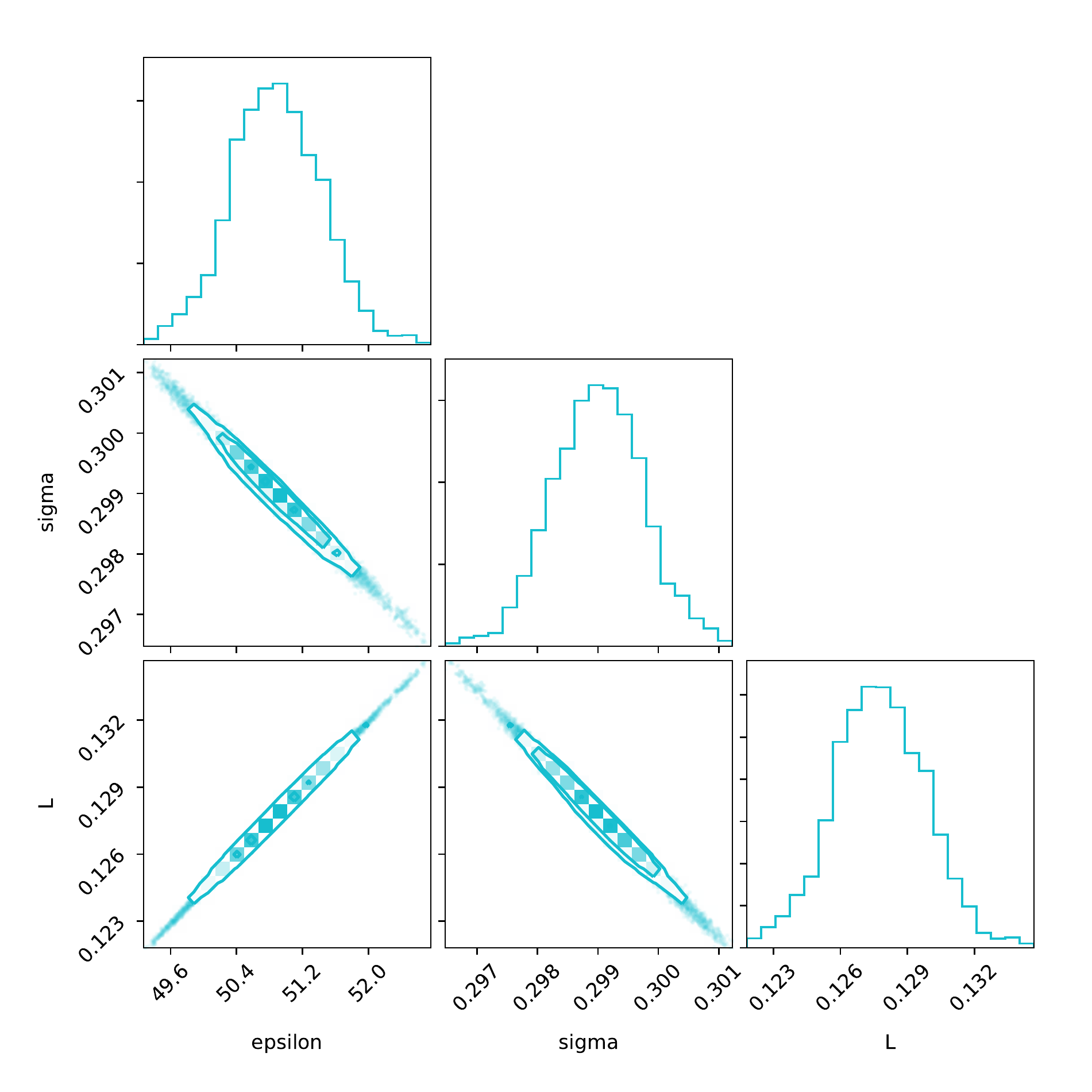}
        \includegraphics[width=0.3\textwidth]{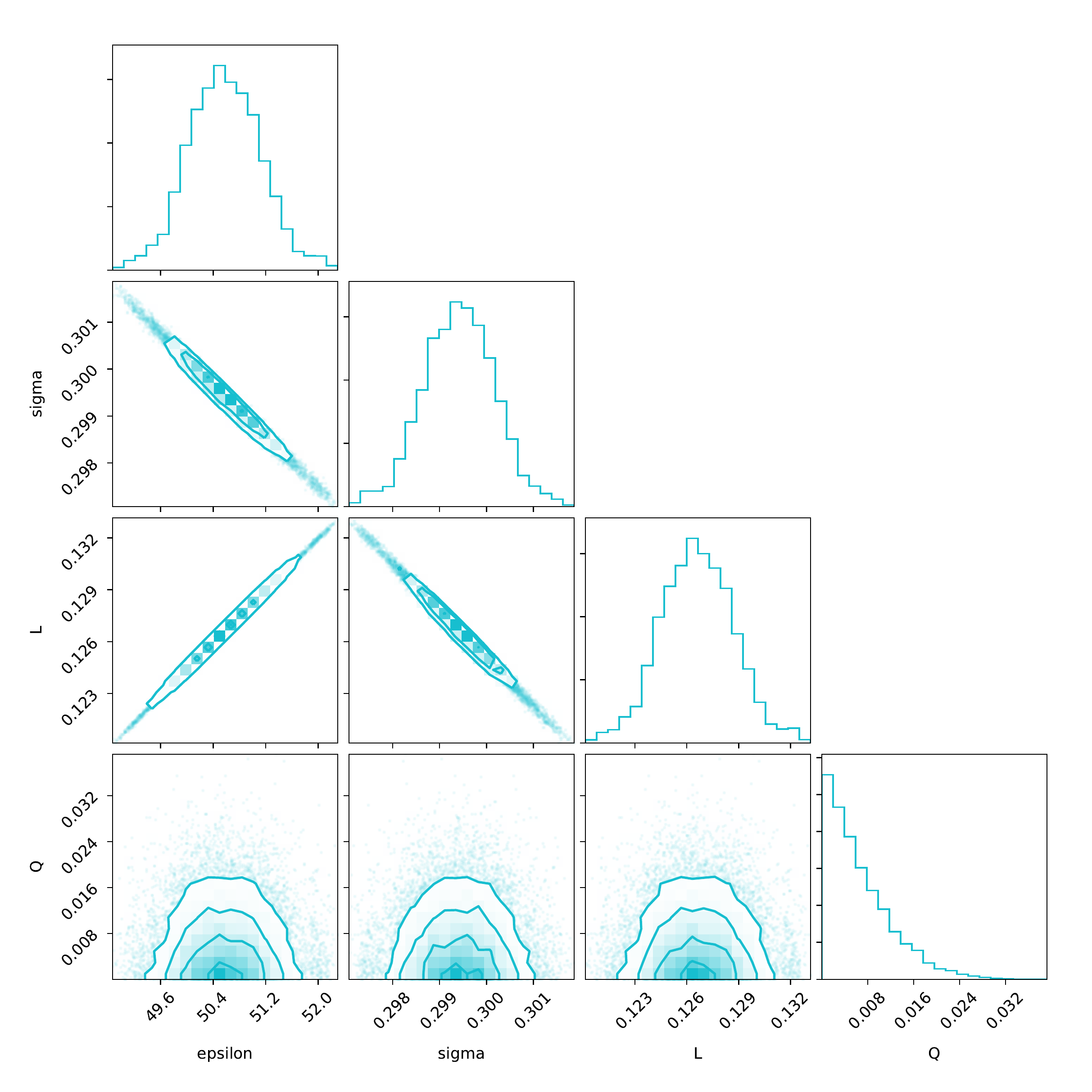}
    \caption{Parameter distributions for O$_2$, $\rho_l, P_{sat}, \gamma$ target. From left to right: UA, AUA, AUA+Q}
    \label{fig:3crit_N2_triangle}
\end{figure}
\newpage
\subsubsection{C$_2$H$_2$}
\begin{figure}[h]
    \includegraphics[width=0.3\textwidth]{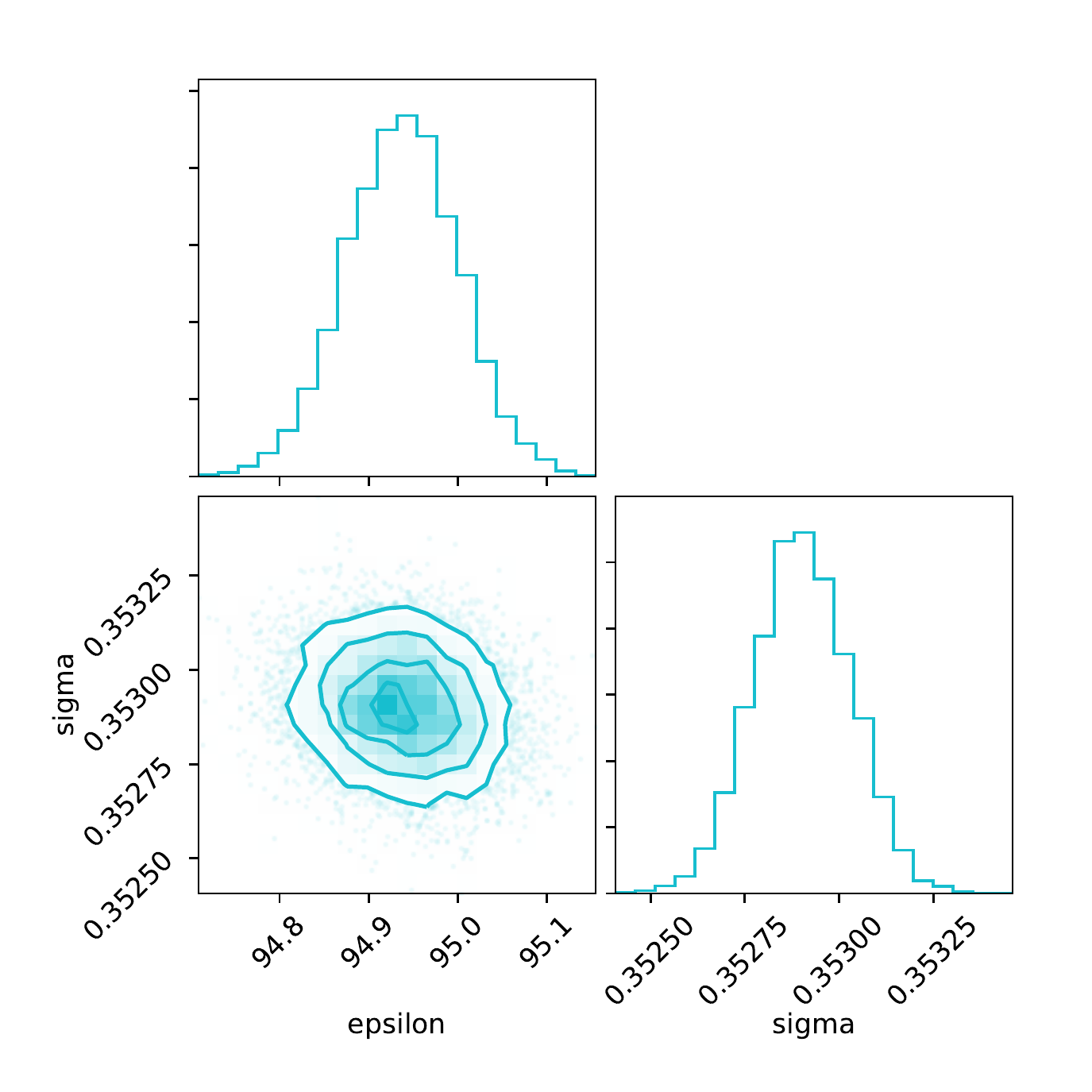}
    \includegraphics[width=0.3\textwidth]{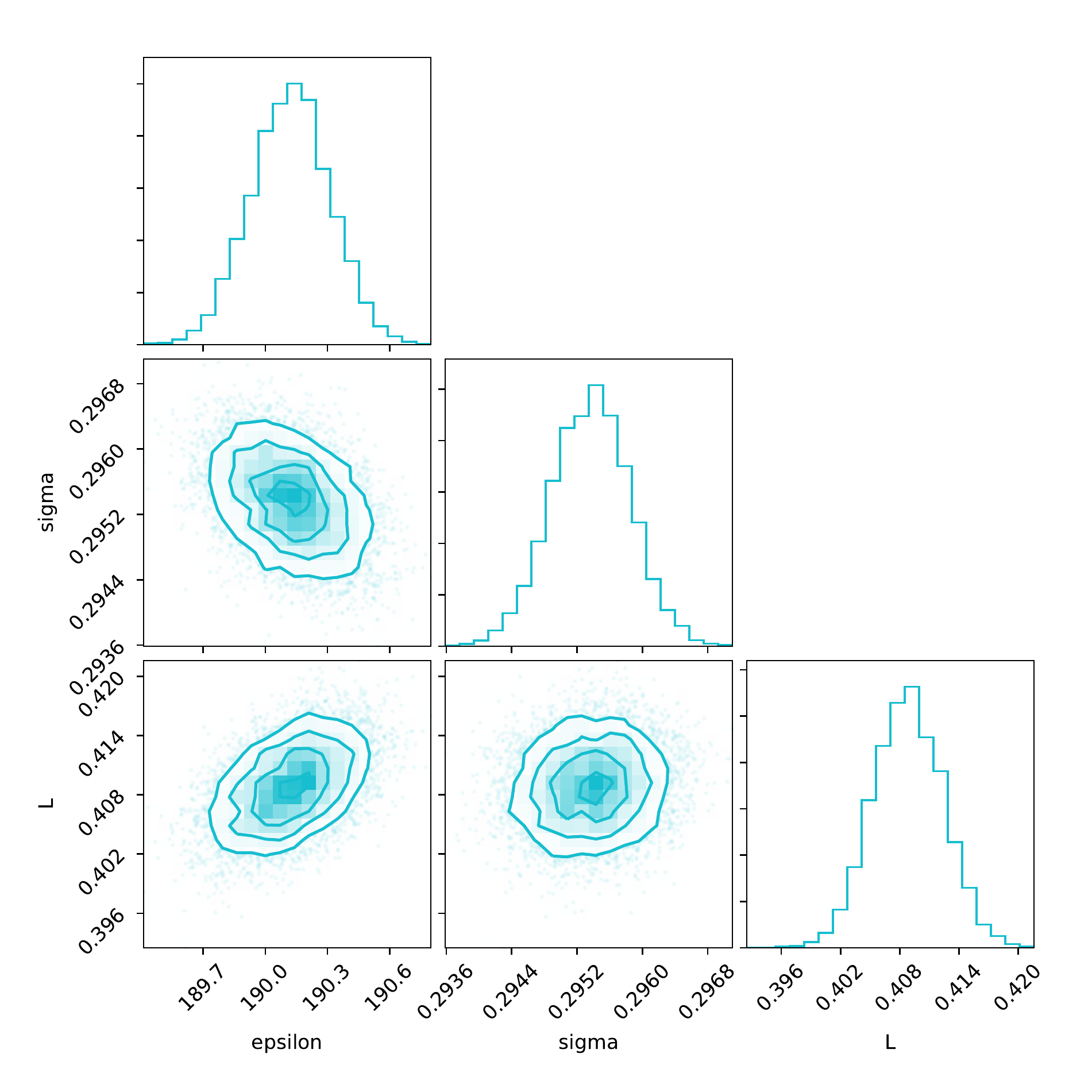}
        \includegraphics[width=0.3\textwidth]{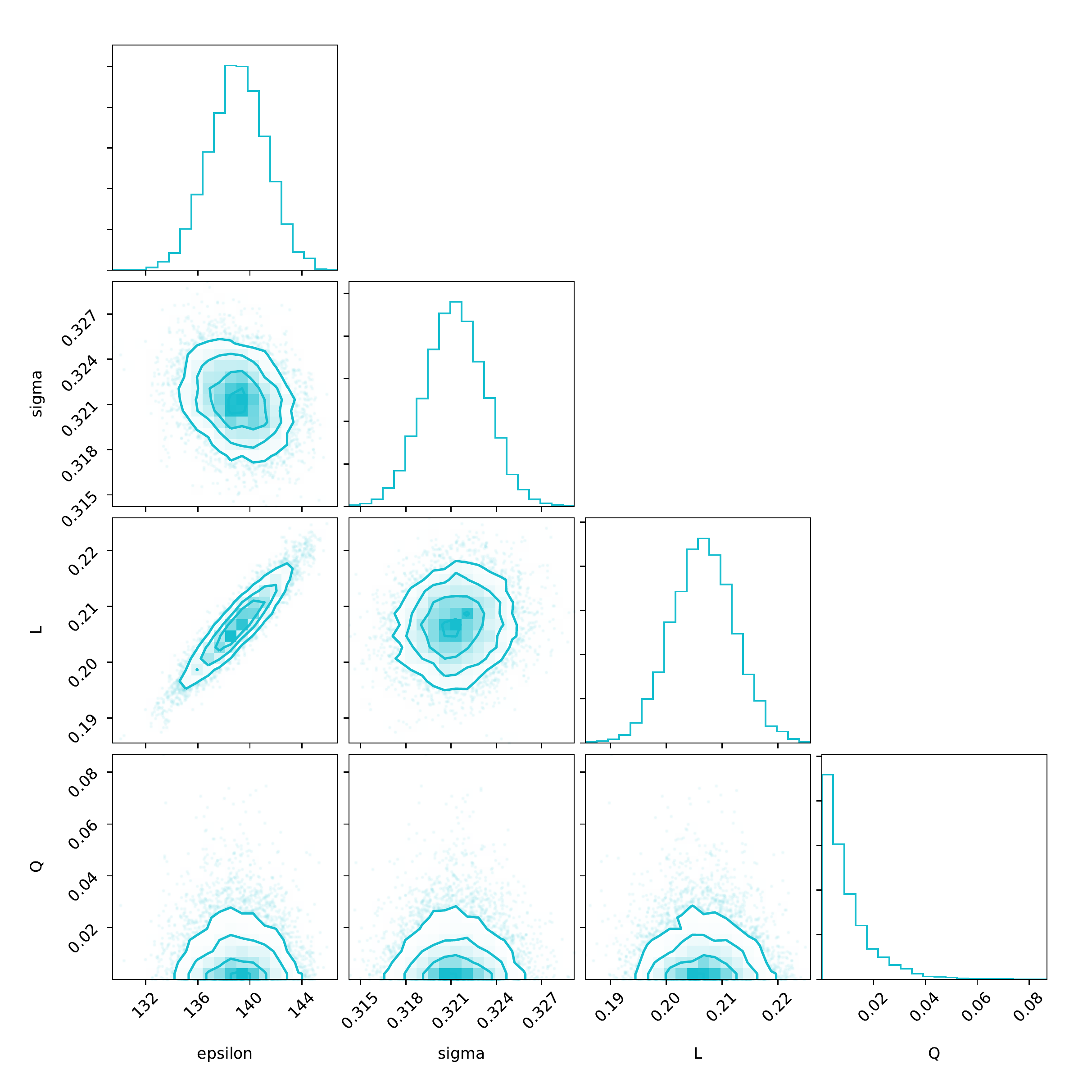}
    \caption{Parameter distributions for C$_2$H$_2$, $\rho_l, P_{sat}, \gamma$ target. From left to right: UA, AUA, AUA+Q}
    \label{fig:3crit_C2H2_triangle}
\end{figure}
\subsubsection{C$_2$H$_4$}
\begin{figure}[h]
    \includegraphics[width=0.3\textwidth]{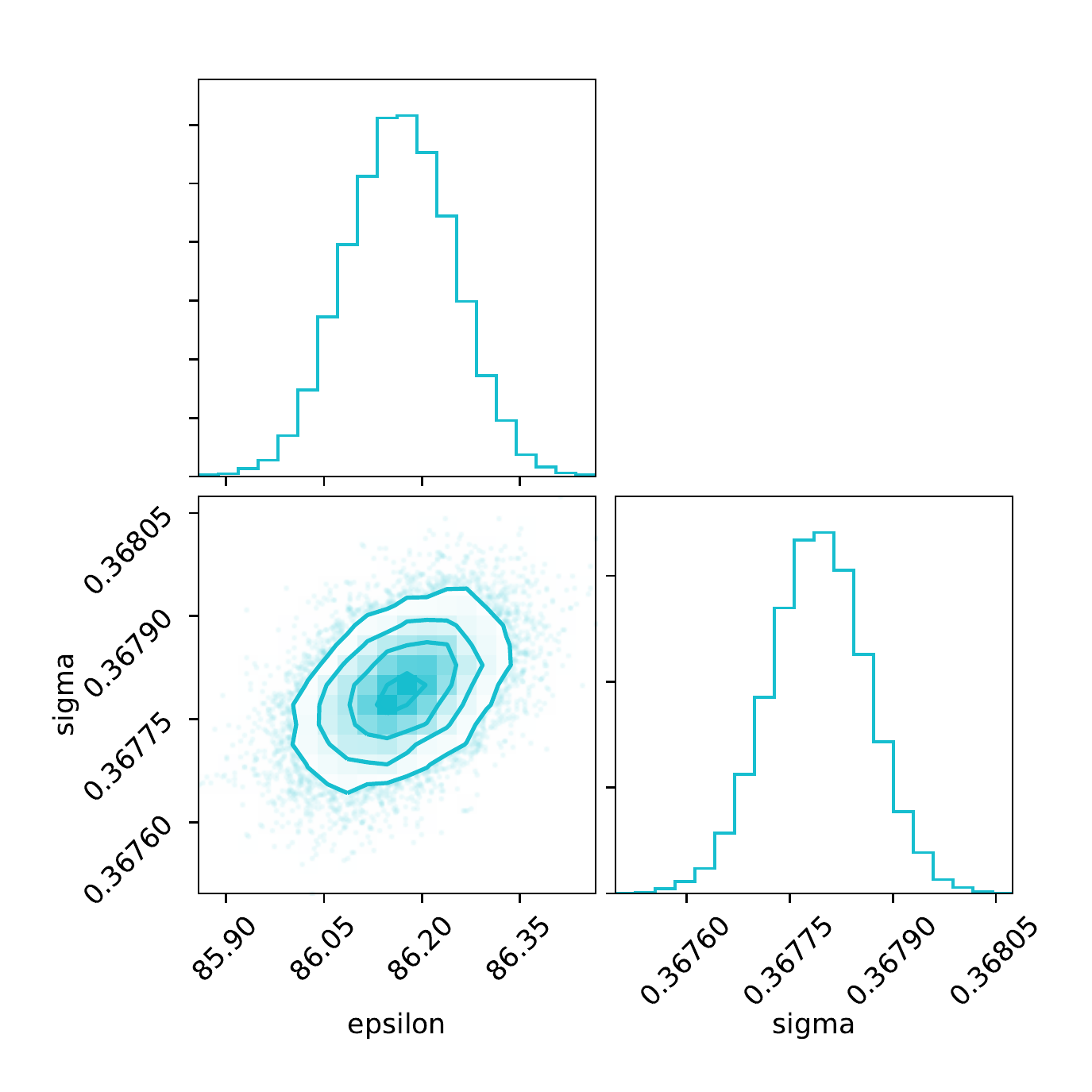}
    \includegraphics[width=0.3\textwidth]{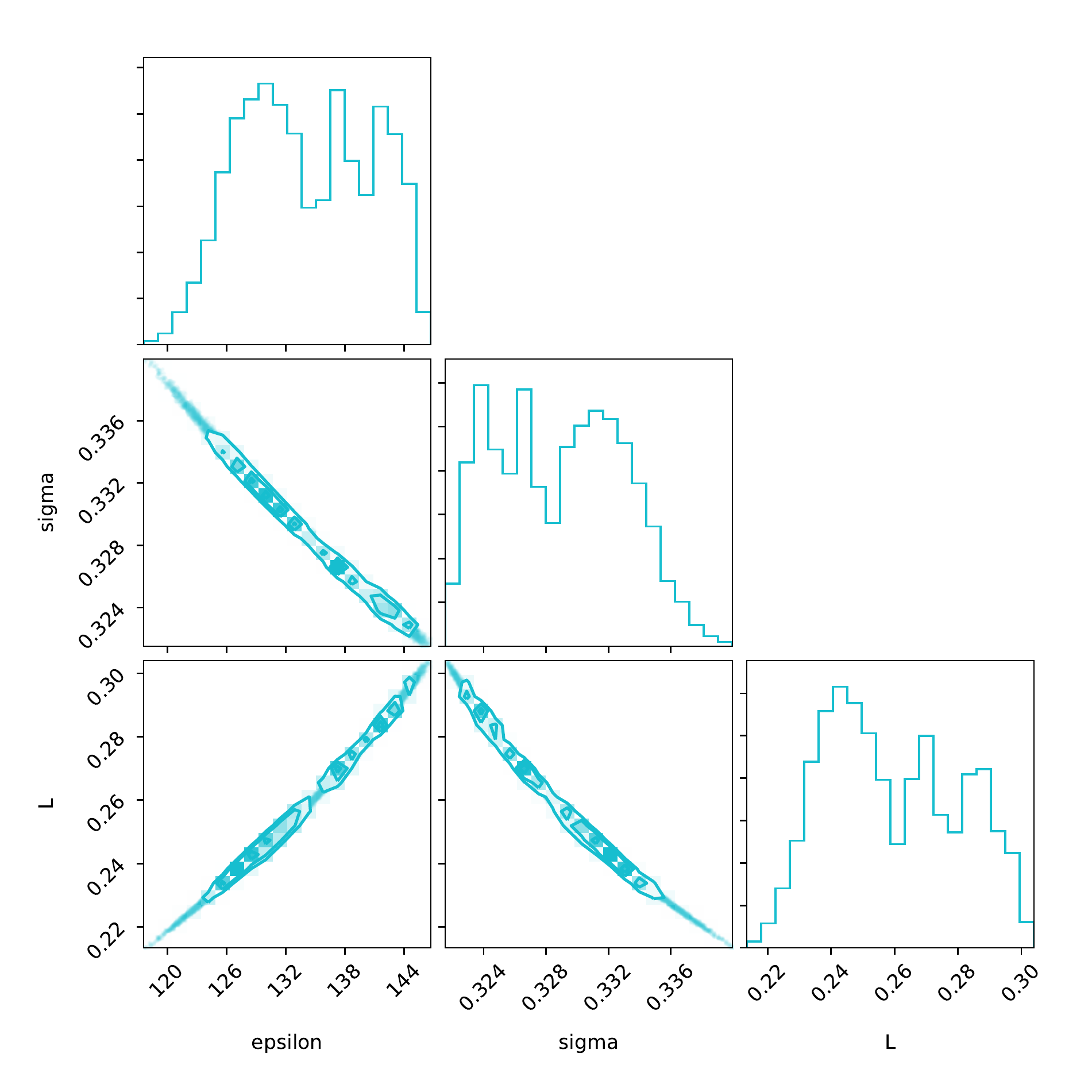}
        \includegraphics[width=0.3\textwidth]{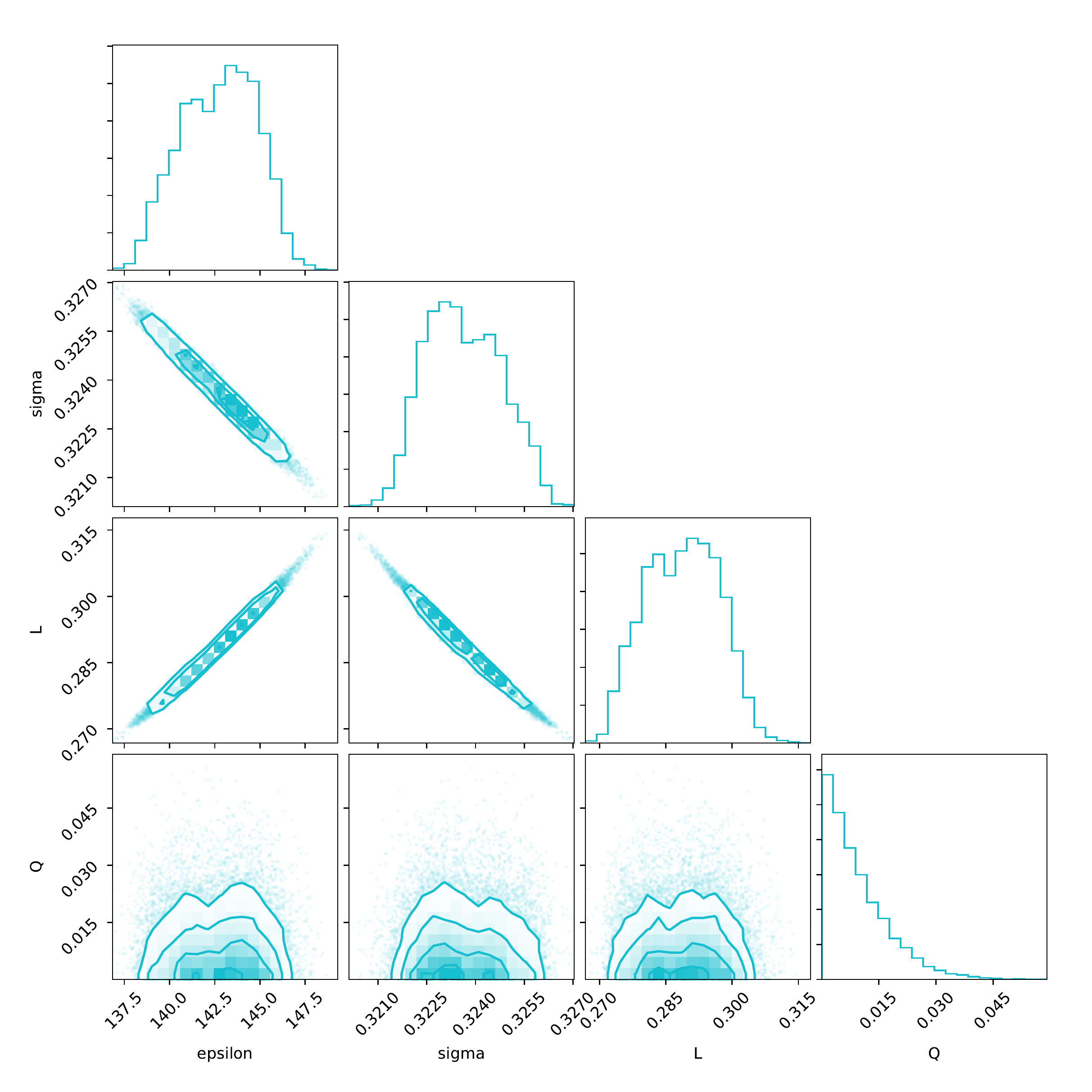}
    \caption{Parameter distributions for C$_2$H$_4$, $\rho_l, P_{sat}, \gamma$ target. From left to right: UA, AUA, AUA+Q}
    \label{fig:3crit_C2H4_triangle}
\end{figure}
\newpage
\subsubsection{C$_2$H$_6$}
\begin{figure}[h]
    \includegraphics[width=0.3\textwidth]{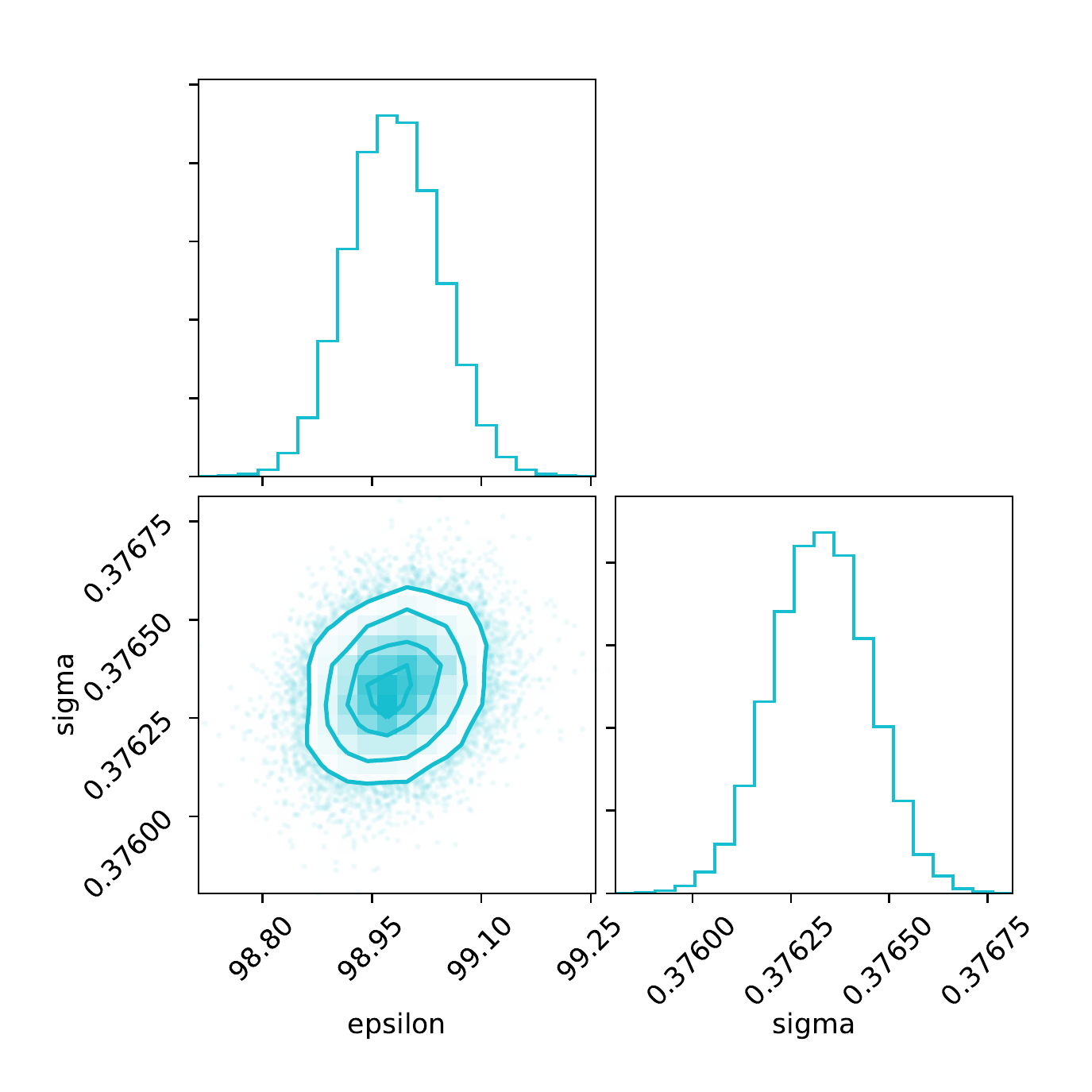}
    \includegraphics[width=0.3\textwidth]{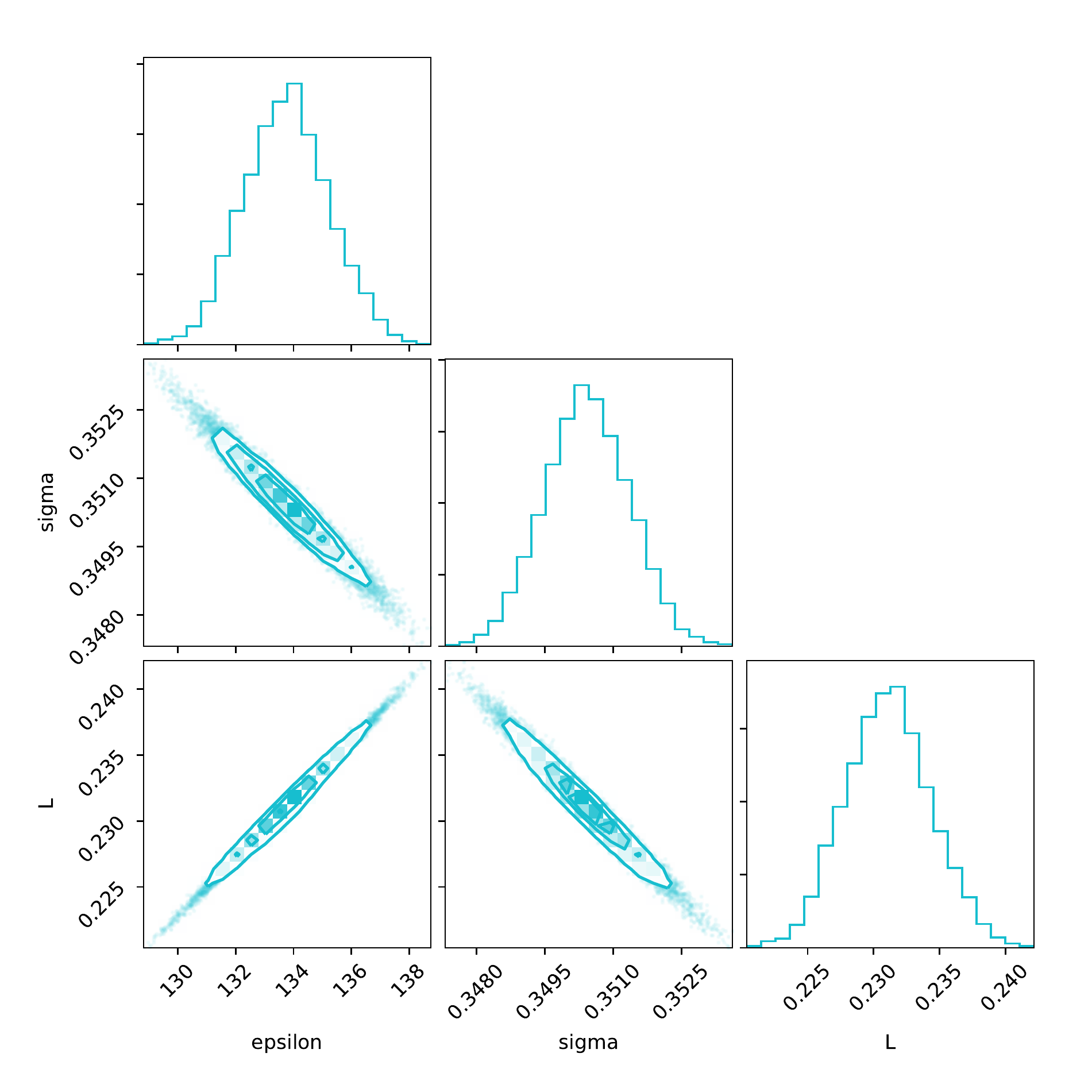}
        \includegraphics[width=0.3\textwidth]{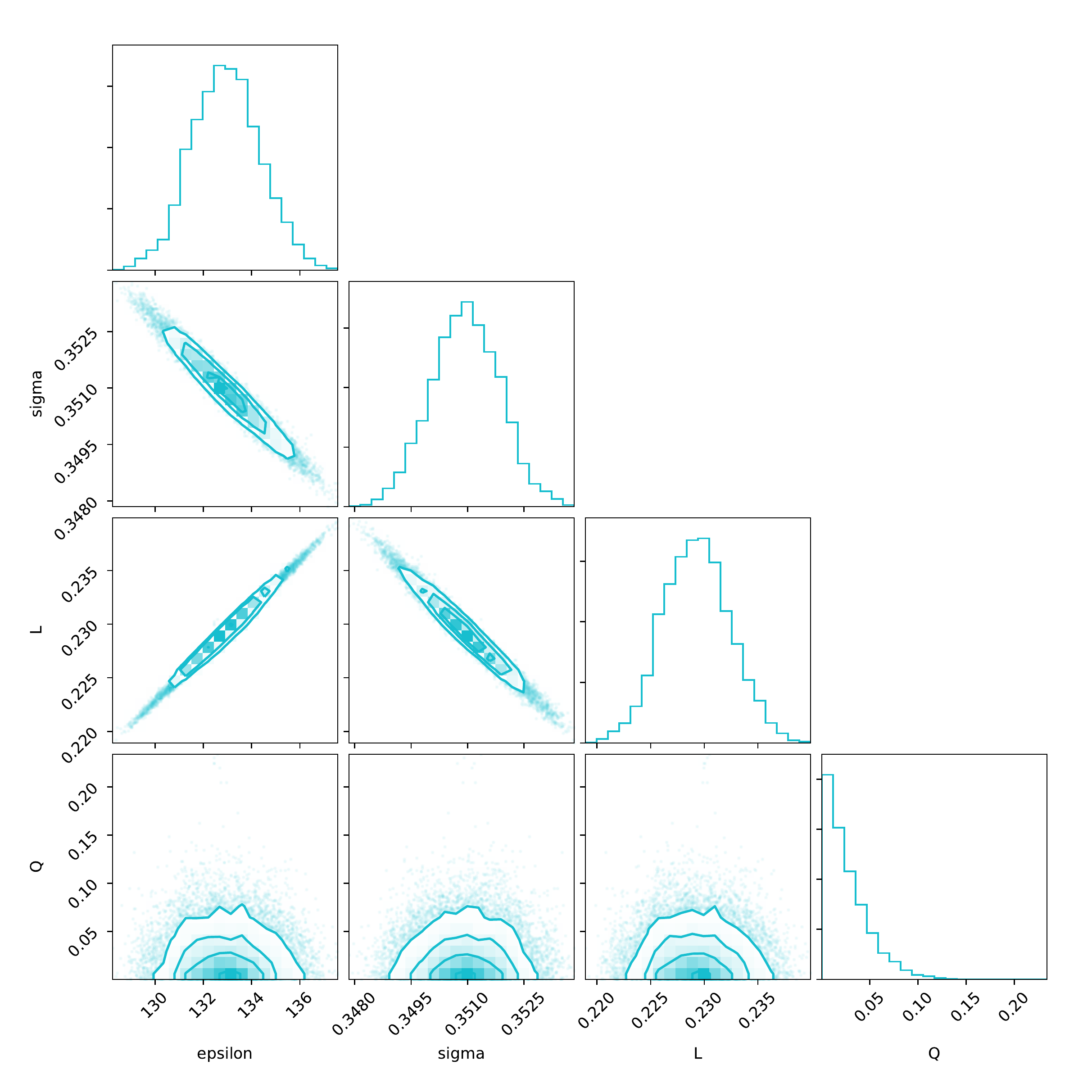}
    \caption{Parameter distributions for C$_2$H$_6$, $\rho_l, P_{sat}, \gamma$ target. From left to right: UA, AUA, AUA+Q}
    \label{fig:3crit_C2H6_triangle}
\end{figure}
\bibliographystyle{ieeetr}
\bibliography{paper}